\documentclass[a4paper,fleqn]{cas-dc}

\usepackage[authoryear,longnamesfirst]{natbib}

\usepackage{amssymb}
\usepackage{amsmath}
\newcommand{\dd}{\mathrm{d}}
\usepackage{float}
\usepackage[T1]{fontenc}
\usepackage[utf8]{inputenc}
\usepackage{subfig}
\usepackage{booktabs}
\usepackage{graphicx}
\usepackage{multirow}
\usepackage{subcaption}
\usepackage[export]{adjustbox}
\usepackage[leftcaption]{sidecap}

\def\tsc#1{\csdef{#1}{\textsc{\lowercase{#1}}\xspace}}
\tsc{WGM}
\tsc{QE}
\begin{document}





\author[1]{Máté István Boros}[
orcid=0009-0000-8347-6087
]

\cormark[1]


\ead{borosm@reak.bme.hu}


\affiliation[1]{organization={Budapest University of Technology and Economics, Institute of Nuclear Techniques},
            addressline={Műegyetem rakpart 3.}, 
            city={Budapest},
          citysep={}, 
            postcode={1111}, 
            country={Hungary}}

\author[1]{Máté Szieberth}





\author[1]{Gergely Klujber}





\author[2]{Imre Pázsit}

\author[1]{István Barth}

\author[3]{Yasunori Kitamura}
\author[3]{Tsuyoshi Misawa}

\affiliation[2]{organization={Chalmers University of Technology,  Department of Physics,  Division of Subatomic, High Energy and Plasma Physics},
            city={G\"oteborg},
          citysep={}, 
            postcode={SE-412 96}, 
            country={Sweden}}

\affiliation[3]{organization={Kyoto University,  Institute for Integrated Radiation and Nuclear Science},
            city={Osaka},
          citysep={}, 
            postcode={590-0494}, 
            country={Japan}}


\shorttitle{Feasibility demonstration of continuous signal-based neutron noise measurements by experiments and simulations}    

\shortauthors{Szieberth, Boros et al.}  

\title [mode = title]{Feasibility demonstration of continuous signal-based neutron noise measurements by experiments and simulations}  



%


\begin{abstract}
Neutron noise methods are used to determine kinetic parameters such as the prompt neutron decay constant, but traditional pulse‑counting suffers from dead‑time and pile‑up at high detection rates. Recent theory shows that analysing the continuous detector current can avoid these limitations if pulse‑shape effects are properly treated. This work presents a feasibility study of continuous‑signal neutron noise analysis based on simulations and experiments performed at two research reactors. The stochastic model of the detector current is applied to derive Rossi‑ and Feynman‑type formulations, and pulse-shape distortions are mitigated using detector pairs or by deconvolving the average pulse-shape through inverse Fourier and Wiener filtering. Simulations demonstrate accurate $\alpha$‑parameter estimation at count rates where pulse counting becomes unusable, and enable evaluation of significantly higher $\alpha$ values. Measurements at KUCA and BME TR confirm that continuous and deconvolved signals provide unbiased results despite dead‑time and electronic artifacts, establishing the method as a practical alternative for high‑rate reactor noise diagnostics.
\end{abstract}




\begin{keywords}
 neutron noise   \sep Rossi-alpha \sep autocorrelation\sep Feynman-alpha  \sep  variance-to-mean 
\end{keywords}

\maketitle


\section{Introduction}
\label{sec1}
Neutron noise techniques (also known as zero power noise) are based on the observation of the fluctuation of the number of neutrons in a subcritical or critical system due to the stochastic nature of the nuclear chain reaction \citep{pazpal08}. Feynman- and Rossi-alpha measurements, as the most important variants, are well-established methods for measuring the reactivity and kinetic parameters of quasi-zero-power reactors since the dawn of nuclear power \citep{feynman1956dispersion,orndoff1957prompt}.  A common characteristic of these techniques is the determination of higher moments of the distribution of detections, from which critical physical quantities can be derived. 

These measurement techniques are traditionally based on registering and evaluating detector pulses as individual detection events, which requires the discrimination of pulses due to neutrons from the continuous signal of the detector. In higher neutron flux environment the increased count rate results in detector-pulse pile-up. This introduces a dead-time effect, leading to lost counts and worse measurement statistics, and eventually rendering the traditional pulse-based counting unusable. 
While the detector dead time effect can be corrected to obtain the real count rate, i.e. the mean value, it is not possible for the higher moments of the distributions, which are involved in the noise methods. Therefore the correction is very complicated if feasible since short-time correlations are affected more significantly than more extended-time correlations. 
In the case of the Rossi-alpha method the dead time results in the loss of counts for the lag times shorter than the typical dead time, while for lag times above that the results are not affected. For the Feynman-alpha method a more sophisticated methodology can be applied  described e.g. in \cite{Calle}  based on the method developed by  \cite{Hazama_deadtime}. Nevertheless, data points belonging to gates with width shorter or close to the dead time, cannot be evaluated.

Due to this nature of the dead time effect it also limits the magnitude of the prompt decay constant $\alpha$ that can be measured, even if the count rate is sufficiently low. This poses further limitation for cases where the determination of high alpha values is necessary. Trivial examples are the fast reactor systems where the critical alpha value is higher than in thermal systems. Further examples are the subcritical systems since the prompt decay constant linearly increases with the subcriticality level and deep subcriticality mean very fast prompt neutron decay time. Furthermore, in source driven systems the higher $\alpha$-modes excited by the external source are also present, which can be orders of magnitude higher than the fundamental $\alpha$ \citep{Delphi_ANucEne}. Since higher modes may introduce a bias to the measurement of the fundamental mode alpha, their accurate measurement is also required to reduce this bias. This is especially important in the case of Accelerator Driven Systems, where neutron noise measurements are considered as a reactivity calibration method.

Although the higher modes die away in stationary critical systems, they appear during fast transients when the flux shape change must be considered, which provides a further incentive for their accurate measurement. The higher modes can describe the spatial and transient behavior of the reactor on such a short time scale (less than 1 ms) for which measurements with any spatial resolution are very complicated, if feasible. 

The dead-time effect primarily restricts the power level where noise methods can be applied,  which is a limitation for the industrial application of the zero power neutron noise methods. 

A practically dead-time free methodology not only opens the way to measurements at higher power, but is crucial to improve the accuracy and reliability of the method by revealing the information carried by the higher alpha values and higher modes even for zero power measurements. 

The present paper investigates a proposed solution to this issue, which is to use the raw continuous voltage signal from the fission chambers.  This approach is inherently resistant to dead-time effects and therefore allows application at higher reactor power. 

The theoretical background of the proposed method is founded on a simple, but very powerful analytical model of the stochastic signals of fission chambers  suggested by \cite{PAL2014}. The model accounts for the random, uncorrelated detection of neutrons, following a simple Poisson statistics, as well as for the random character of the individual pulse shapes for each detection. Originally this model was suggested for a simple and transparent derivation of the traditional (second order) and higher order Campbell formulas, which were treated earlier by an different approach \citep{lux1982higher}. However, soon it became clear that the model was also suitable for generalizing it to account for the more advanced scenario of correlated detector counts. This way stochastic methods of reactivity measurements, as well as multiplicity counting, which both have been made traditionally by counting discrete pulses, could also be achieved by analyzing the continuous signals of fission chambers. One potential advantage of using continuous signals in the above applications is that it would be free from the problems of the dead time. 

To extend the original model of \cite{PAL2014} to derive the analogue of the Feynman- and Rossi-alpha methods of reactivity measurements from continuous detector signals, one had to abandon the assumption of particles arriving to the detector following Poisson statistics. Instead, similarly to the starting point of the pulse counting methods, a multiplying subcritical medium driven by an extraneous Poissonian source was assumed, and the detections were assumed to be made from the chains induced by the source. Calculation of the auto-covariance of the detector signal so induced, which is an analogue of the traditional Rossi-alpha formula, showed that the prompt neutron decay constant $\alpha$ was possible to extract. This extension was first made with a simple extraneous Poisson source and without delayed neutrons \citep{PAL201590}, then by a compound Poisson source for both the the Rossi- and Feymnan-alpha methods but still without delayed neutrons \citep{KITAMURA2018691}, and finally with the inclusion of delayed neutrons \citep{Kitamura2019}.

Using fission chambers in the current mode to estimate reactor kinetic parameters via neutron noise measurements has previously been investigated by others, with promising results using the Power Spectral Density (PSD) method \citep{SPECTRON,CROCUS}. The main difference in our approach is the use of the continuous signal-applicable form Rossi- and Feynman-$\alpha$ methods and the application of higher sampling rates to resolve the detector pulse-shape, enabling advanced signal processing and allowing the measurement of higher $\alpha$ values relevant to higher-order kinetic modes and fast systems. We have also conducted detailed simulation-based investigations into the measurement conditions under which the method based on the continuous detector signal is feasible and provides an advantage over pulsed-based methods. 

The aim of this paper is to assess the practical feasibility of continuous‑signal neutron noise analysis through simulations and experiments. We focus on the usable detection‑rate range, the achievable $\alpha$‑values, and the impact of electronic noise and frequency‑transfer characteristics. The results show that continuous‑signal methods remain reliable under conditions where pulse counting fails, and that pulse‑shape deconvolution further extends their applicability.

\section{Theoretical background}

A simple theory of continuous signals of fission chambers, which provides an elegant derivation of the classic Campbell theorems for uncorrelated detection events was elaborated by \cite{PAL2014}. 
In the formalism, the detector pulse-shape was written as the product of a deterministic shape $f(t)$ and a random amplitude $\eta$ with probability distribution $w(\eta)$. 
In this way, the probability density function of the pulse value $u$ at time $t$ is written as: 

\begin{equation}
h(u, t) = \int_0^\infty \delta[u - \eta \, f(t)] \, w(\eta) \, {\dd}\eta.
\label{eq:pulse_PDF}
\end{equation}

The formalism was thereafter extended for correlated detection events from a fission chain by \cite{PAL201590},
which is true for multiplying systems and is the essence of noise techniques. 

\begin{figure}
    \centering
    \includegraphics[width=0.9\columnwidth]{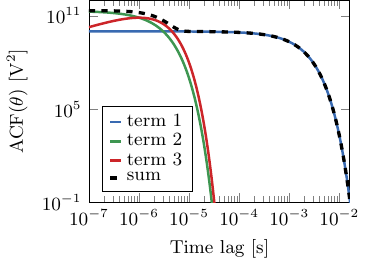}
    \includegraphics[width=0.9\columnwidth]{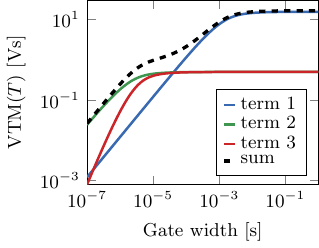}
    \caption{The ACF and VTM functions of the continuous detector signal together with the additional terms  caused by the finite decay speed of the detector pulses.}
    \label{fig:AC_VTM_with_pulse_effect}
\end{figure}

With these assumptions, the auto-covariance (ACF) and variance-to-mean ratio (VTM) functions were derived for the continuous signal, which asymptotically matched the well-established Rossi-$\alpha$ and Feynman-$\alpha$ formulas known for discrete, pulse-based signals. However, the ACF includes additional terms, containing the detector pulse decay constant ($\alpha_e$), which also distorts the VTM function. Specifically, for a given pulse-shape $f(t)\sim t\cdot \exp(-\alpha_e\cdot t)$, the ACF and the VTM of the detector signal in a multiplying system as of \cite{KITAMURA2018691}, read as

\begin{equation}
\begin{split}
    \text{ACF}(\theta)&=\phi\cdot \exp(-\alpha\cdot|\theta|)+\psi_1\cdot \exp(-\alpha_e\cdot|\theta|)\\
    &+\psi_2\cdot|\theta|\cdot \exp(-\alpha_e\cdot|\theta|)\\    
    \text{VTM}(T)&=\Phi\cdot f_1(\alpha\cdot T)+\Psi_1\cdot f_1(\alpha_e\cdot T)\\
    &+\Psi_2\cdot f_2(\alpha_e\cdot T)\\
    f_1(x)&=1-\frac{1-\exp(-x)}{x}\\
    f_2(x)&=1+\exp(-x)-2\cdot\dfrac{1-\exp(-x)}{x}
    \label{eq:AC_VTM_with_pulse_effect}
\end{split}
\end{equation}
The full forms of the ACF and VTM functions, including the extra terms corresponding to the pulse-shape are shown on Figure \ref{fig:AC_VTM_with_pulse_effect}.
Luckily, in most cases, the fundamental prompt decay constant of the multiplying system is by several orders of magnitude slower than the decay constant of the detector pulse ($\alpha\ll\alpha_e$), meaning that these distortion effects are significant only in a very short initial section of the relevant region of these functions, and these short sections can simply be ignored without significantly impacting the accuracy of the evaluation. However, when trying to determine higher-order $\alpha$-modes of the system, $\alpha_e$ and the sought $\alpha$ value can become more comparable, causing problems.

As the simulation results presented in the following section show, the distortion of the detector pulse-shape can be mitigated when using pairs of detectors and computing the cross-covariance (CCF) and cross-covariance to mean ratio (CTM) functions of the two signals.

Another possible way of mitigating the problem caused by the finite time constant of the detector pulse is by the deconvolution of the average pulse-shape from the signal. If the assumption that the pulses can be written as the product of the deterministic average pulse-shape $f(t)$ and a random amplitude $\eta$, then the continuous signal is the convolution of the average pulse-shape and Dirac delta pulses:
\begin{figure}
    \centering
    \includegraphics[width=0.9\columnwidth]{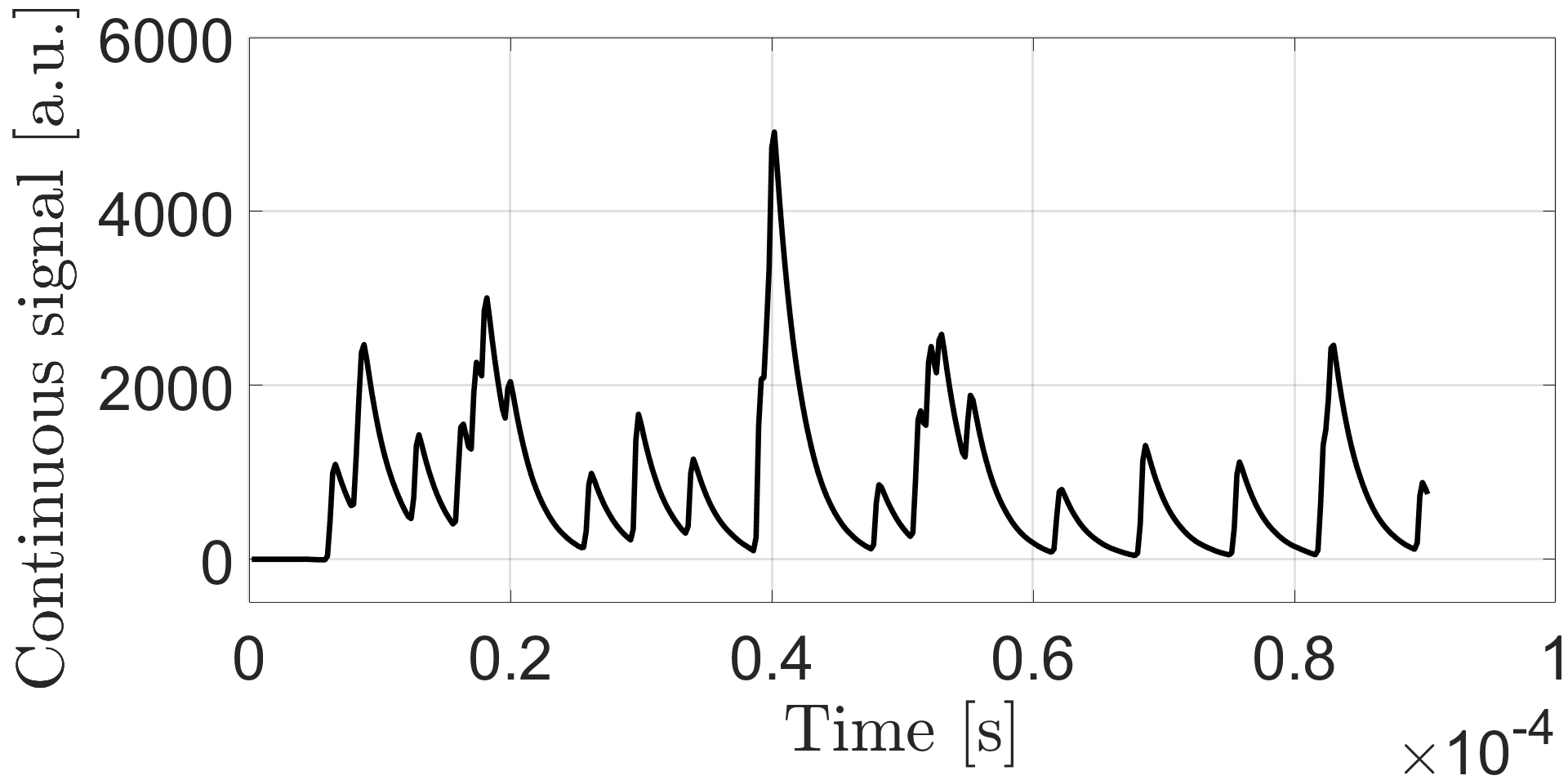}
    \includegraphics[width=0.9\columnwidth]{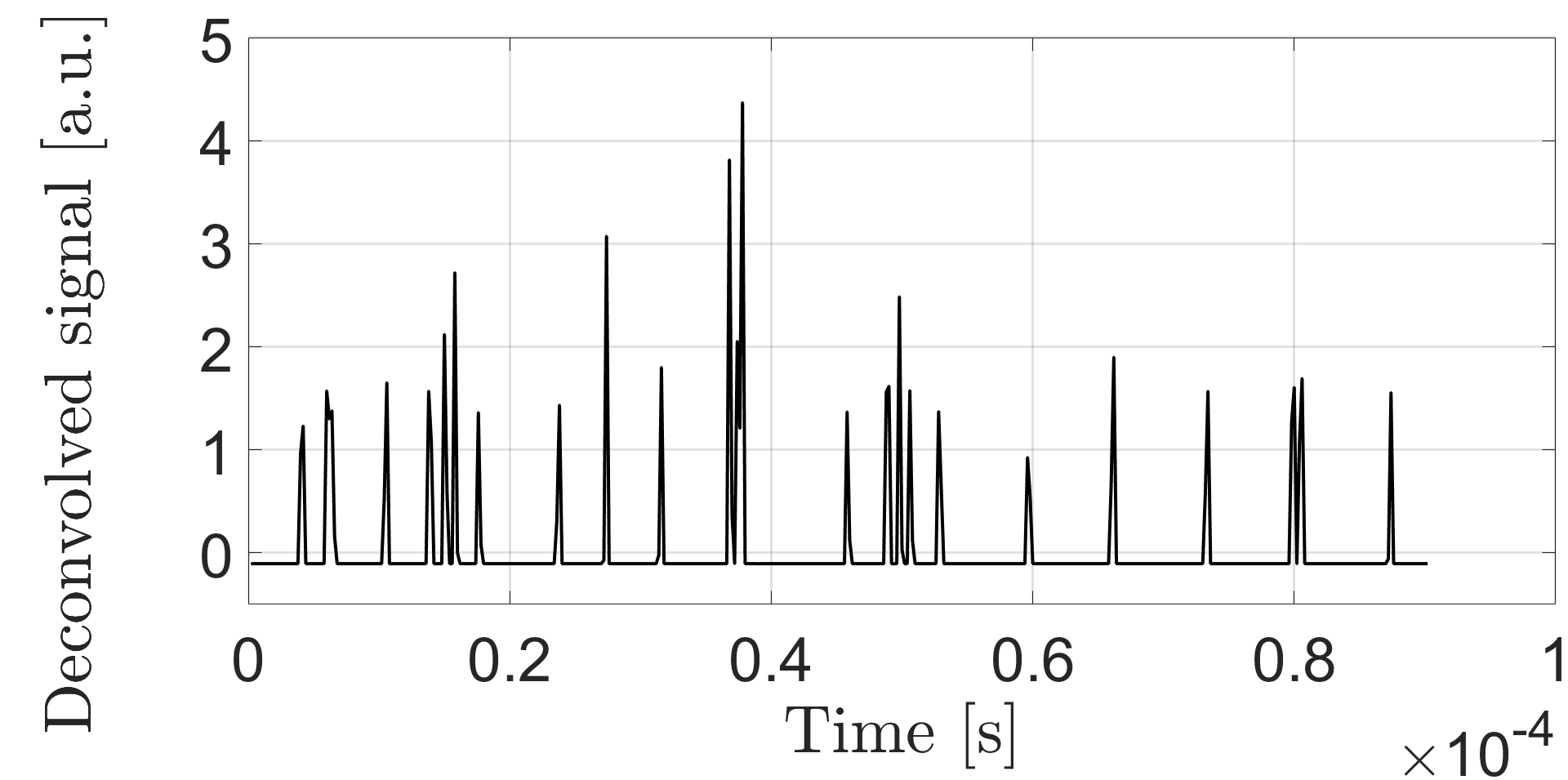}
    \caption{Short section of a simulated detector signal before and after the deconvolution of the average pulse-shape.}
    \label{fig:deconv_noiseless}
\end{figure}
\begin{equation}
\begin{split}
    c(t)&=\sum_{n=1}^N\eta_n\cdot f(t-t_n)\\
    &=\int_{-\infty}^{\infty}f(t')\cdot\left(\sum_{n=1}^N\eta_n\cdot\delta(t-t_n-t')\right)\dd t'\\
     &=\sum_{n=1}^N\eta_n\cdot\int_{-\infty}^{\infty}f(t')\cdot\delta(t-t_n-t')\dd t'\\
     &=\sum_{n=1}^N\eta_n\cdot[f*\delta](t-t_n)
\end{split}
\end{equation}
Here, there are $N$ detected neutrons during the measurement period, and $t_n$ is the time of the detection of the $n$th neutron. The average pulse-shape $f(t)$ can easily be measured, therefore it can be assumed to be known. In that case, we can deconvolve it from the recorded continuous signal, which can be performed in the Fourier space:
\begin{equation}
   \mathcal{F}^{-1} \left\{ \frac{\mathcal{F}\left\{c\right\}}{\mathcal{F}\left\{f\right\}}\right\}=\sum_{n=1}^N\eta_n\cdot\delta(t-t_n)
\end{equation}
This, without the $\eta_n$ random amplitudes would be the ideal form of the continuous signal, as ideally, we would get an infinitely short pulse at the time of each neutron detection. The integral of this ideal continuous function would simply be the counting function of the neutron detection events.

The effect of the deconvolution of the average pulse-shape from the continuous signal in the ideal case is illustrated in Figure \ref{fig:deconv_noiseless}.
\section{Simulations}
To determine and compare the domain of usability of both the traditional pulse-based and continuous signals, simulations of measurements were performed. The simulations were based on the Training Reactor of the Budapest University of Technology and Economics (BME TR) \cite{BMETR}, and were performed using a simple in-house program implementing a 1-speed point Monte Carlo model of a subcritical multiplying system and an external source.

Results of the different signal types were compared while increasing the detection rate through the intensity of the external source, and while increasing the $\alpha$ theoretical prompt decay constant. The $\alpha$ value was changed by increasing all the $\lambda_x$ intensities of all the reaction types, which can be interpreted as changing the speed of the neutrons.

\subsection{The simulation of detector signals}
\begin{figure}
    \centering
    \includegraphics[width=\columnwidth]{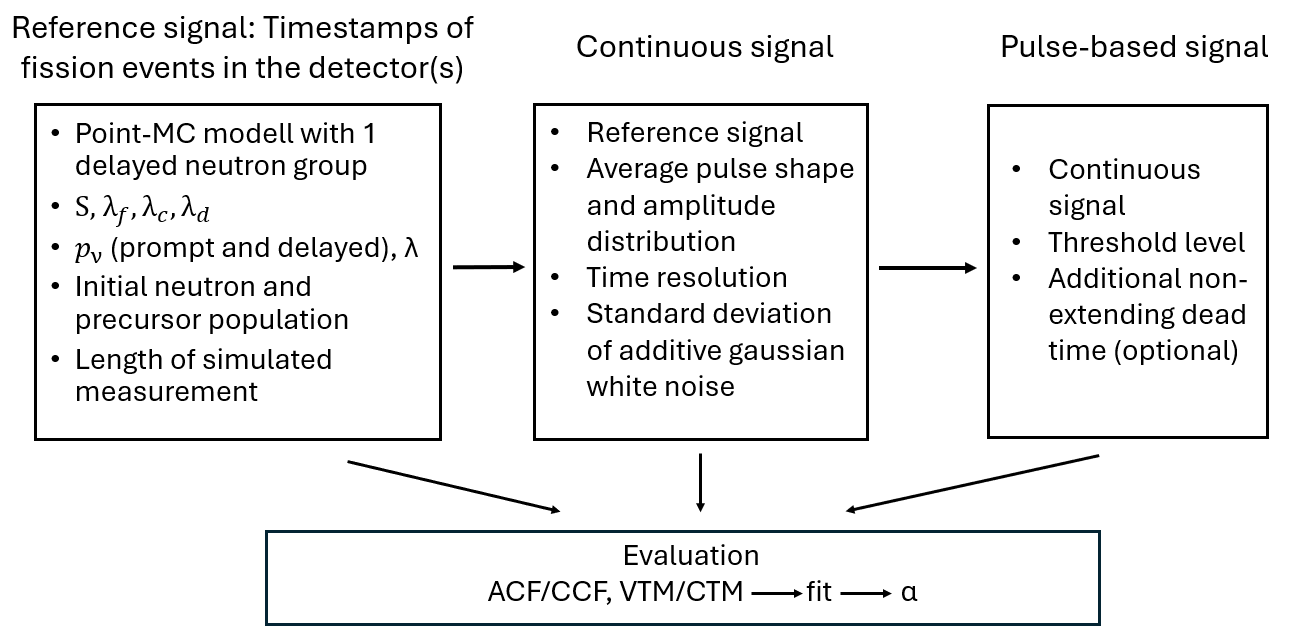}
    \caption{The process of simulating the different signal types.}
    \label{fig:sim_process}
\end{figure}

First, the timestamps of each neutron detection event (a fission reaction in the fission chamber detector) were produced by the Monte Carlo simulation. Then, using previously measured $f(t)$ average pulse-shapes and $w(\eta)$ amplitude (maximum voltage value) distributions, the continuous voltage signal of the detector was produced with some additive Gaussian noise component. Finally, using this continuous signal, the pulse-based signal was produced using an appropriately chosen threshold level. Figure \ref{fig:sim_process} shows the schematic of the simulation process of the different signal types.
The fission intensity $\lambda_f$ was obtained from the MCNP model \citep{MCNP6} of the BME Training Reactor. The model yielded a generation time of $\Lambda\approx 6.9\cdot10^{-5}$ s, from which the fission intensity is $\lambda_f=\frac{1}{\Lambda\cdot\Bar{\nu}}\approx5870\text{~s}^{-1}$.
\begin{table}
\caption{The $p_\nu$ probability mass function of fission neutron multiplicity ($\nu$) used for the simulations, and the $\Bar{\nu}$ expected value. The data was taken from \cite{Kitamura2019}.}
\centering
\begin{tabular}{|l|l|lllll}
\hline
$\nu$    & 0        & \multicolumn{1}{l|}{1}        & \multicolumn{1}{l|}{2}        & \multicolumn{1}{l|}{3}        & \multicolumn{1}{l|}{4}        & \multicolumn{1}{l|}{5}        \\ \hline
$p_\nu$ & 0,0341 & \multicolumn{1}{l|}{0,1645} & \multicolumn{1}{l|}{0,3177} & \multicolumn{1}{l|}{0,3069} & \multicolumn{1}{l|}{0,1482} & \multicolumn{1}{l|}{0,0286} \\ \hline
$\Bar{\nu}$ & 2,4564 &                               &                               &                               &                               &                               \\ \cline{1-2}
\end{tabular}
\end{table}
Delayed neutrons (precursors) were omitted from the simulations. This way, the simulated system at steady-state acts as a critical system, the constant-intensity external source replacing the role of the delayed neutron precursors at equilibrium population.

\begin{figure}
\centering
    \subfloat[Average pulse-shape.]{\includegraphics[width=0.49\columnwidth]{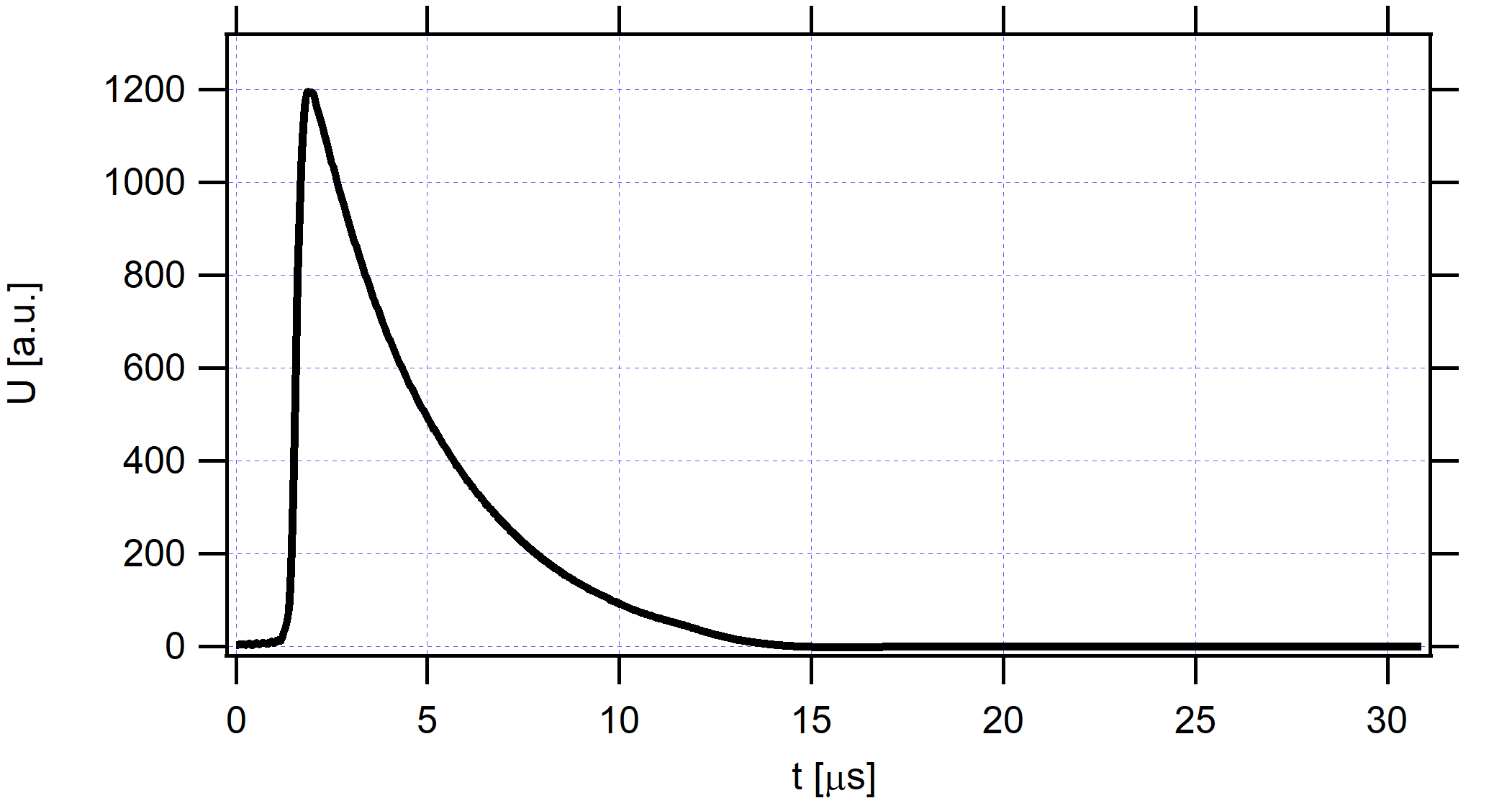}}
    \subfloat[Probability density of the amplitude.]{\includegraphics[width=0.49\columnwidth]{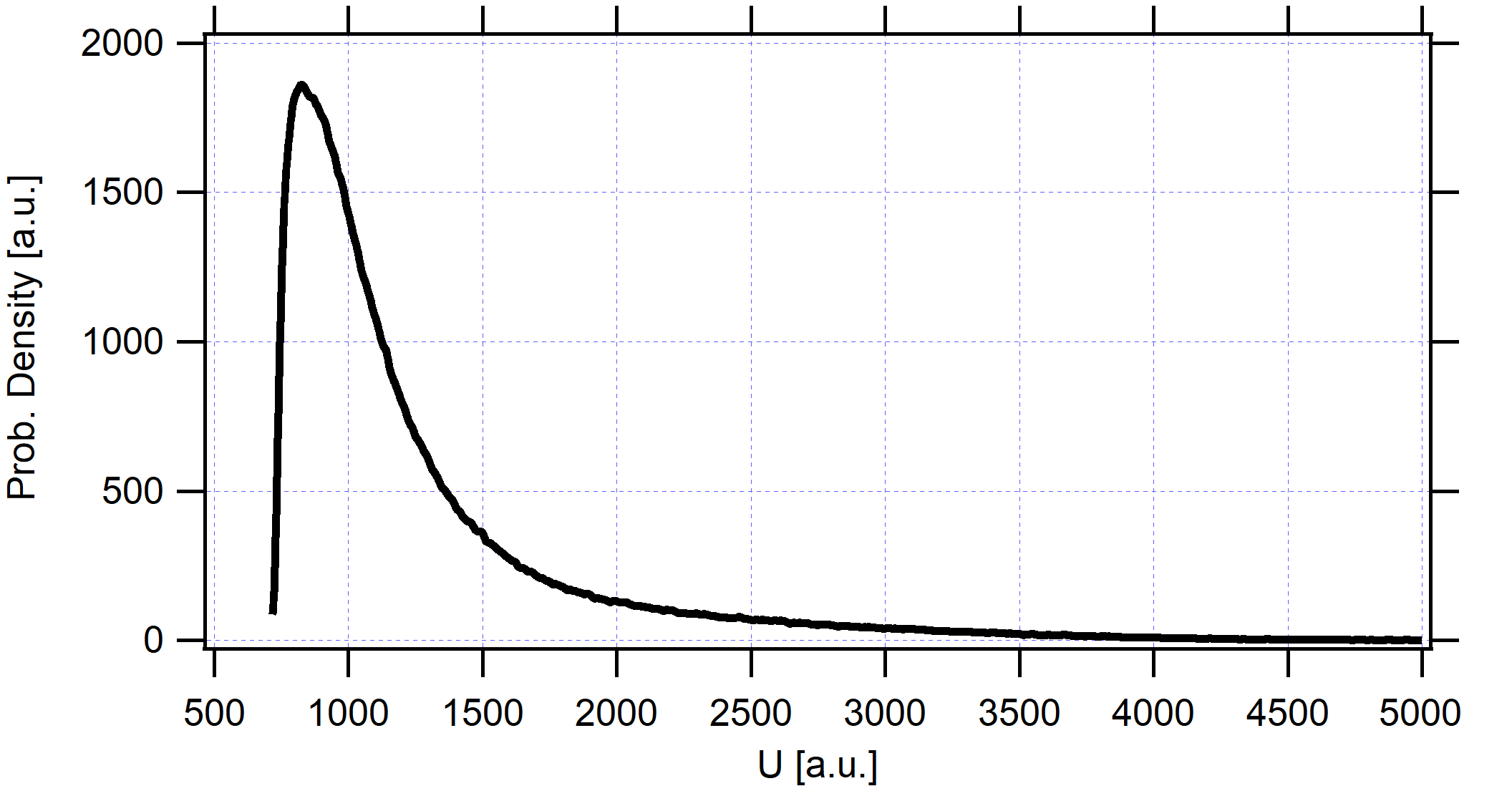}}
    \caption{The average pulse-shape and the the probability density function (without normalization) of the amplitude used for the simulations (KNT-31 fission chamber).}
    \label{fig:Sim_shape_ampl_distr}
\end{figure}

Figure \ref{fig:Sim_shape_ampl_distr} shows the $f(t)$ average pulse-shape and the $w(\eta)$ amplitude distribution, taken from older measurements \citep{avershape} using KNT-31 fission chambers at the BME TR.
For relatively low detection rate scenarios, the threshold for pulse detection was set to 715 units, as 715 units was the lowest possible amplitude of a pulse. Generally, it is beneficial to set the threshold as high as possible without losing counts, as the effective width of pulses is smaller for higher threshold levels. In higher detection rate cases, when the pile-up of pulses became significant, this threshold was raised to a level where the number of detected pulses reached a maximum value.
\subsection{The use of detector pairs}
\begin{figure*}
    \centering
    \subfloat[Continuous Rossi-evaluation of a single detector.]{\includegraphics[width=0.45\textwidth]{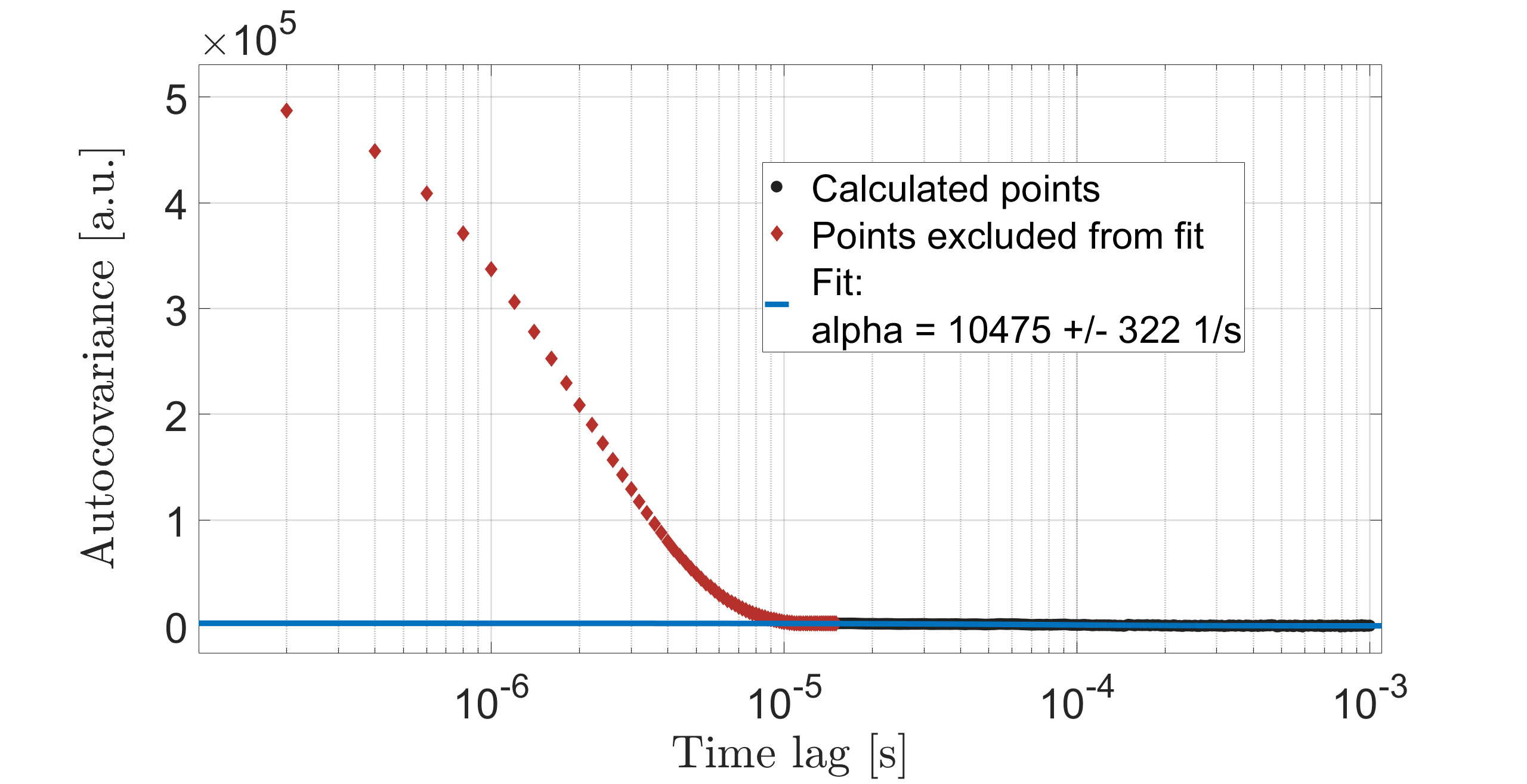}}
    \subfloat[Continuous Rossi-evaluation of a detector pair.]{\includegraphics[width=0.45\textwidth]{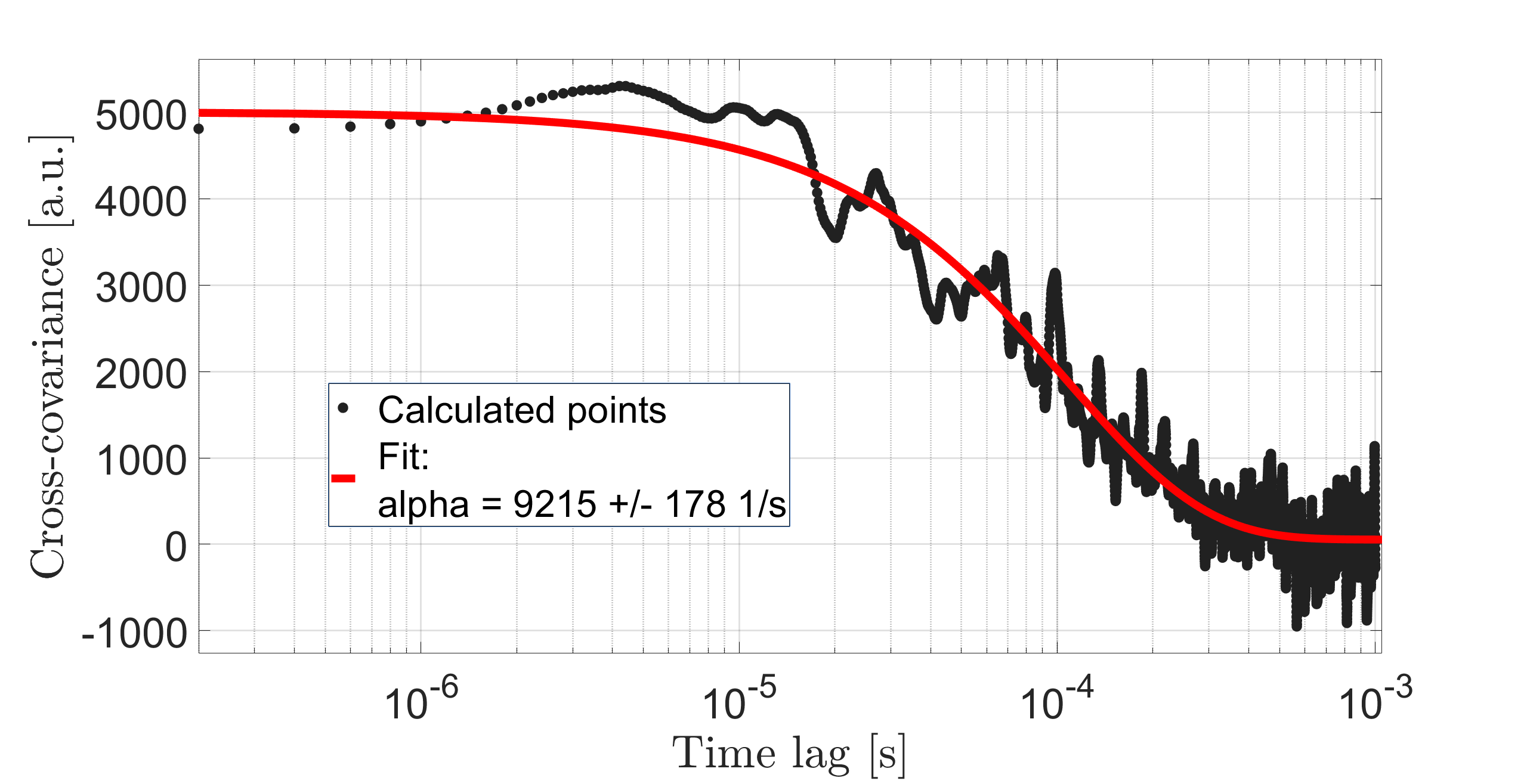}}\\
    \subfloat[Continuous Feynman-evaluation of a single detector.]{\includegraphics[width=0.45\textwidth]{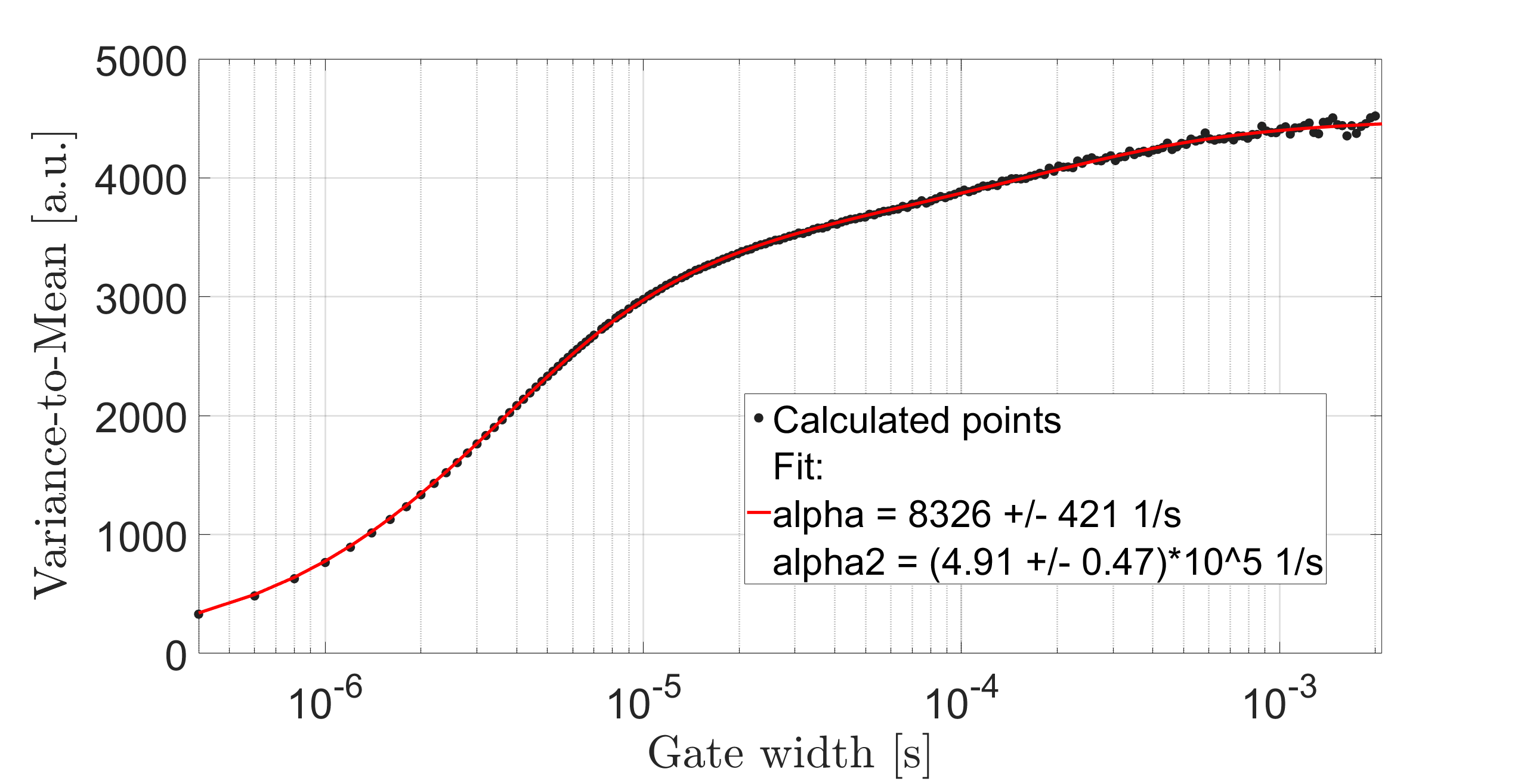}}
    \subfloat[Continuous Feynman-evaluation of a detector pair.]{\includegraphics[width=0.45\textwidth]{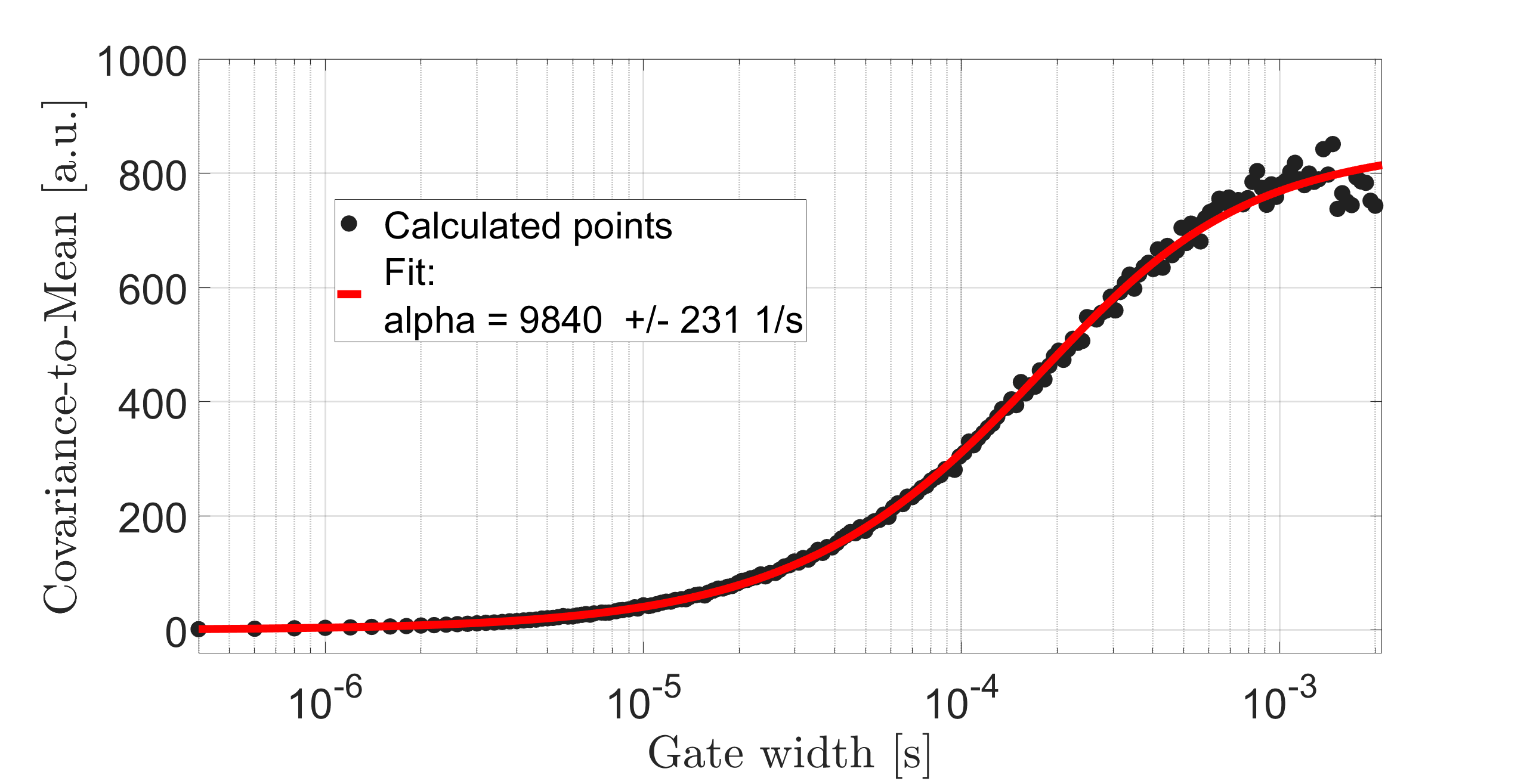}}\\
    \subfloat[Pulse-based Rossi-evaluation of a single detector.]{\includegraphics[width=0.45\textwidth]{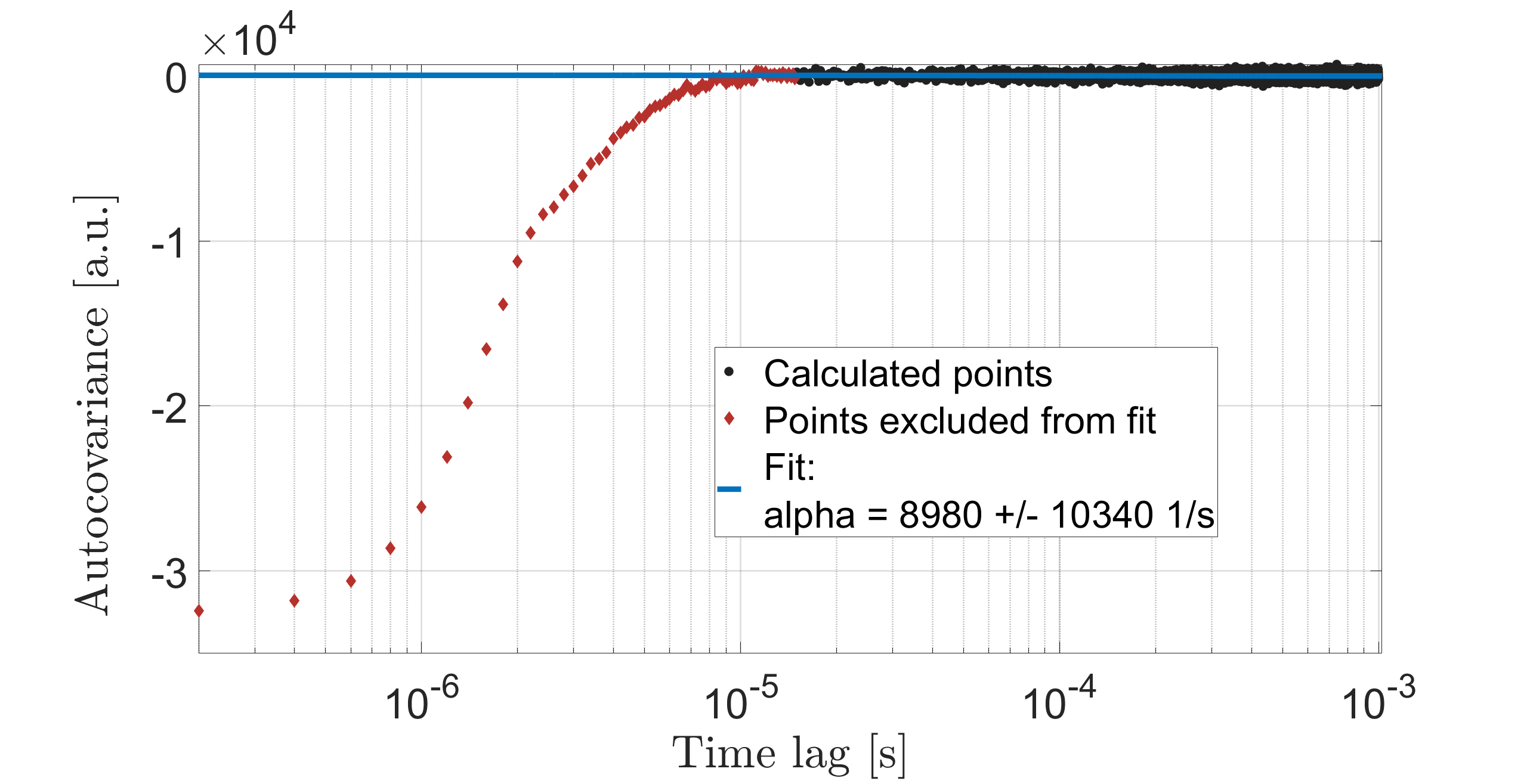}}
    \subfloat[Pulse-based Rossi-evaluation of a detector pair.]{\includegraphics[width=0.45\textwidth]{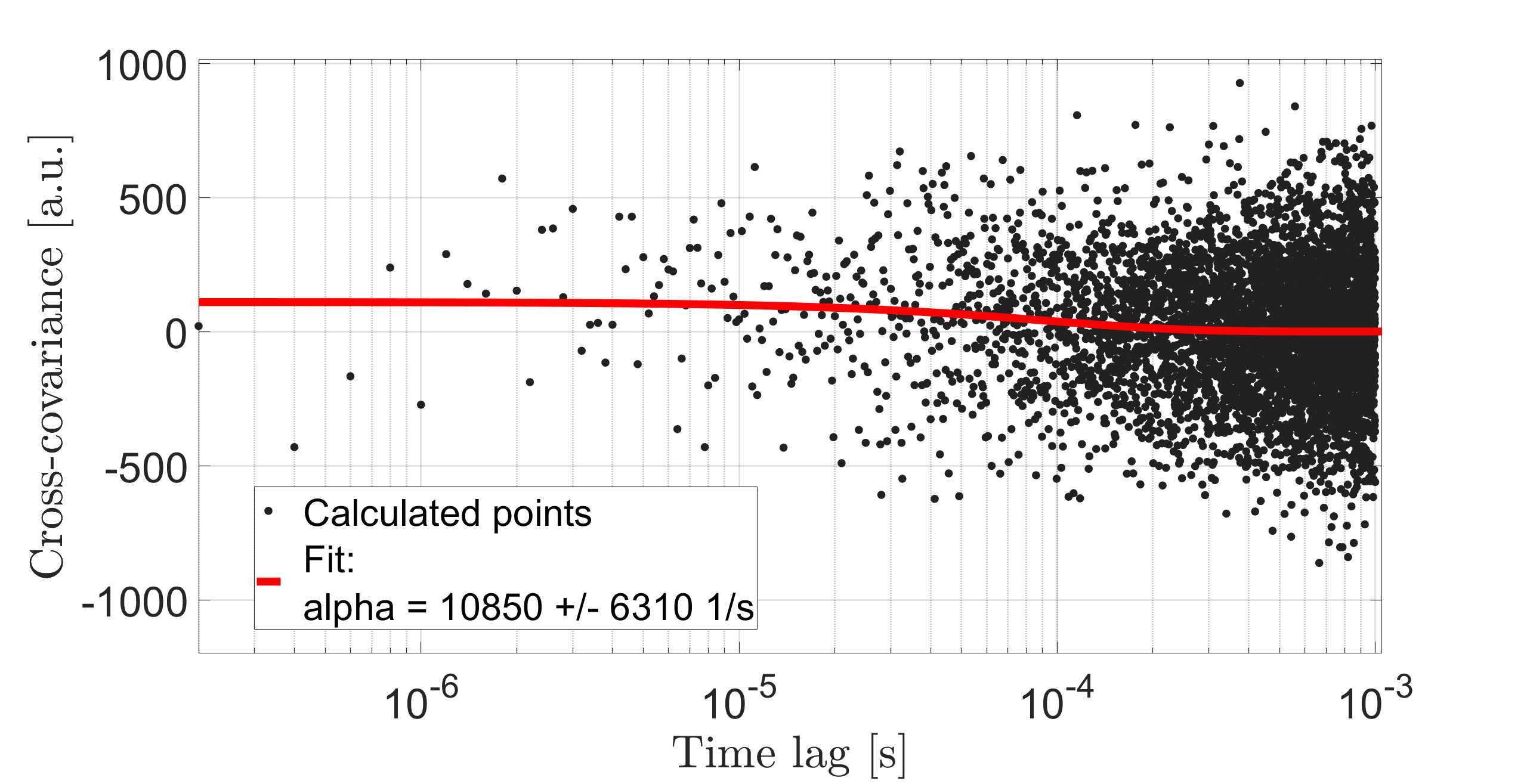}}\\
    \subfloat[Pulse-based Feynman-evaluation of a single detector.]{\includegraphics[width=0.45\textwidth]{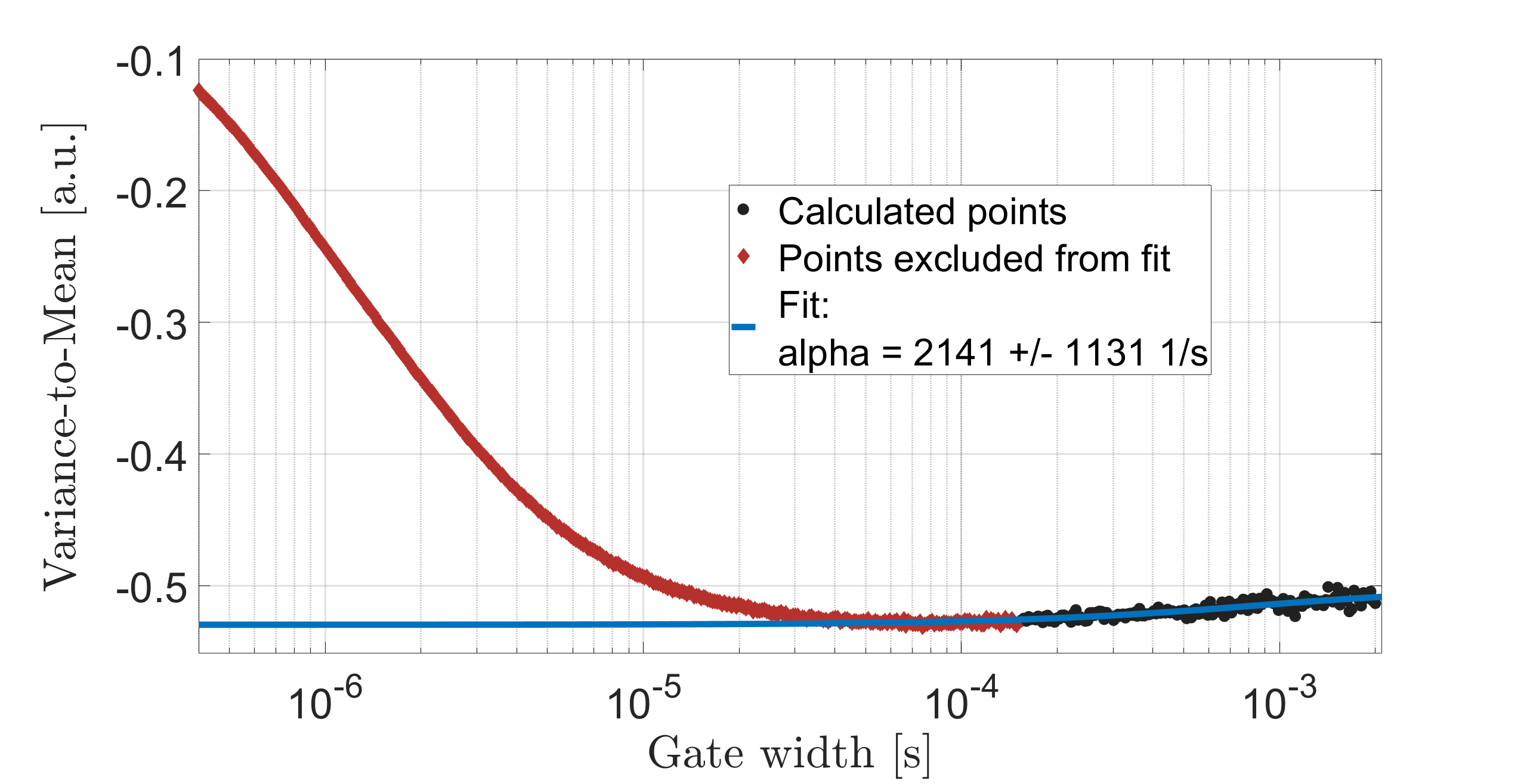}}
    \subfloat[Pulse-based Feynman-evaluation of a detector pair.]{\includegraphics[width=0.45\textwidth]{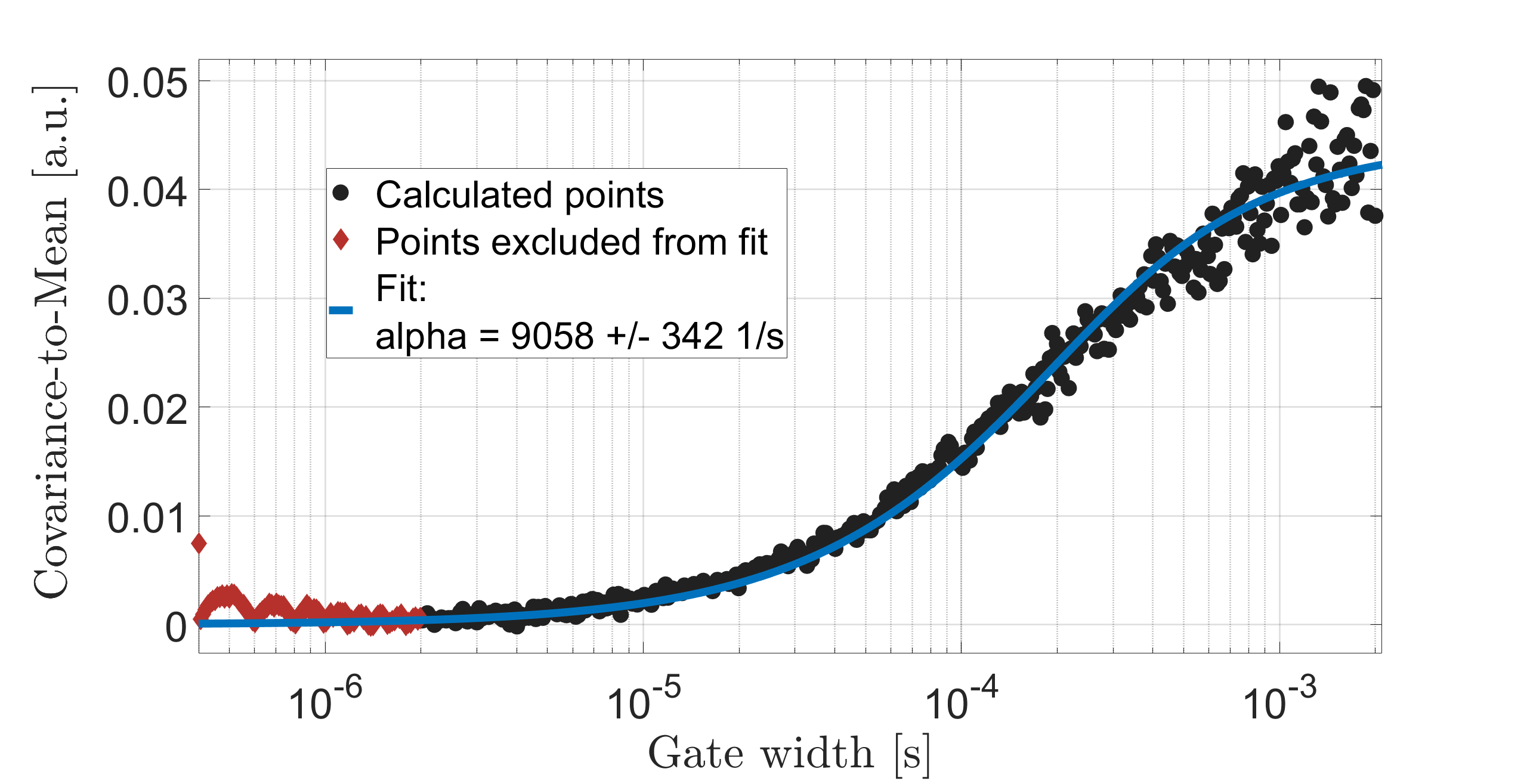}}
    \caption{Rossi-$\alpha$ and Feynman-$\alpha$ evaluations of the continuous and pulse-based detector signal, using a single detector and a pair of detectors. The theoretical $\alpha$ value of the system is 10000~s$^{-1}$, and the detection rate is 100000~neutrons/s/detector.}
    \label{fig:onevs2det}
\end{figure*}
Using pairs of detectors instead of a single one is a well-known enhancement of Rossi-$\alpha$ and Feynman-$\alpha$ techniques. The benefits of using detector pairs are even more apparent when using the continuous voltage signal, as the distortion effect of the finite pulse width can be negated this way.

The benefit of using detector pairs for the evaluation of both the continuous and pulse-based signals is illustrated in Figure \ref{fig:onevs2det}, with a relatively high theoretical $\alpha$ value and an average detection rate of 100000~neutrons/s/detector. In the case of the variance-to-mean function, the sum of two traditional Feynman-$\alpha$ terms with different $\alpha$ parameters produced a high quality fit. This yielded a second, higher $\alpha$ value introduced by the decay of the detector pulses.

The use of detector pairs successfully suppresses the effect of the finite pulse width of the pulses for low time lag and gate width, leading to higher quality evaluations and more accurate estimations of the real $\alpha$ value of the system.

Similarly to the continuous signal, the use of detector pairs effectively mitigates the problem of losing detections of the correlated neutrons coming in quick succession because of the dead time (resulting in negative correlations for low time lags in the autocovariance function). Still, the continuous signal provides more accurate results than the pulse-based one, as some counts are still lost due to the dead time yielding worse statistics than the continuous signal.

\subsection{Increasing the detection intensity}
One of the main goals of this research is to demonstrate the usability of the continuous signal for higher detection rates compared to the traditional pulse-based signal. In the simulations, the detection rate was varied by changing the intensity of the external source in the subcritical system. This source intensity was changed between $5\cdot10^6\text{~s}^{-1}$ and $2\cdot10^8\text{~s}^{-1}$ while keeping all other system parameters the same, resulting in a theoretical $\alpha\approx107.2$~s$^{-1}$. Higher source intensities have not been tested because the computation time required for both the Monte Carlo simulation and the evaluation of the pulse-based and reference signals became too long. The ratio between the detected neutrons per detector and the source neutrons was approximately $\lambda_d/\alpha\approx0.01$ detected neutron/source neutron.
\begin{figure*}
    \centering
    \subfloat[Rossi-evaluation of the reference signal.]{\includegraphics[width=0.4\textwidth]{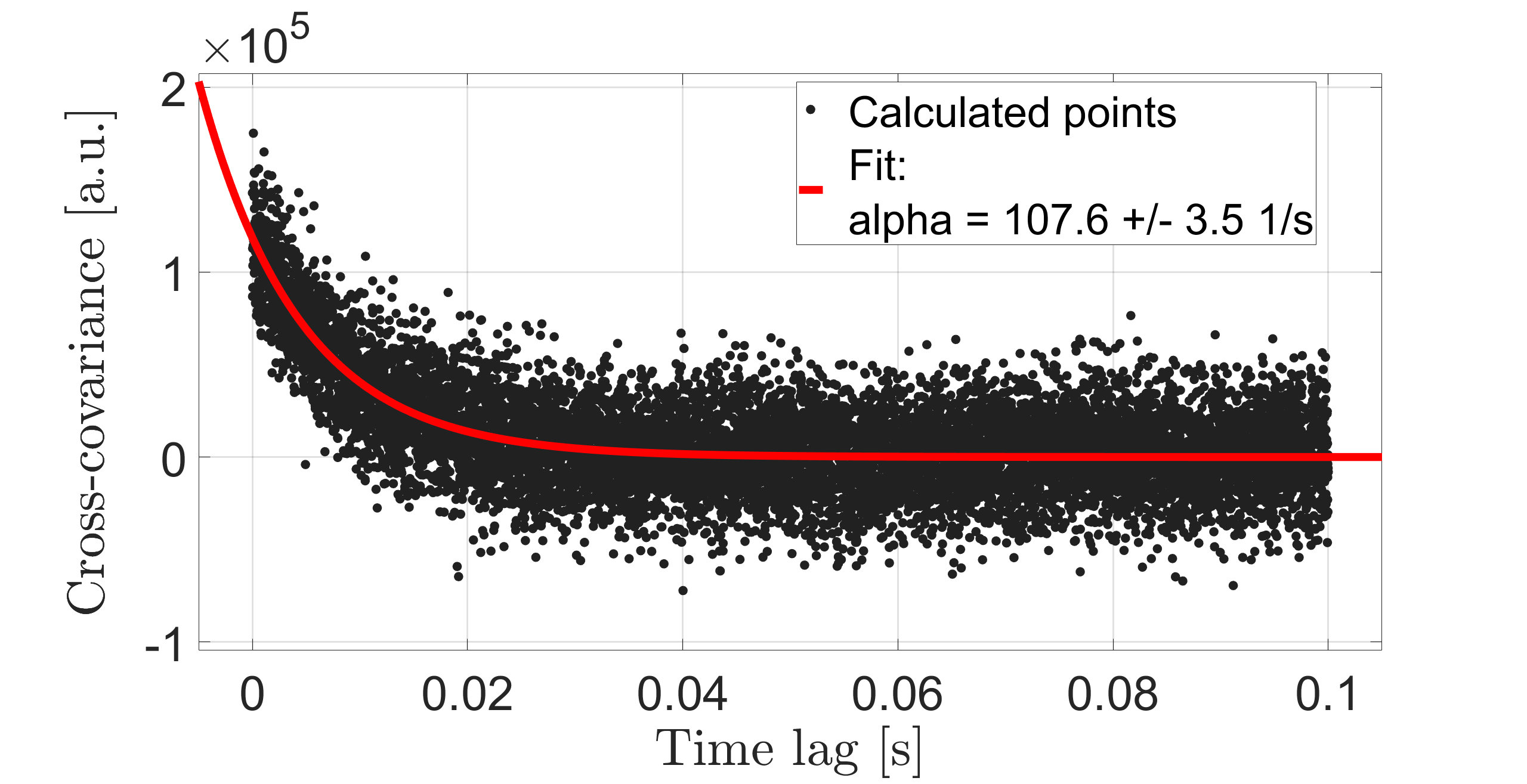}}
    \subfloat[Feynman-evaluation of the reference signal.]{\includegraphics[width=0.4\textwidth]{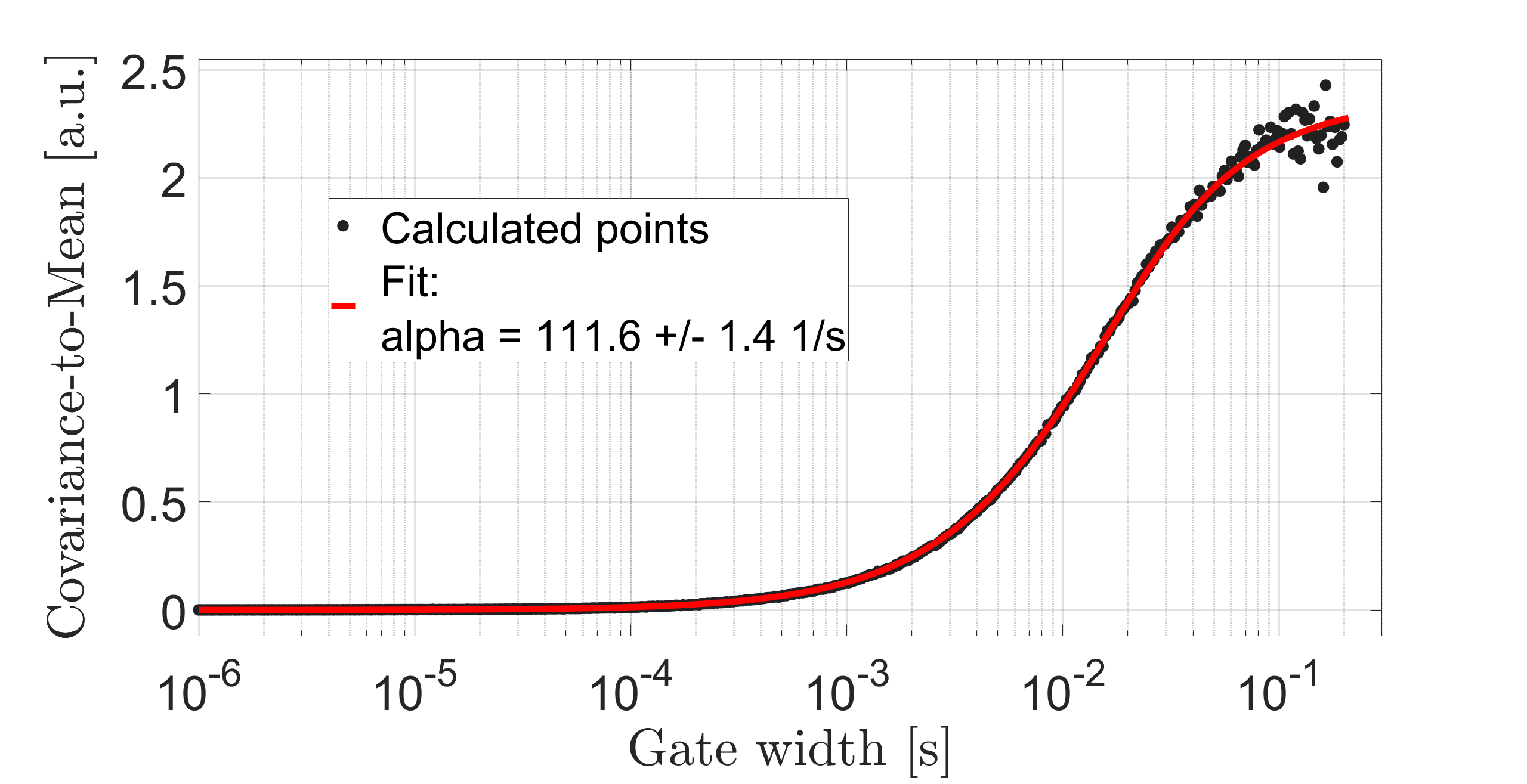}}\\
    \subfloat[Rossi-evaluation of the pulse-based signal.]{\includegraphics[width=0.4\textwidth]{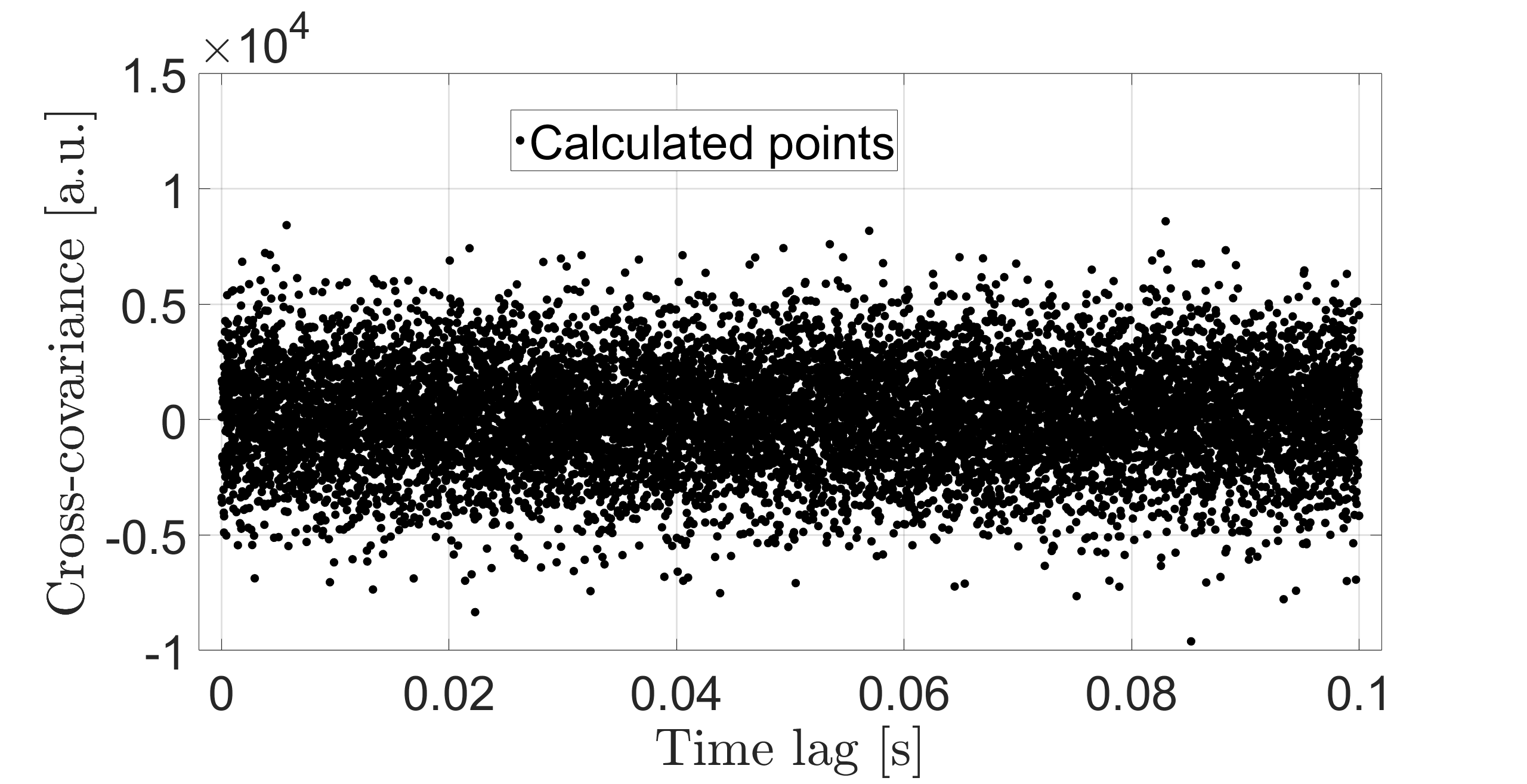}}
    \subfloat[Feynman-evaluation of the pulse-based signal.]{\includegraphics[width=0.4\textwidth]{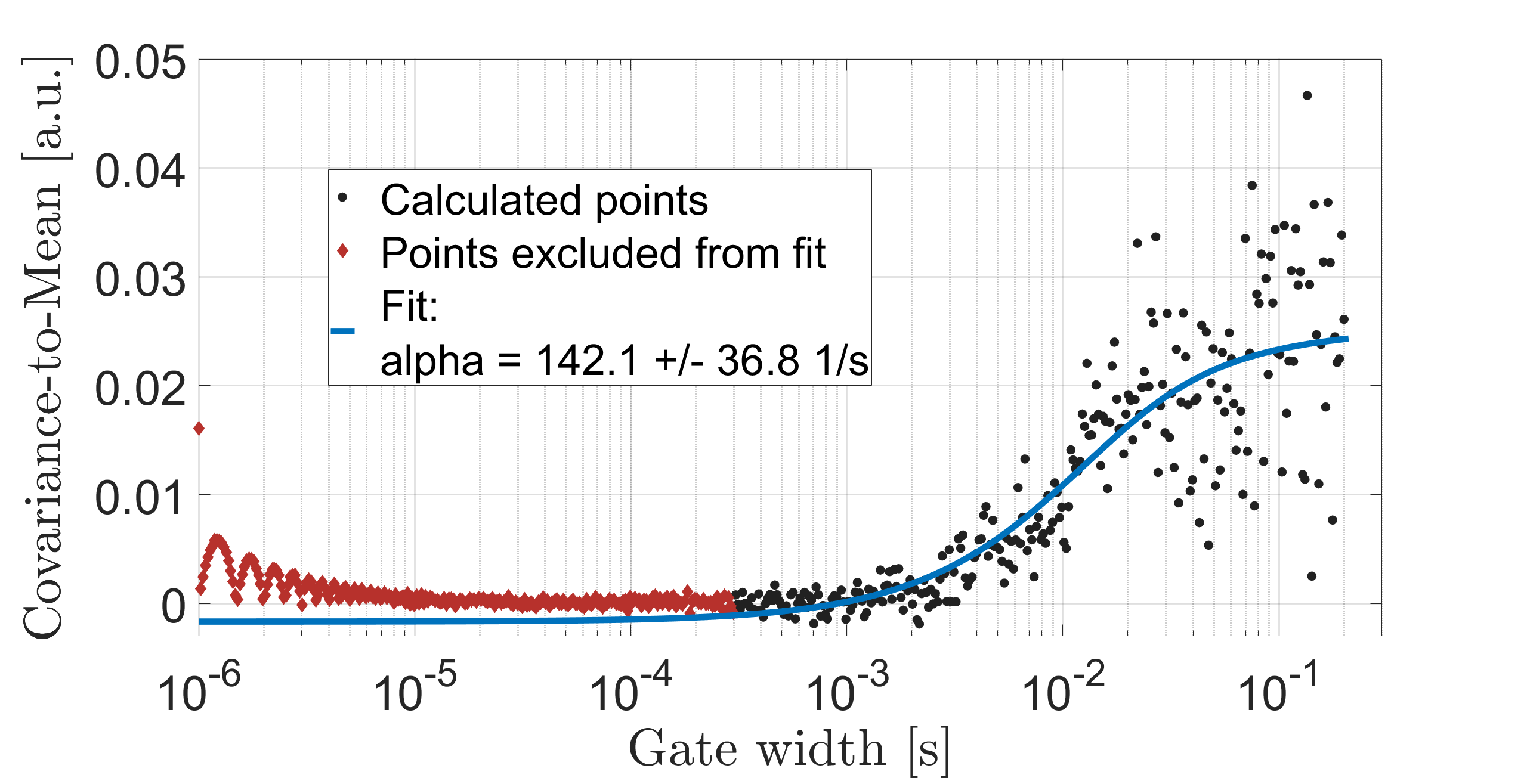}}\\
    \subfloat[Rossi-evaluation of the continuous signal.]{\includegraphics[width=0.4\textwidth]{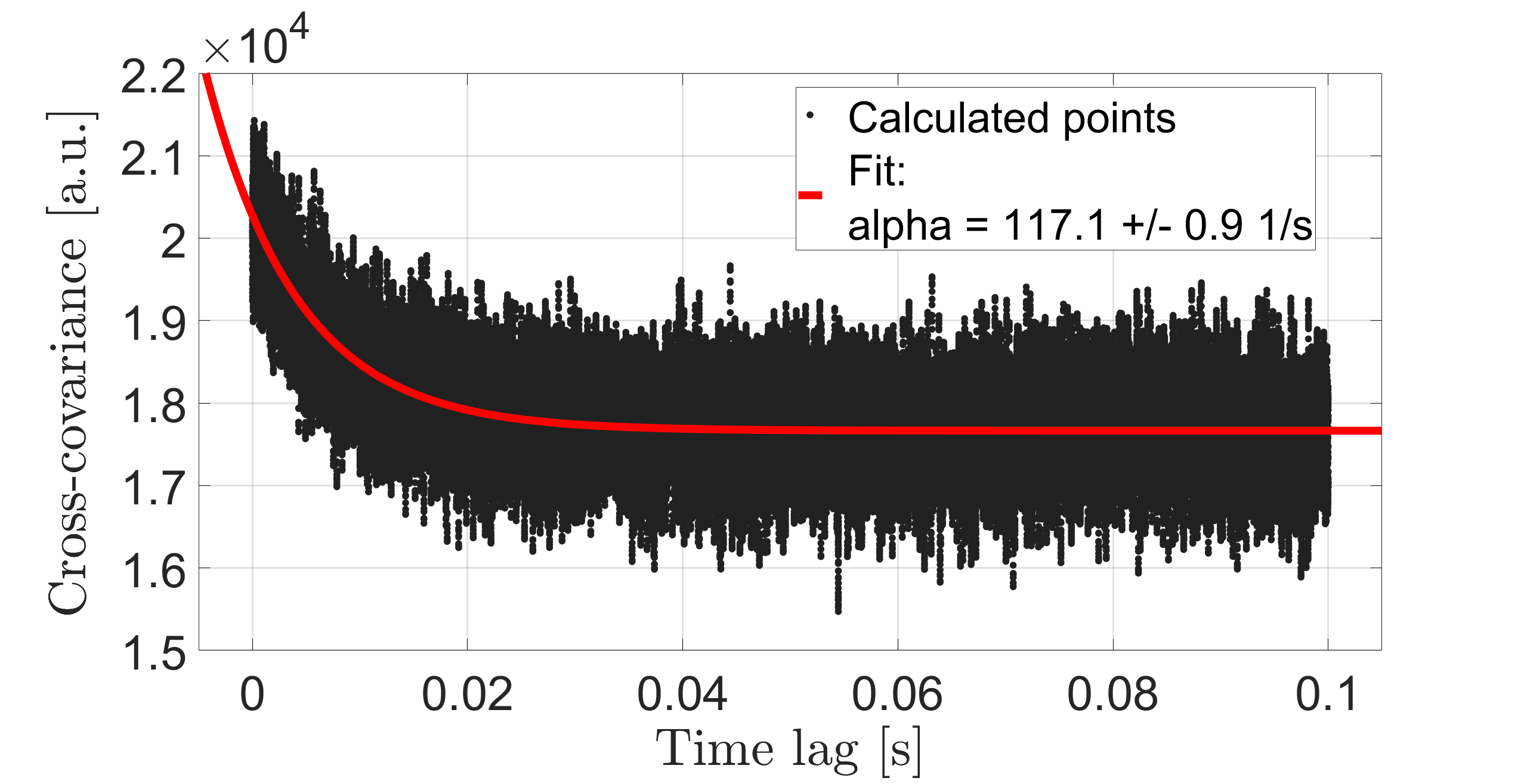}}
    \subfloat[Feynman-evaluation of the continuous signal.]{\includegraphics[width=0.4\textwidth]{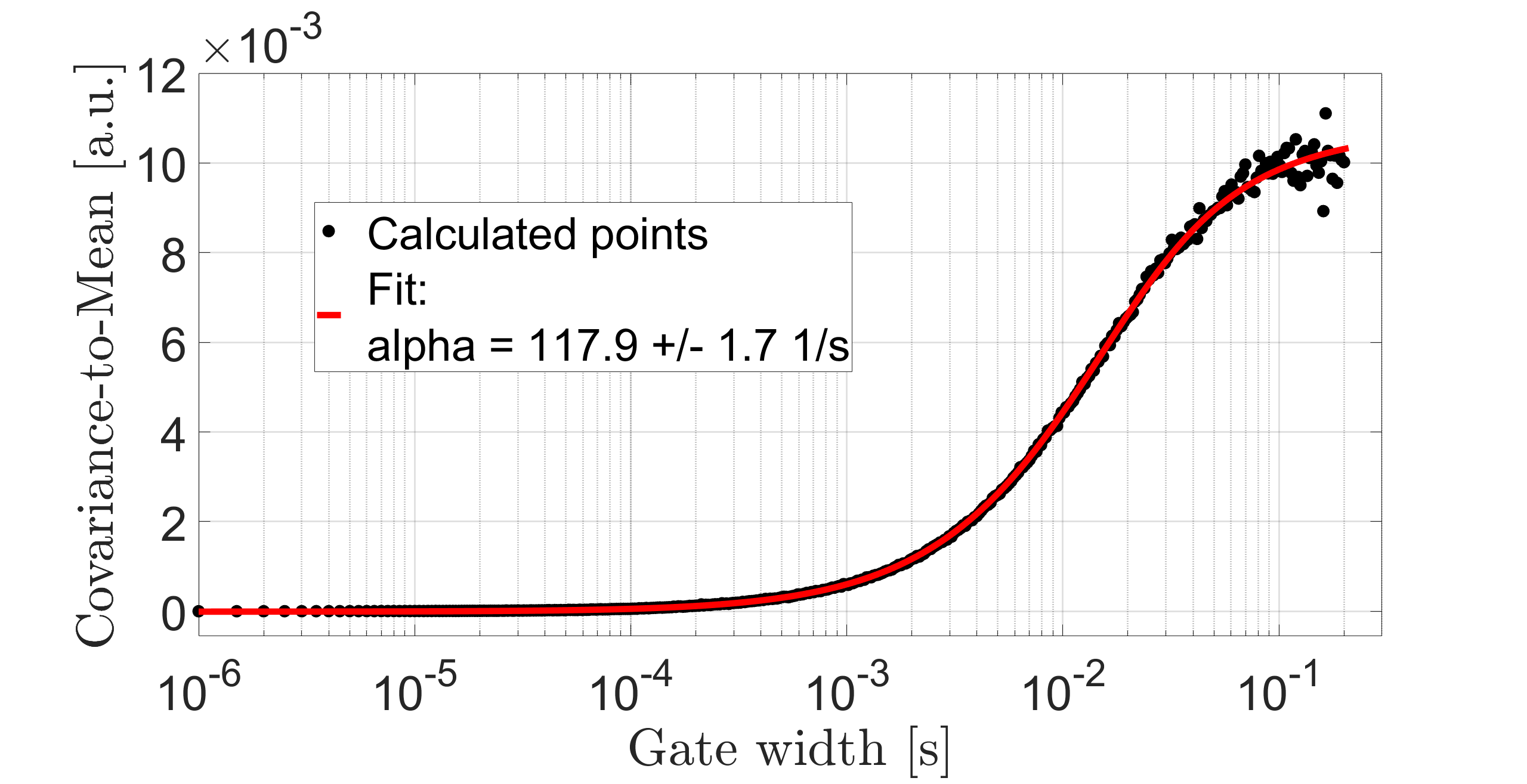}}
    \caption{Rossi-$\alpha$ (cross-covariance) and Feynman-$\alpha$ (cross-covariance-to-mean) results obtained from the \emph{Reference signal}, the \emph{Pulse-based signal} and the \emph{Continuous signal} based on a simulation with $5\cdot10^7\text{~s}^{-1}$ external source intensity.}
    \label{fig:comparison_int5e7}
\end{figure*}
\begin{figure*}
    \centering
    \subfloat[Rossi-$\alpha$ results.]{\includegraphics[width=0.4\textwidth]{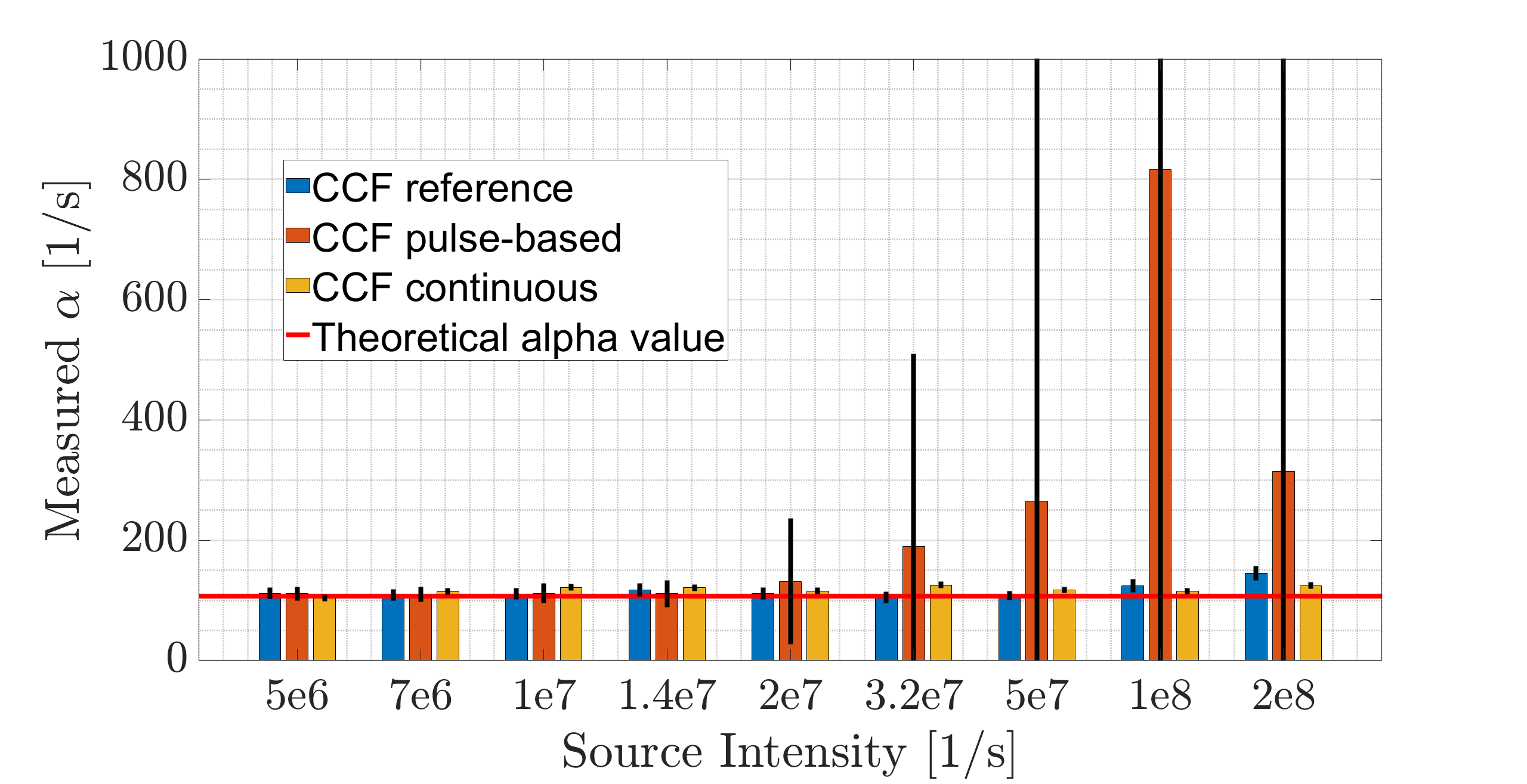}}
    \subfloat[Feynman-$\alpha$ results.]{\includegraphics[width=0.4\textwidth]{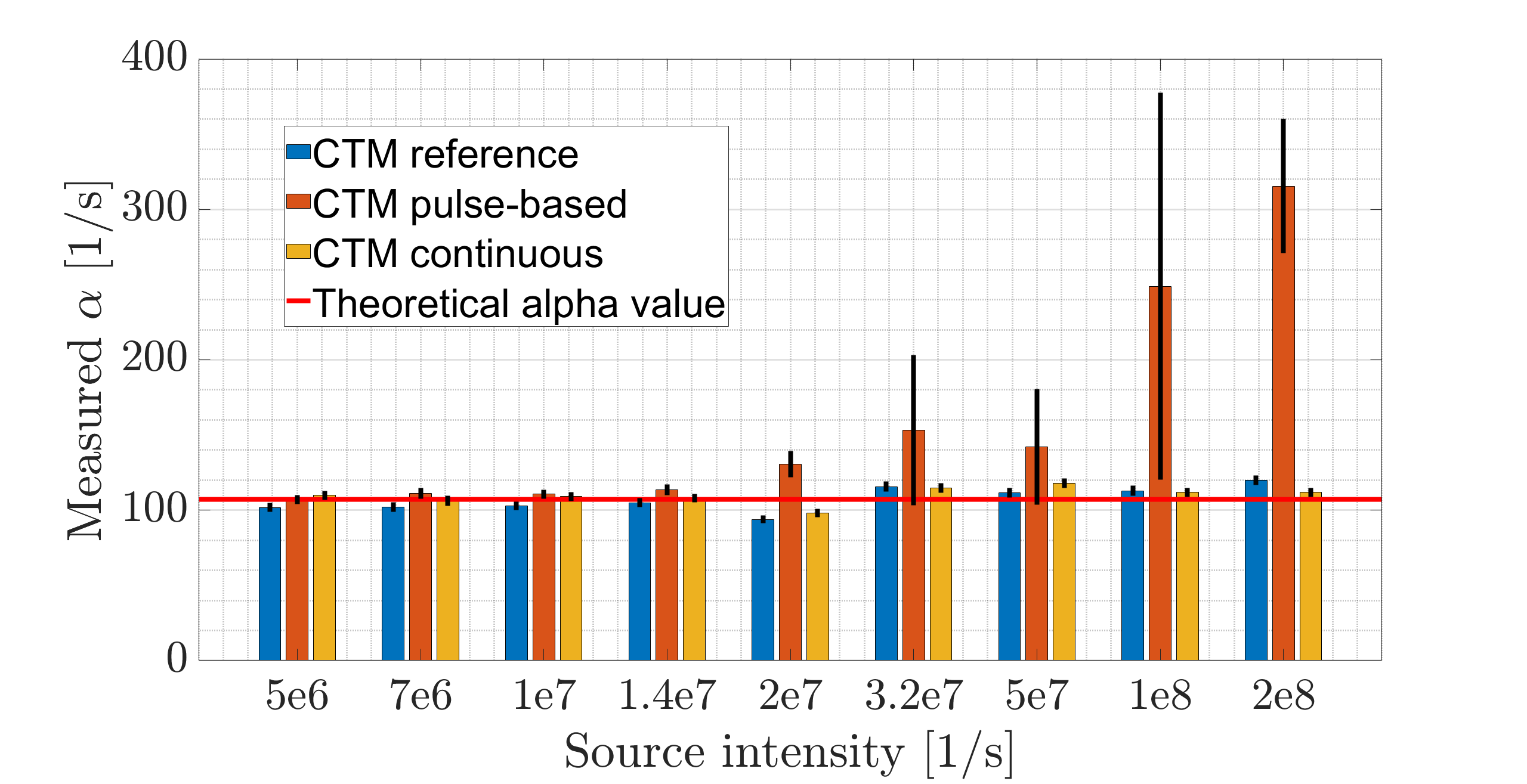}}
    \caption{Rossi-$\alpha$ (cross-covariance) and Feynman-$\alpha$ (cross-covariance-to-mean) results obtained from the \emph{Reference signal}, the \emph{Pulse-based signal} and the \emph{Continuous signal}.}
    \label{fig:int_summary}
\end{figure*}
For all signal types, the 2-detector variants of the Rossi-$\alpha$ and Feynman-$\alpha$ methods were used, resulting in a best-case scenario for both continuous and pulse-based signals. Figure \ref{fig:comparison_int5e7} shows the results of all 3 signal types of a simulation with $5\cdot10^7\text{~s}^{-1}$ source intensity (approximately $5\cdot10^5\text{~s}^{-1}$~neutron/s/detector)

At this level of source intensity, the traditional pulse-based signal is on the edge of usability, as only the cross-covariance to mean function can be fitted successfully with the theoretical Feynman-$\alpha$ function, and with much higher uncertainties compared to the reference and continuous signals.

The results obtained from all the simulations with source intensities ranging from $5\cdot10^6\text{~s}^{-1}$ and $2\cdot10^8\text{~s}^{-1}$ are shown in Figure \ref{fig:int_summary}.

Across all simulations, the continuous signal reproduces the correct $\alpha$‑value over a much wider count‑rate range than the traditional pulse‑based signal. Once the detection rate reaches several $10^5$ events/s per detector, pulse‑counting begins to lose correlated events, and Rossi‑ and Feynman‑$\alpha$ fits degrade significantly. In contrast, the continuous‑signal method remains stable and accurate because it does not require resolving individual pulses.
\subsection{Increasing the $\alpha$ value of the system}
\begin{figure*}
    \centering
    \subfloat[$\text{NSR}=0.1\%$.]{\includegraphics[width=0.42\textwidth]{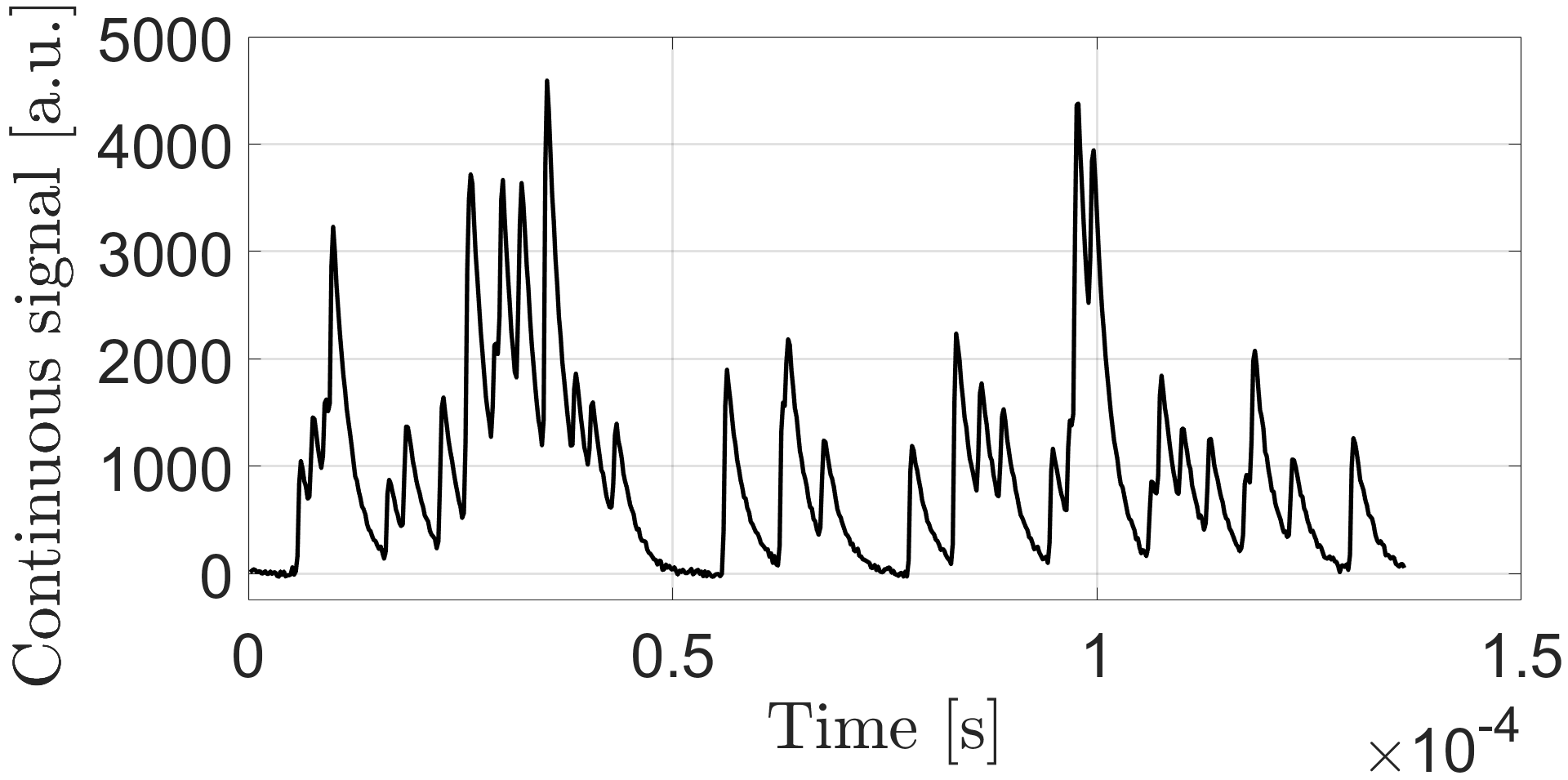}}
    \subfloat[$\text{NSR}=2.5\%$.]{\includegraphics[width=0.42\textwidth]{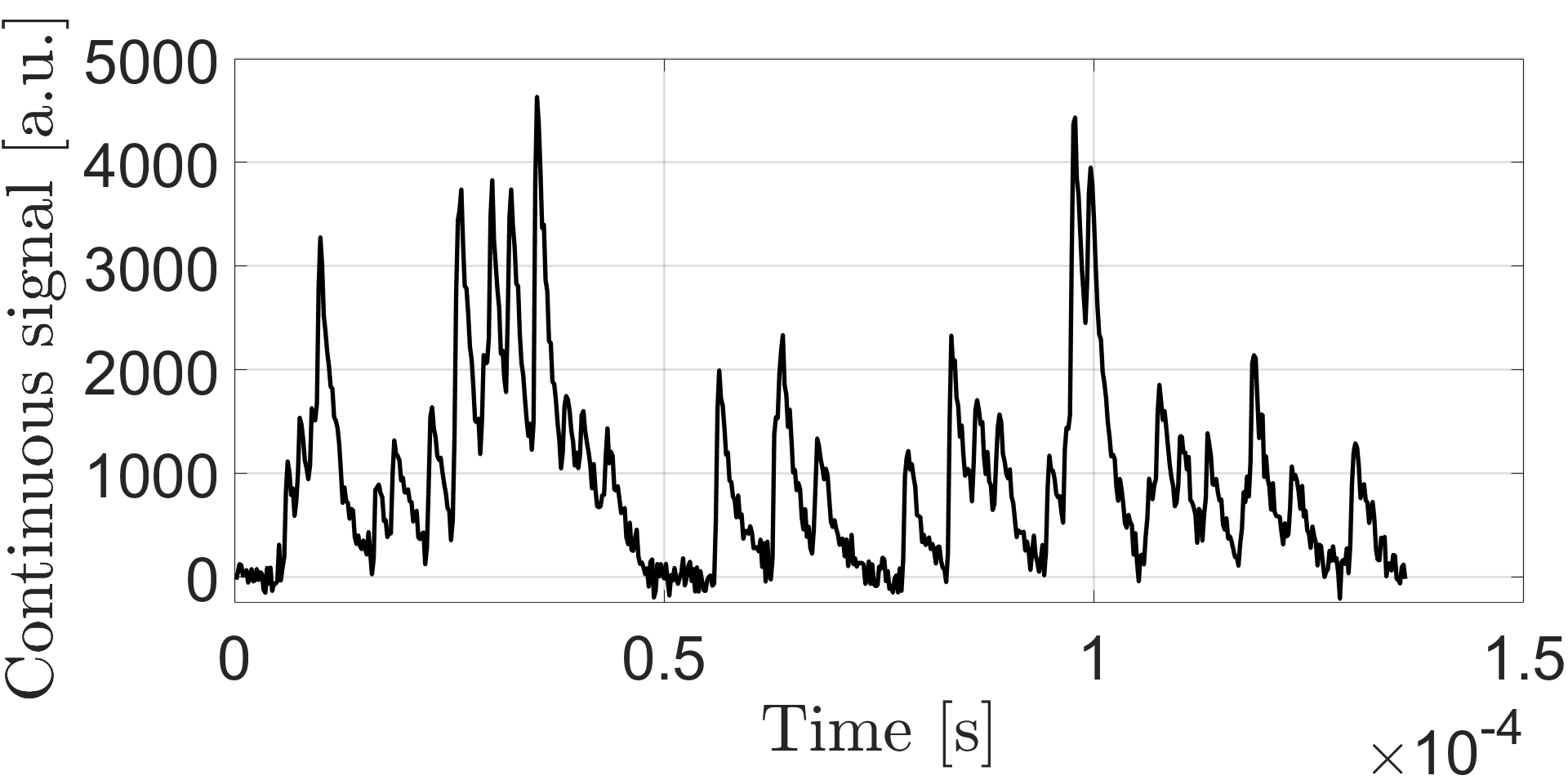}}\\
    \subfloat[Inverse Fourier deconvolution for $\text{NSR}=0.1\%$ and $\text{NSR}=2.5\%$.]{\includegraphics[width=0.42\textwidth]{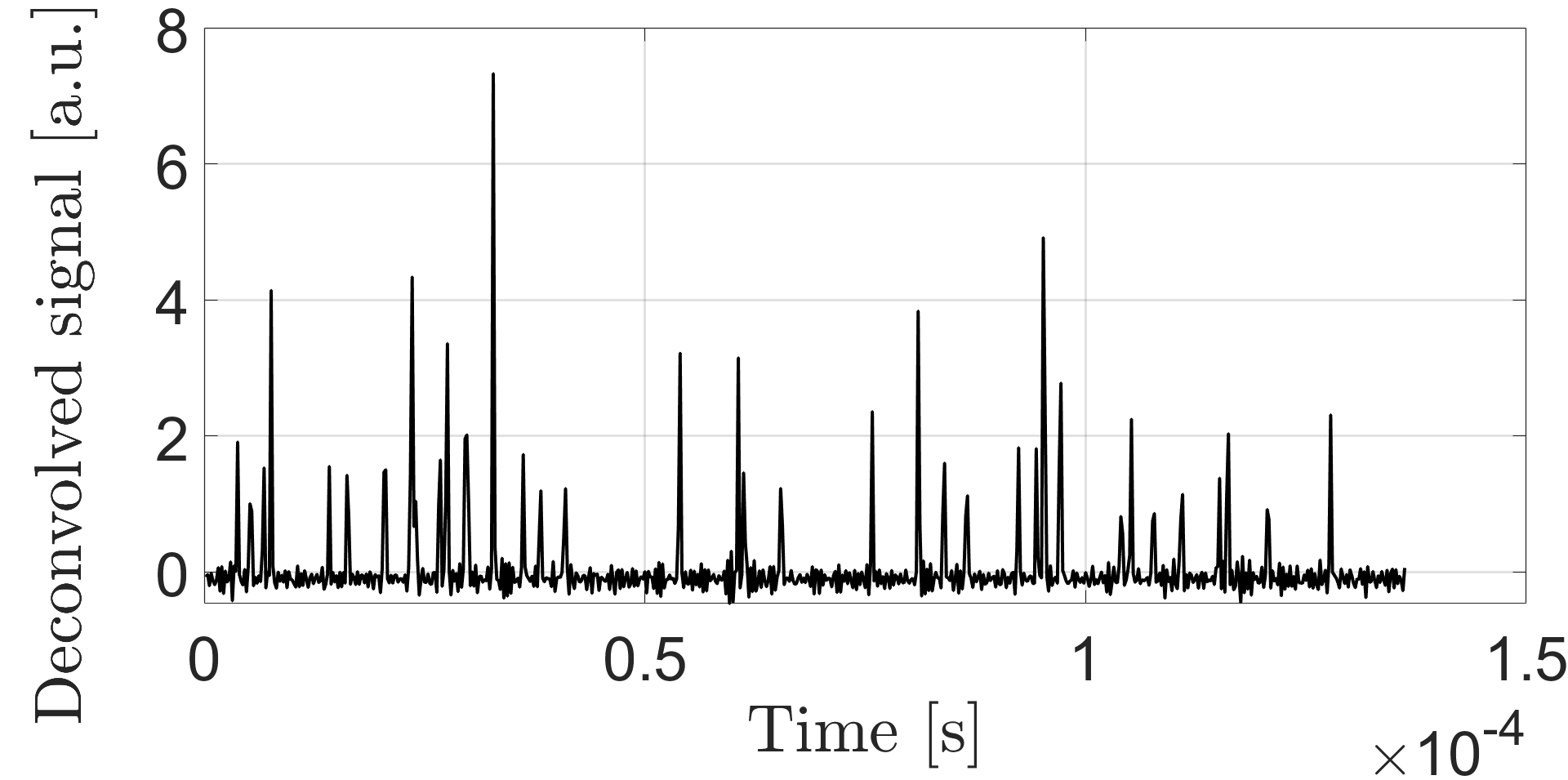}
\includegraphics[width=0.42\textwidth]{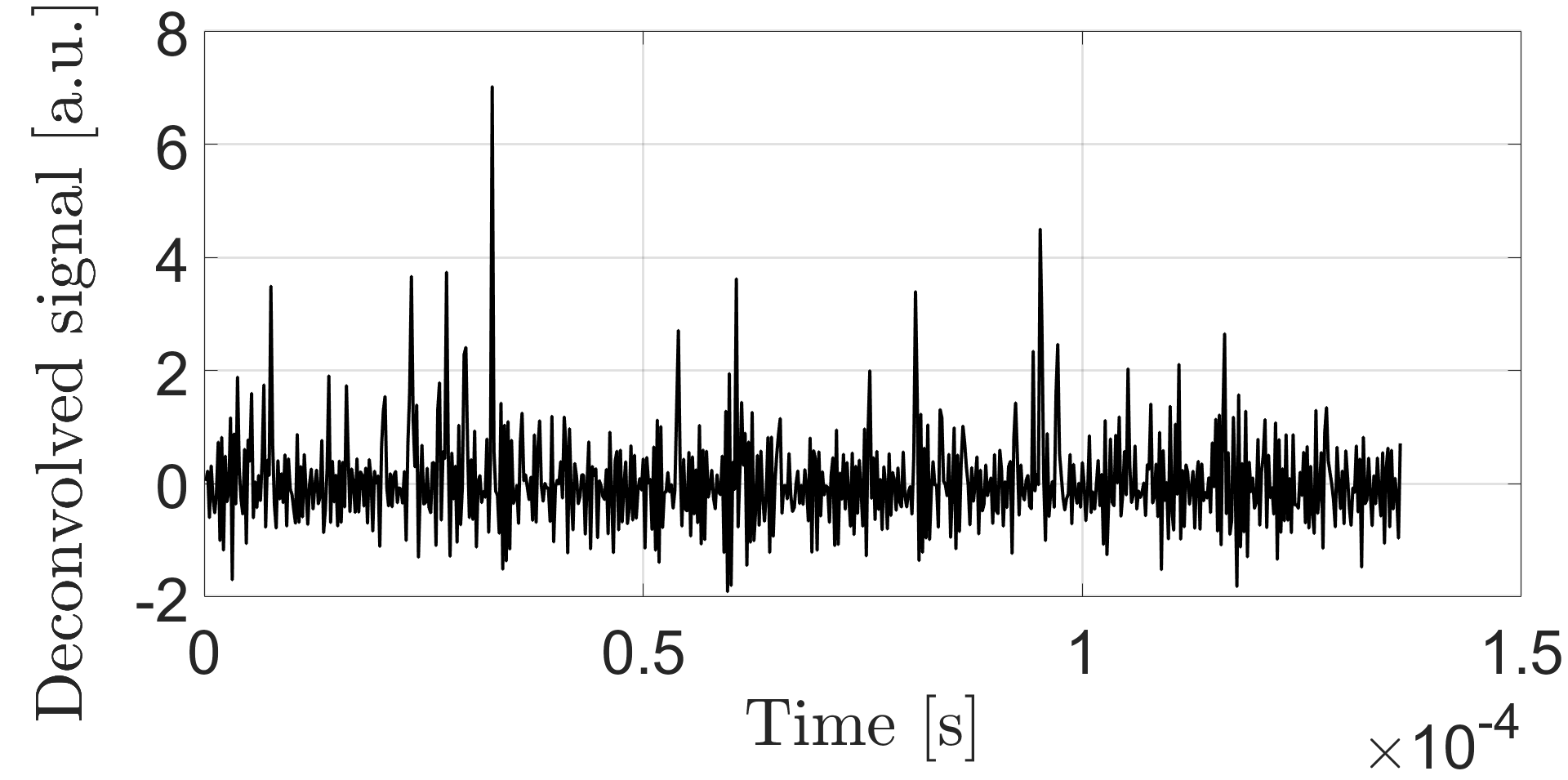}}\\
    \hspace{7.3cm}
    \subfloat[Wiener deconvolution with $\text{NSR}=2.5\%$.]{\includegraphics[width=0.42\textwidth]{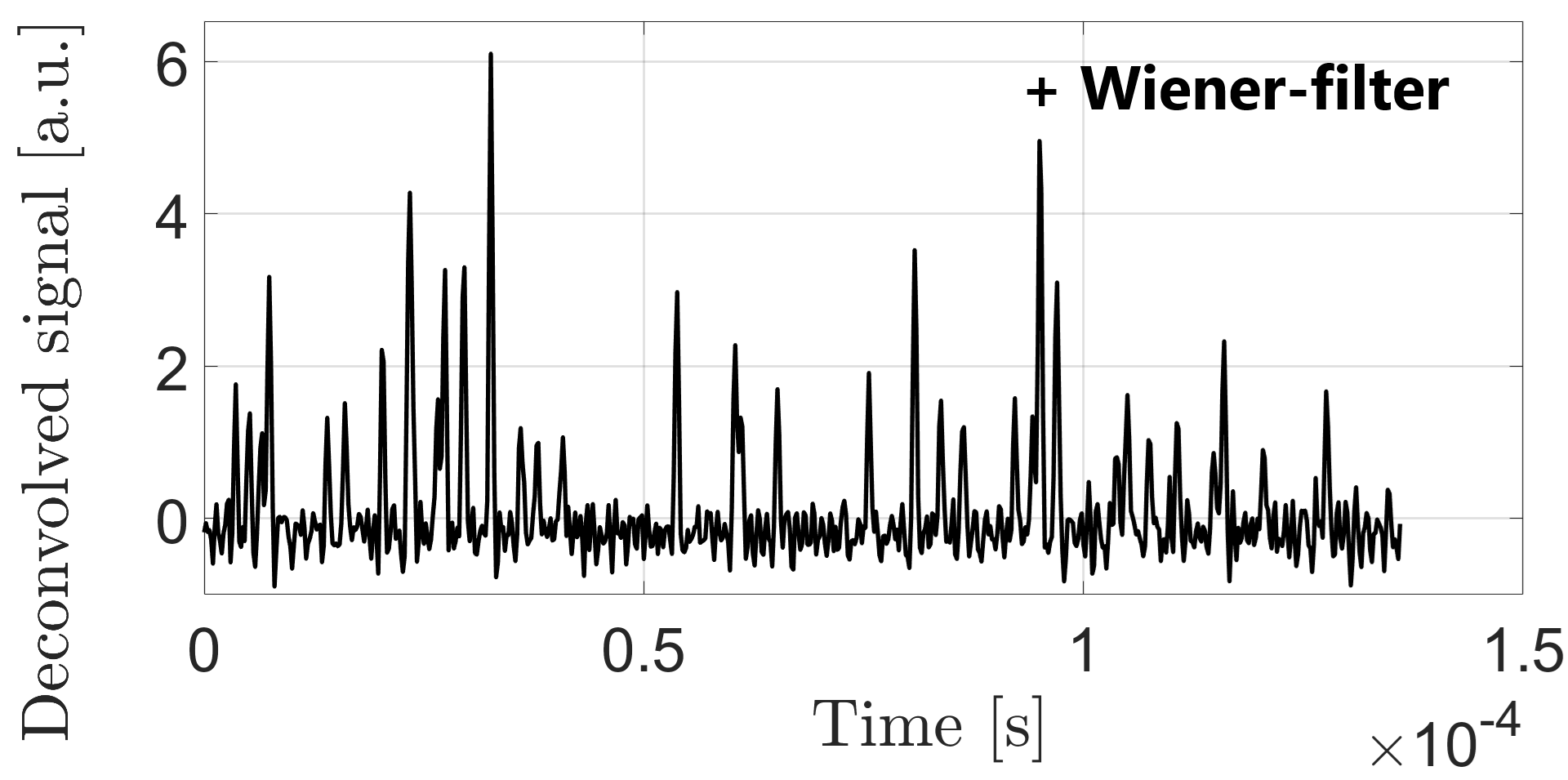}}
    \caption{Simulated continuous signals and the result of the deconvolution, of the same simulation, but with different levels of additive Gaussian white noise (0.1\% and 2.5\% NSR). With the noisier signal, the deconvolution was also performed using a frequency-dependent Wiener-filter, significantly improving the stability of the deconvolution in exchange for some sharpness.}
    \label{fig:deconv_noise}
\end{figure*}
\begin{figure}
    \centering
    \subfloat[Rossi-$\alpha$ results.]{\includegraphics[width=0.9\columnwidth]{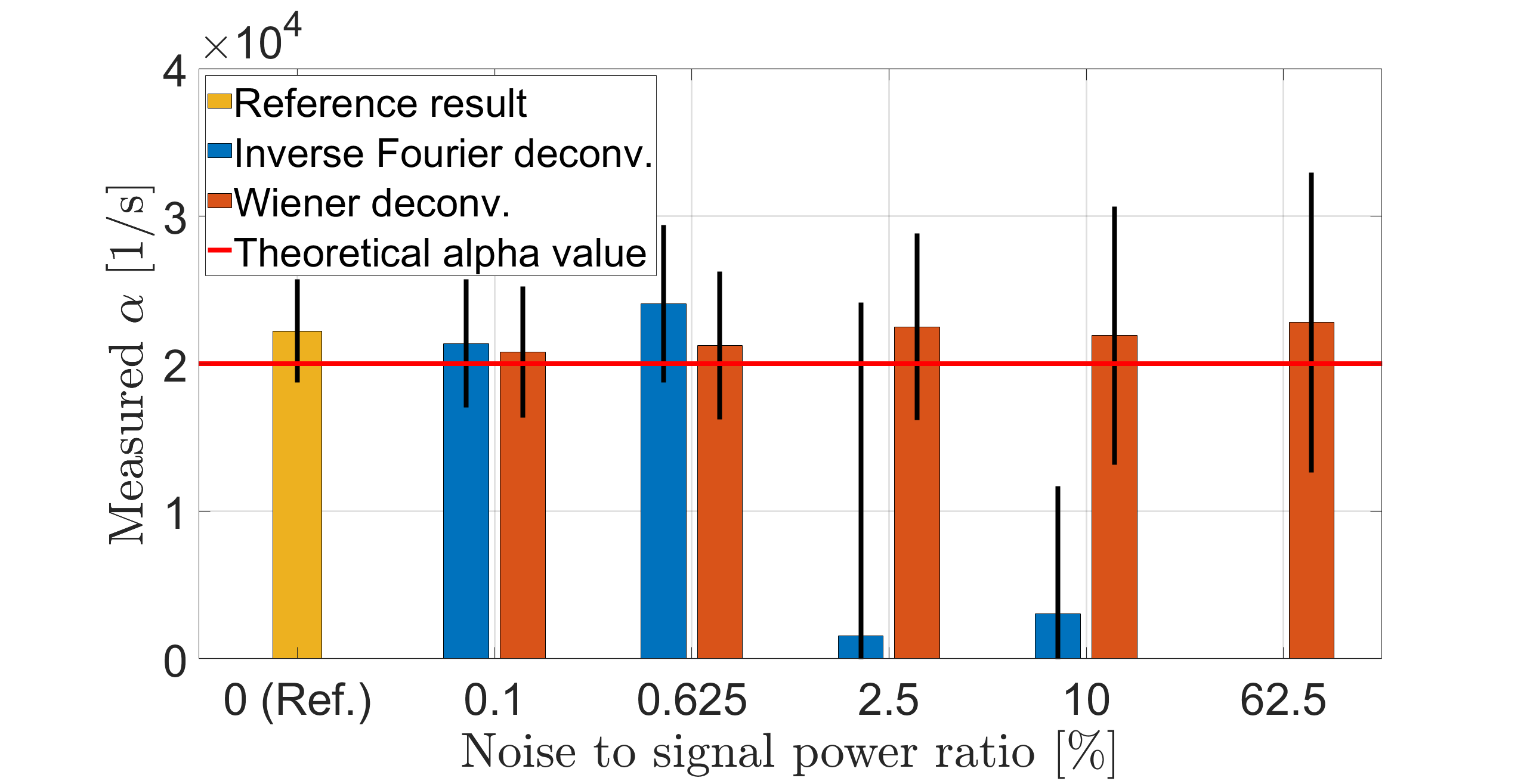}}\\
    \subfloat[Feynman-$\alpha$ results.]{\includegraphics[width=0.9\columnwidth]{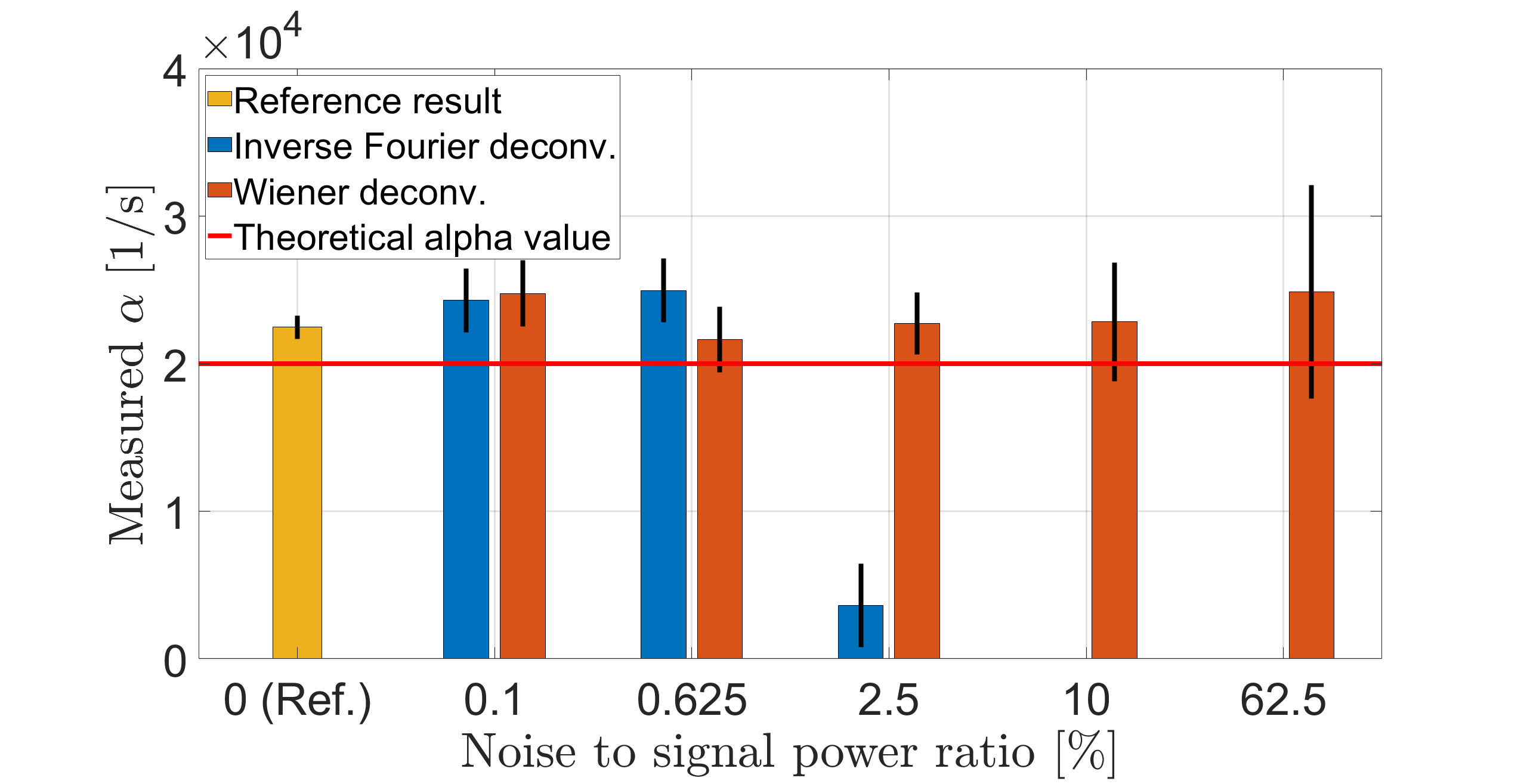}}
    \caption{Rossi- and Feynman-$\alpha$ values estimated using signals obtained by performing a pulse-shape deconvolution from raw voltage signals having NSR ranging from 0 to 62.5\%}
    \label{fig:wiener_comparison}
\end{figure}
The second parameter changed in the simulations was the system's prompt decay constant $\alpha$, up to $10^5\text{~s}^{-1}$. Higher $\alpha$ values are relevant in fast or deeply subcritical systems, or when one tries to estimate the decay constant corresponding to higher-order kinetic modes of the system. Accurate measurement of higher-order $\alpha$ modes is well known to be challenging with traditional pulse-based signals.

The simulations were tuned to yield an average detection rate of $10^5$ neutrons/s/detector. This is an intensity for which the traditional pulse-based signal gave satisfactory results with $\alpha\approx100\text{~s}^{-1}$.

Alongside the pulse-based and continuous signals, the possible benefits of deconvolving the average pulse-shape from the continuous signal were also investigated. Namely, after deconvolving the continuous signal, the resulting deconvolved continuous signal was transformed into a pulse-based signal, similar to how the traditional pulse-based signal is produced from the original continuous signal, using an appropriate threshold level. 

First, the sensitivity of the proposed inverse Fourier deconvolution technique with regard to additive white noise was investigated. Short sections of the simulated continuous signal and its deconvolved form are shown in Figure \ref{fig:deconv_noise}, with different Noise to Signal Power Ratios (NSR) (approximately 0.1\% and 2.5\%). 
\begin{figure}
    \centering
    \subfloat[Rossi-$\alpha$ evaluation.]{\includegraphics[width=0.9\columnwidth]{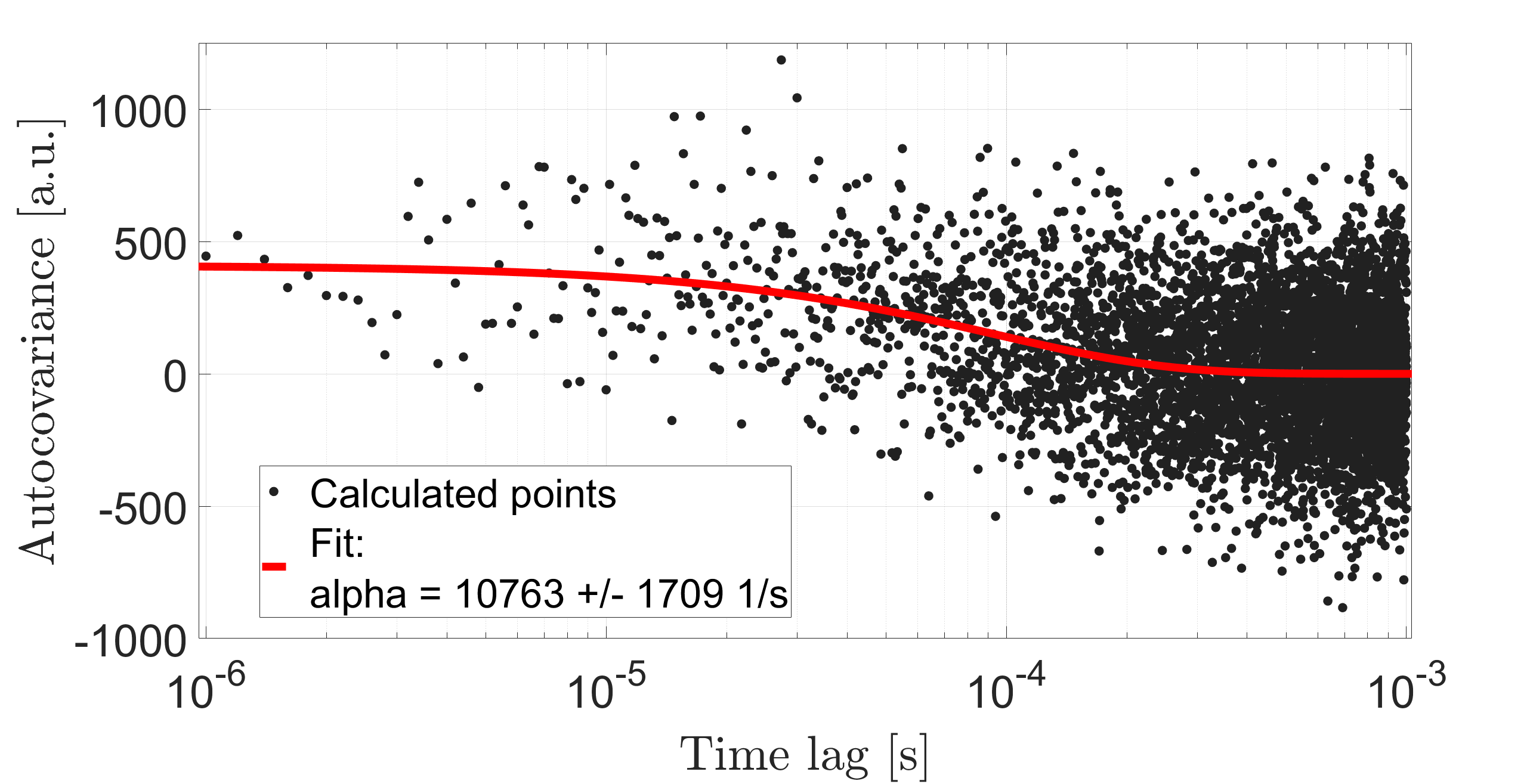}}\\
    \subfloat[Feynman-$\alpha$ evaluation.]{\includegraphics[width=0.9\columnwidth]{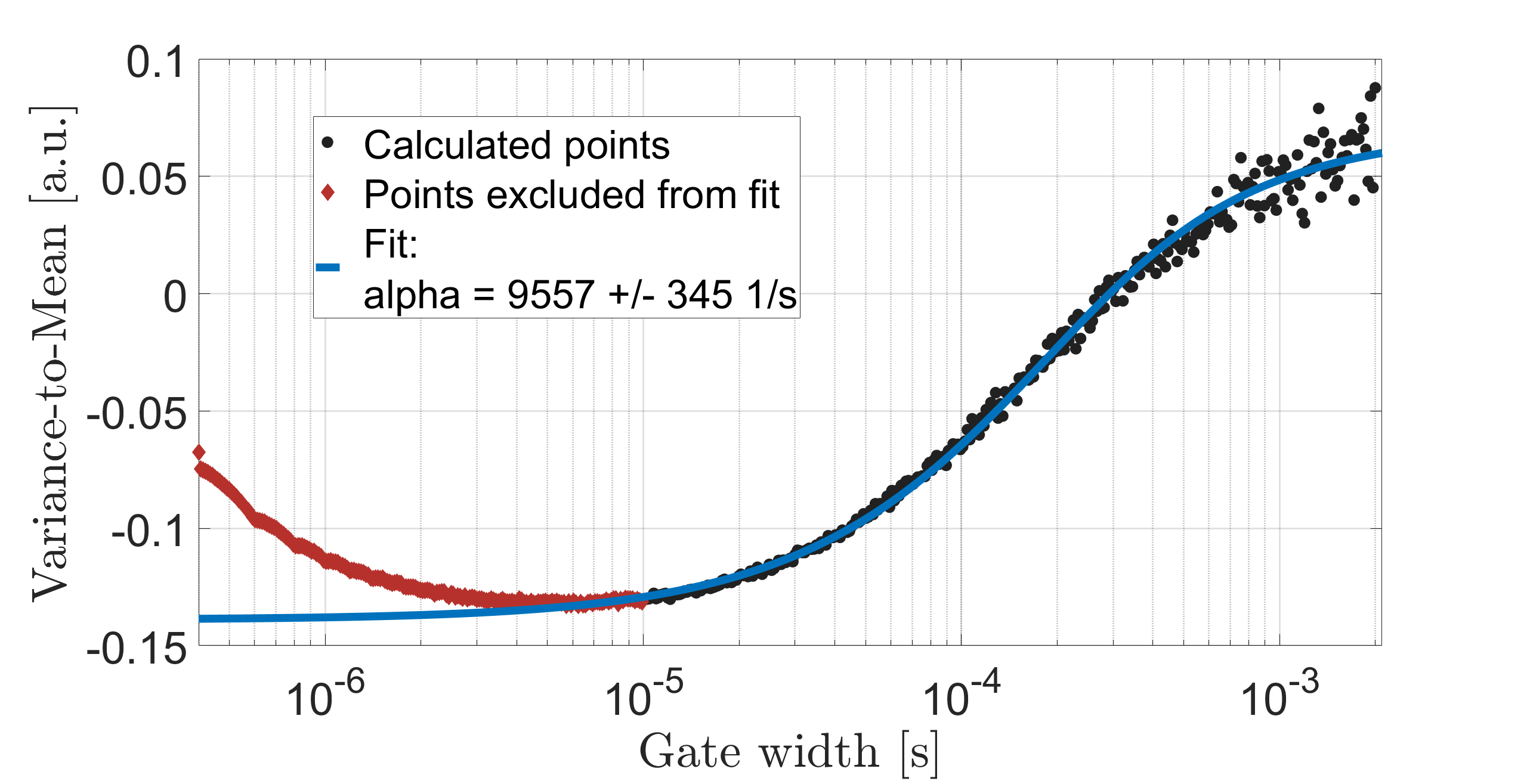}}
    \caption{Rossi-$\alpha$ (ACF) and Feynman-$\alpha$ (VTM) evaluation of the deconvolved pulse-based signal.}
    \label{fig:deconv_eval_example}
\end{figure}
\begin{figure}
    \centering
    \subfloat[Rossi-$\alpha$ results.]{\includegraphics[width=0.9\columnwidth]{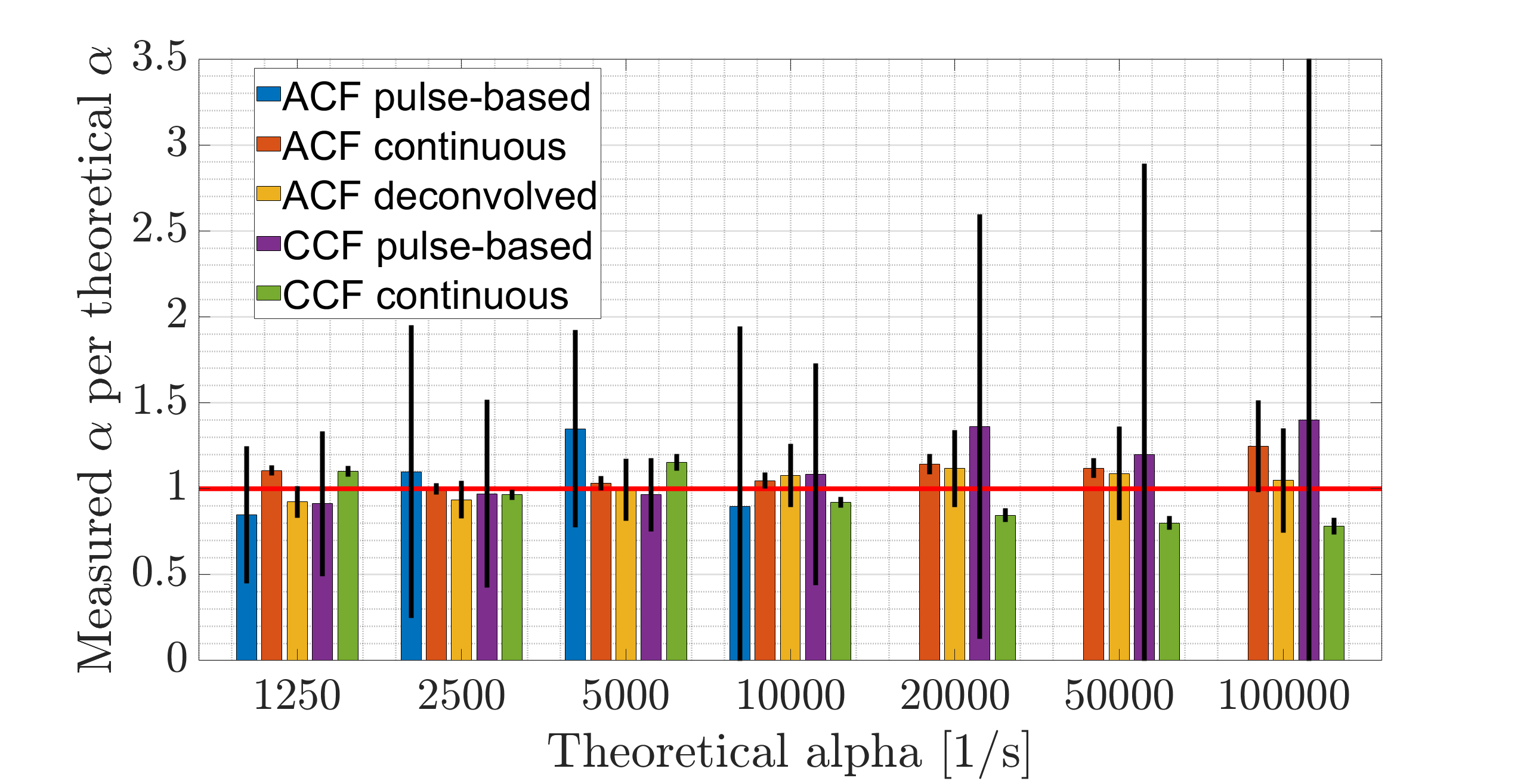}}\\
    \subfloat[Feynman-$\alpha$ results.]{\includegraphics[width=0.9\columnwidth]{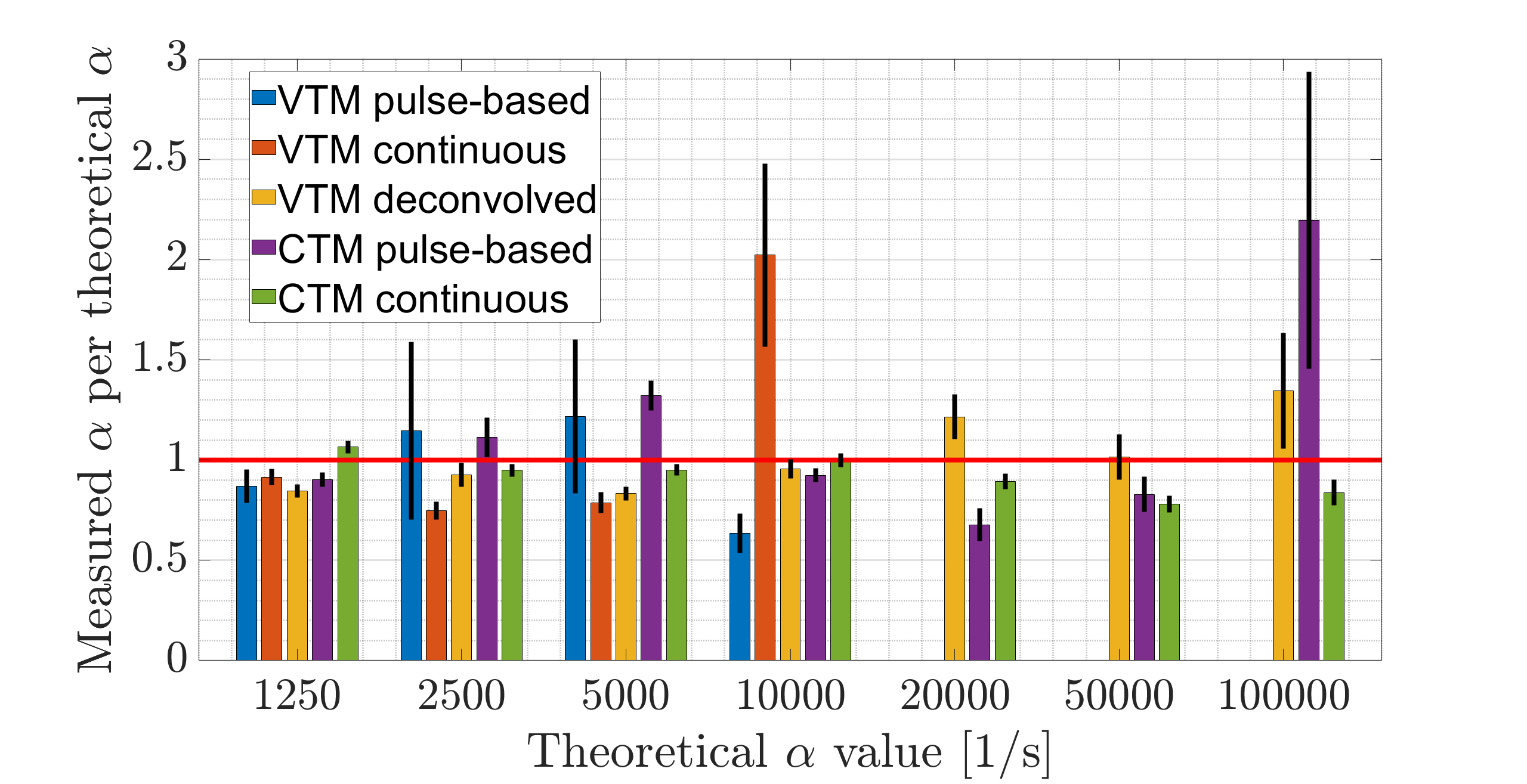}}
    \caption{Rossi-$\alpha$ and Feynman-$\alpha$ evaluations of different signal types with different theoretical $\alpha$ values. For the continuous and (traditional) pulse-based signals, the charts contain the results obtained from using a single detector (ACF, VTM) and detector pairs (CCF, CTM).}
    \label{fig:alpha_summary}
\end{figure}
As expected, the results show that this simple inverse Fourier deconvolution technique is quite sensitive to electronic noise. However, as long as the NSR remains well under 1\%, even this gives satisfactory results. If the too high NSR or other undesirable effects render the inverse Fourier deconvolution overly unstable, a Wiener filter might be used that can be generated from an estimation of the frequency dependence of the NSR of the recorded signal.

In the case of Wiener deconvolution \cite{WIENERBOOK}, the raw signal $c$ is not simply divided by average pulse-shape $f$ to obtain the deconvolved signal $d$ in the frequency space, but multiplied with the Wiener filter $W(\omega)$ containing the frequency dependent noise to signal power ratio NSR$(\omega)$:
\begin{equation}
\begin{split}
    \mathcal{F}\{d(t)\}(\omega)&=\mathcal{F}\{c(t)\}(\omega)\cdot W(\omega)\\
    =\mathcal{F}\{c(t)\}&(\omega)\cdot\frac{\mathcal{F}^*\{f(t)\}(\omega)}{|\mathcal{F}\{f(t)\}(\omega)|^2+\gamma\cdot\text{NSR}(\omega)},
\end{split}
\end{equation}
where $\gamma$ is called the Lagrange constant or regularization parameter \citep{Gunawan01092016}. $\gamma=0$ results in the simple inverse Fourier deconvolution, while $\gamma=1$ would yield what is usually referred to as Wiener deconvolution. However, it is usually beneficial to find an optimal value for $\gamma$ that might differ from unity, especially when NSR$(\omega)$ is not explicitly known but estimated only. Increasing $\gamma$ can help with noise suppression, but it also leads to the widening of pulses in the deconvolved signal, which can make pulse detection more difficult. In our case, the optimal value of $\gamma$ was determined empirically for each case.

The performance of the inverse Fourier and Wiener deconvolution was compared by investigating the ultimately estimated $\alpha$ values using a pulse detection-based evaluation of the signals resulting from the two deconvolution techniques using raw voltage signals with various noise to signal power ratios (NSR) ranging from 0 to 62.5\%. The results are collected in Figure \ref{fig:wiener_comparison}.

Figure \ref{fig:wiener_comparison} shows that the Wiener deconvolution can be used rather reliably even with raw signals having very high NSR, where the inverese Fourier deconvolution can be very unstable even with signals having little noise.

An example of the results obtained from the deconvolved pulse-based signal of a \emph{single} detector is shown in Figure \ref{fig:deconv_eval_example}, with theoretical $\alpha$ value of 10000~s$^{-1}$. These results are directly comparable with those presented in Figure \ref{fig:onevs2det}.

Figure \ref{fig:deconv_eval_example} shows that the deconvolution of the average pulse-shape provides a substantial improvement compared to both the continuous and the traditional pulse-based signals when using only a single detector. The results of all the simulations with $\alpha$ values ranging from 1250~s$^{-1}$ to $10^5$~s$^{-1}$ are collected in Figure \ref{fig:alpha_summary}.

With the continuous and traditional pulse-based signals, the theoretical Rossi-$\alpha$ and Feynman-$\alpha$ functions could not be fitted successfully for larger $\alpha$ values. These results are missing from Figure \ref{fig:alpha_summary}. One can see that using \emph{detector pairs} instead of a \emph{single detector} benefits the traditional pulse-based signal, but especially the continuous signal in terms of evaluability and/or uncertainty of the estimation. The continuous signal is a clear improvement over the traditional pulse-based signal, whether using a single detector or a pair. 
At high $\alpha$‑values (above $2\cdot10^4$~s$^{-1}$), traditional pulse‑based methods fail due to overlapping pulses, while continuous‑signal methods remain evaluable, although with a slight systematic underestimation. This suggests that continuous‑signal techniques can be used to access higher kinetic modes or fast‑system $\alpha$‑ranges that are inaccessible by pulse counting.
\section{Measurements}
To complement the simulation results, measurements were also performed to demonstrate the advantages of using the continuous signal and the feasibility of the deconvolution of the average pulse-shape. Results of two different sets of measurements will be presented in the following, conducted at the Kyoto University Critical Assembly (KUCA) and at the BME TR.
\subsection{Measurements at KUCA}
The set of measurements presented in this section was performed in 2019, but the results were published only in a conference paper of the authors \cite{fullexperimental}. With the use of the new pulse-shape deconvolution technique, the evaluation of some additional measurement configurations could be performed successfully. In addition, the cause of some unexpected results was identified which led to the successful Feynman-evaluation of the continuous detector signal in subcritical and low power critical configurations. All the data of this measurement campaign was re-evaluated with different approaches, meaning that the results presented in the following are slightly different from those published in \cite{fullexperimental}.
\subsubsection{Instrumentation}
The KUCA core A configuration used for the presented measurements is a modular system with solid moderation and reflection. The fuel consists of 93.2\% enriched $^{235}$U, embedded in an aluminium alloy, forming metallic plates with dimensions of 5.08~cm~$\times$~5.08~cm~$\times$~0.1587~cm. 
Assemblies are made by placing these fuel plates horizontally inside rectangular aluminium tubes, together with polyethylene moderator blocks of similar size.
Thanks to their modular design, the assemblies can be arranged in numerous configurations to meet the requirements of different experiments. Further details on the fuel, moderator, and other assembly components, as well as the facility itself can be found in \cite{KURRI-TR-444}.

\begin{figure}
    \centering
    \includegraphics[width=0.9\columnwidth]{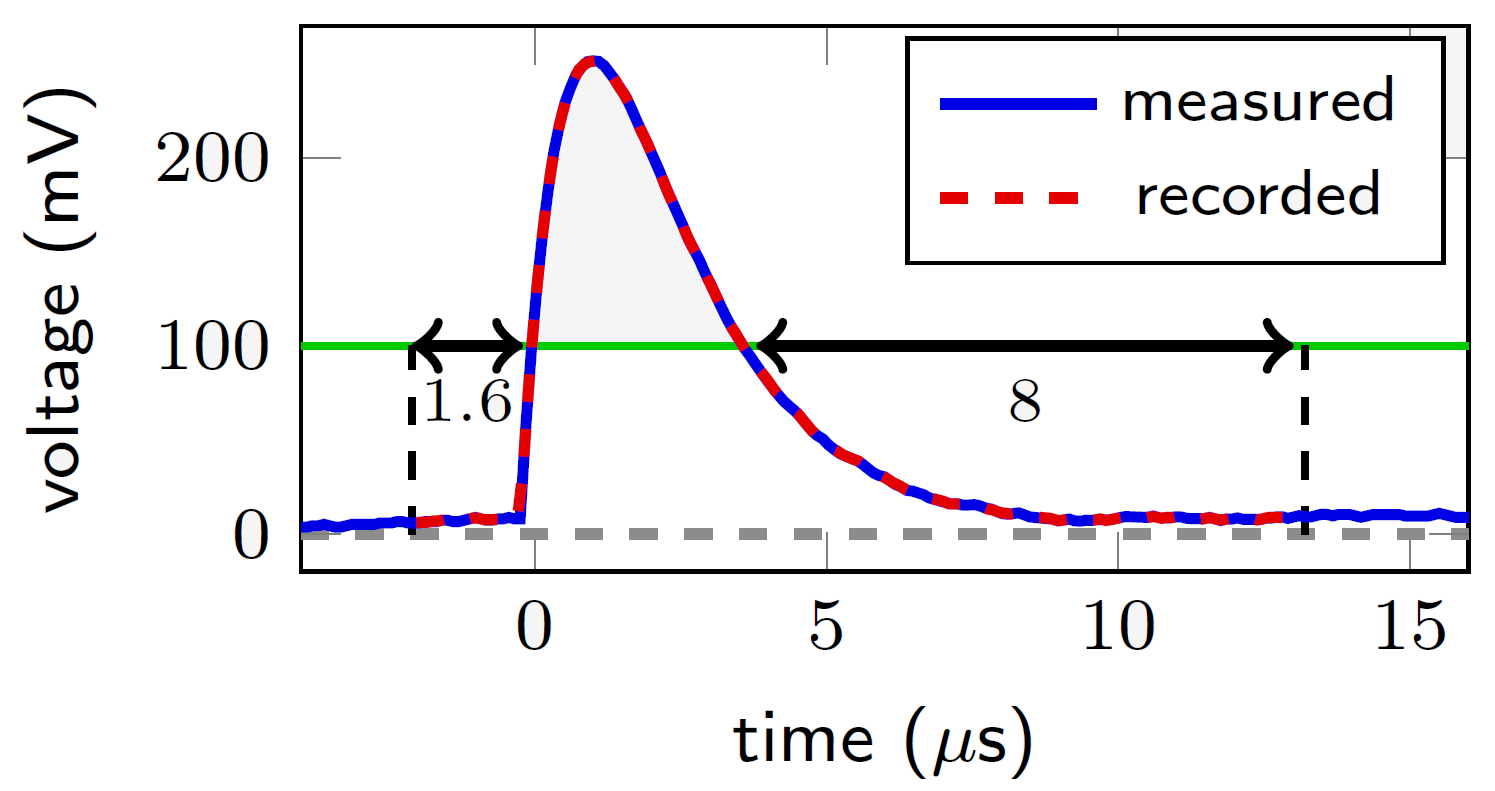}
    \caption{Illustration of the compressed recording of the continuous signal.}
    \label{fig:compressed_recording}
\end{figure}

The data acquisition system developed at BME was designed to simultaneously record continuous detector signals with sufficient time resolution and capture the timing of individual pulses for pulse-based recording. Each detector signal was directed to a dedicated, in-house-built high-speed preamplifier, producing voltage signals in the range of $-1$ to $1$~V. The preamplifiers were designed with a short time constant relative to the detector charge collection time, ensuring that the amplified pulse shapes accurately reflected the original detector signals. 
Signal acquisition was performed using two Red Pitaya STEMLab 125–14 FPGA-digitizers, each equipped with two analog input channels with 14-bit vertical resolution, and a maximum sampling frequency of 125~MHz (corresponding to 8~ns time resolution). To determine individual pulse arrival times, the preamplified signals were also processed through a separate chain consisting of an amplifier, a single-channel analyser, and a NI myRIO system for time-stamped data acquisition. Both the digitized continuous signals and the extracted pulse-time information were then saved to binary files on a computer. To reduce storage and bandwidth requirements, the continuous signals were recorded in a \emph{compressed} format: data points were preserved only for segments where the signal exceeded a predefined low threshold level, along with their immediate surroundings. For the remaining intervals, only their durations were stored, and these sections were later reconstructed as zero-valued data during analysis, as illustrated on Figure \ref{fig:compressed_recording}. An extra feature of this compression technique is that it effectively suppresses low frequency noise and artifacts related to the frequency transfer of the data acquisition chain, which had a big impact in the evaluability of the data, as presented in the following.
To be able to compare the continuous and pulse-based signals well, measurements were performed on configurations with a low enough detection rate so that the pulse pile-up and the induced dead time was expected to be insignificant. 
This condition is fulfilled in sufficiently subcritical, given the intensity of the available neutron source. 
On the other hand, measurements were also performed in critical configurations at several power levels in order to do the comparison with pulse counting at low count rates and test also the method at higher count rates. In these cases, the objective was to examine the anticipated differences between the methods, arising from the fact that pulse-counting techniques are influenced by dead time, whereas approaches based on continuous fission chamber signals are not.

\begin{figure}
    \centering
    \includegraphics[width=0.8\columnwidth]{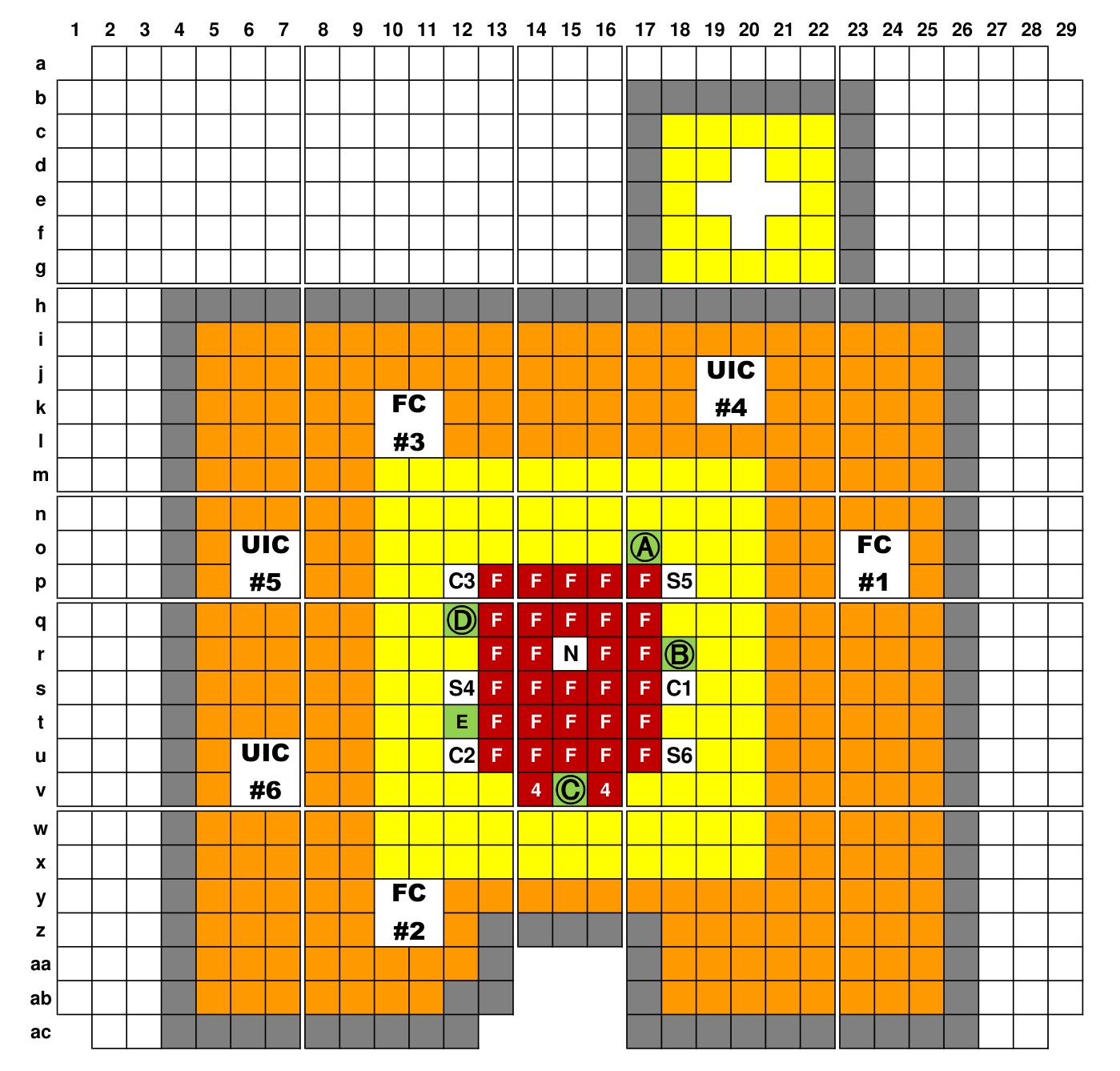}
    \caption{KUCA A-core configuration. Red = fuel; yellow = polyethylene moderator; orange = polyethylene reflector; gray = graphite. C1, C2 and C3 denote the control rods; S3, S4, and S5 refer to safety rods. A, B, C and D in green cells show the detector positions.}
    \label{fig:KUCA_core}
\end{figure}

The core configuration used for the measurements can be seen in Figure \ref{fig:KUCA_core}, where the four assemblies accommodating the fission chambers are also indicated. This core configuration had 170~pcm excess reactivity.
In total, three critical measurements were performed at different power levels listed in Table \ref{tab:critical}, and two subcritical ones at subcriticality levels listed in Table \ref{tab:subcritical}, together with the corresponding control rod positions.

\begin{table}
\caption{
Power level of the KUCA A-core during the different critical measurements.
         Detector labels refer to the fission chambers where voltage signal data was recorded.\label{tab:critical}}
\centering
\begin{tabular}[t]{c c c}
  \toprule
  Measurement name & Power [W] & Detector labels \\
  \midrule
  CR-1 & 5.46e-1 & A \\
  CR-2 & 1.84e-2 & A, D \\
  CR-3 & 1.84e-3 & A, D \\
  \bottomrule
\end{tabular}
\end{table}

\begin{table}[H]
\caption{Estimated $k_\text{eff}$, detectors used and control rod positions in the subcritical measurement configurations. The $k_\text{eff}$ value was estimated from the reactivity worth of the control rods.}\label{tab:subcritical}
\centering
\begin{tabular}[t]{c c c c}
  \toprule
  Configuration & $k_\text{eff}$ & Detectors used & Inserted rod(s)\\
  \midrule
  SCR-1 & 0.9906 & A, D & C1 \\
  SCR-2 & 0.978 & A, B, C, D & C1, C2, C3 \\
  \bottomrule
\end{tabular}
\end{table}
\subsubsection{Results}
As expected, the recorded signals were affected by superimposed high-frequency electronic noise. 
To mitigate the effect of this, the signals were smoothed before analysis offline. Smoothing was performed using a simple moving average, in which each data point was replaced by the unweighted mean of 21 consecutive points symmetrically distributed around the original point in measurements CR-3, SCR-1 and SCR-2 and using 9 consecutive points in CR-1 and CR-2. 
The time resolution of the datasets is 40~ns, 40~ns, 48~ns, 104~ns and 104~ns, respectively.
This resulted in averaging windows of $\sim1~\mu$s, which is long enough to suppress most of the high-frequency electronic noise, but sufficiently short not to distort the shape of the pulses too much. 
\begin{figure*}
    \centering
    \subfloat[The shape of the PSD of the continuous signal acquired during the CR-3 measurement.]{\includegraphics[width=0.45\textwidth]{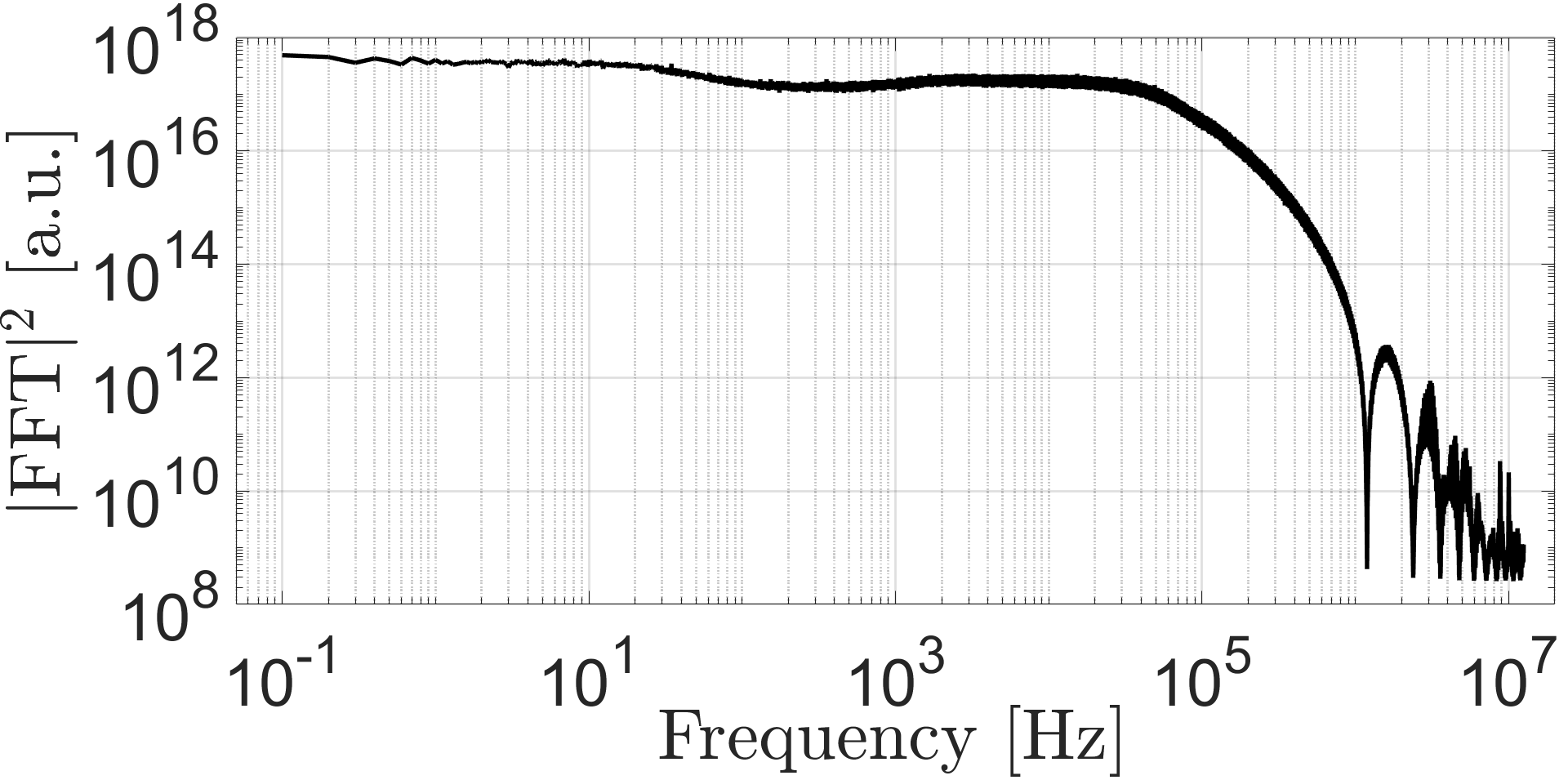}}
    \subfloat[The shape of the PSD of the continuous signal acquired during the CR-2 measurement.]{\includegraphics[width=0.45\textwidth]{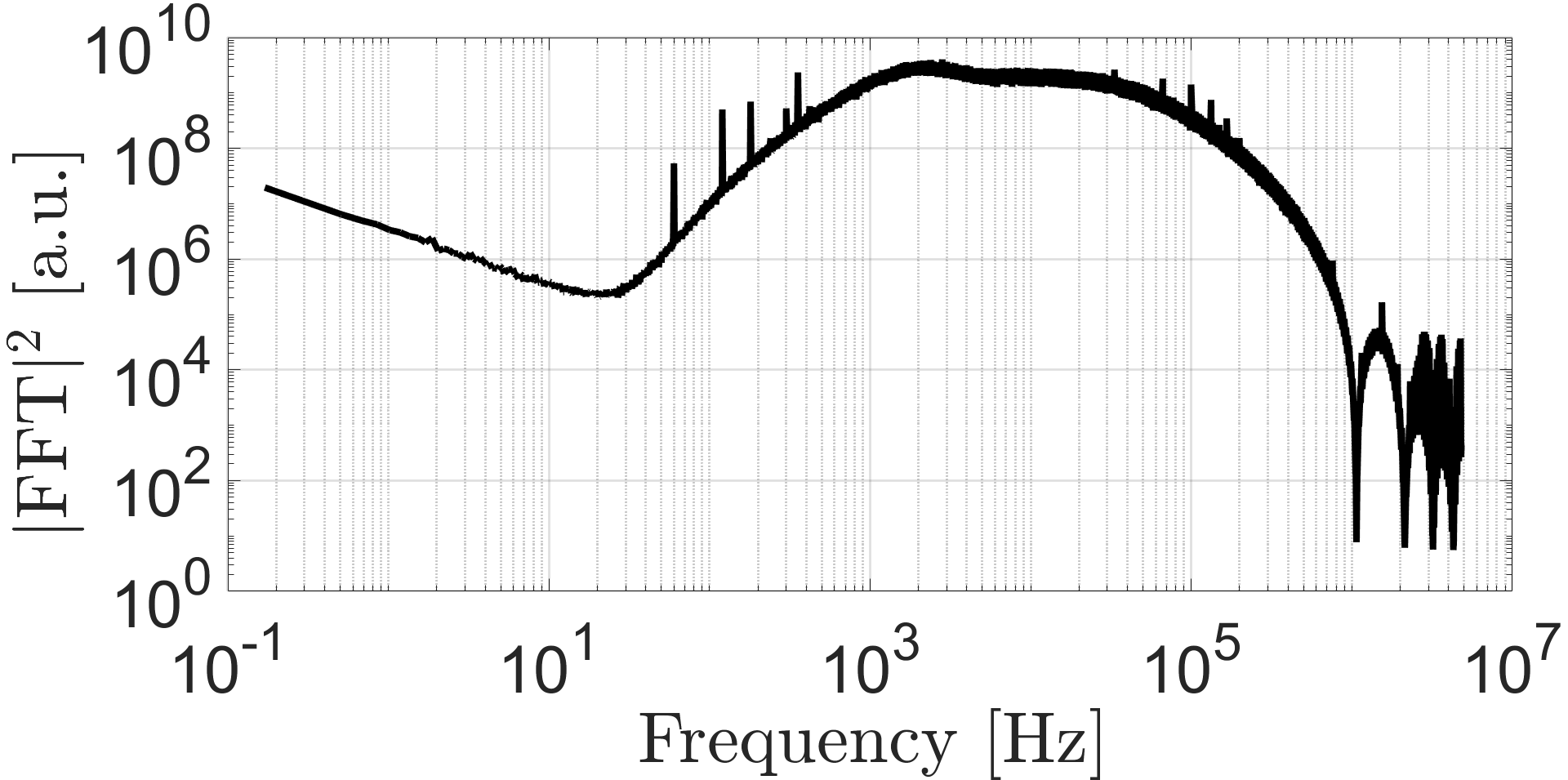}}\\
    \subfloat[Relevant section of the PSD from CR-3 measurement.]{\includegraphics[width=0.45\textwidth]{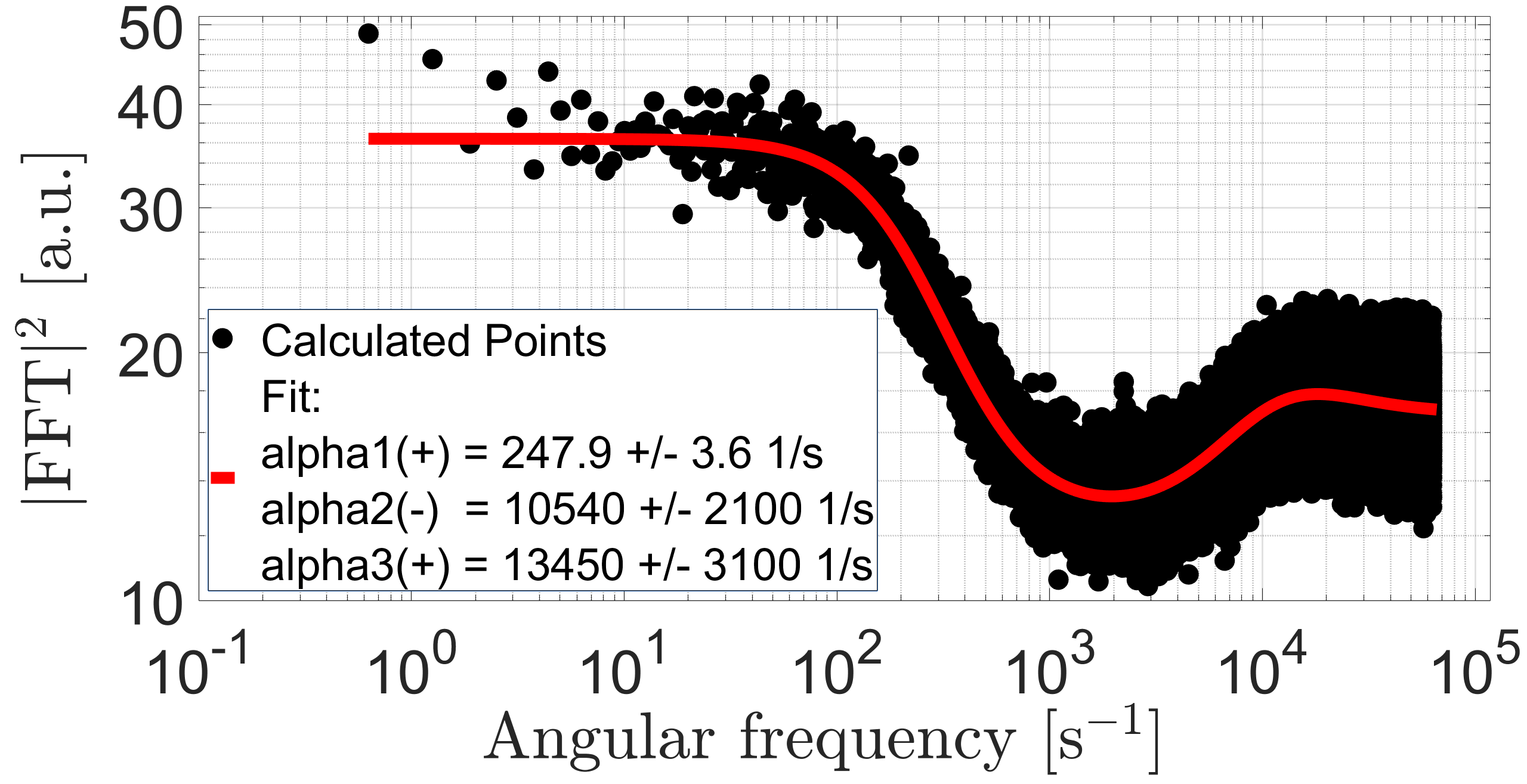}}
    \subfloat[Relevant section of the PSD from CR-2 measurement.]{\includegraphics[width=0.45\textwidth]{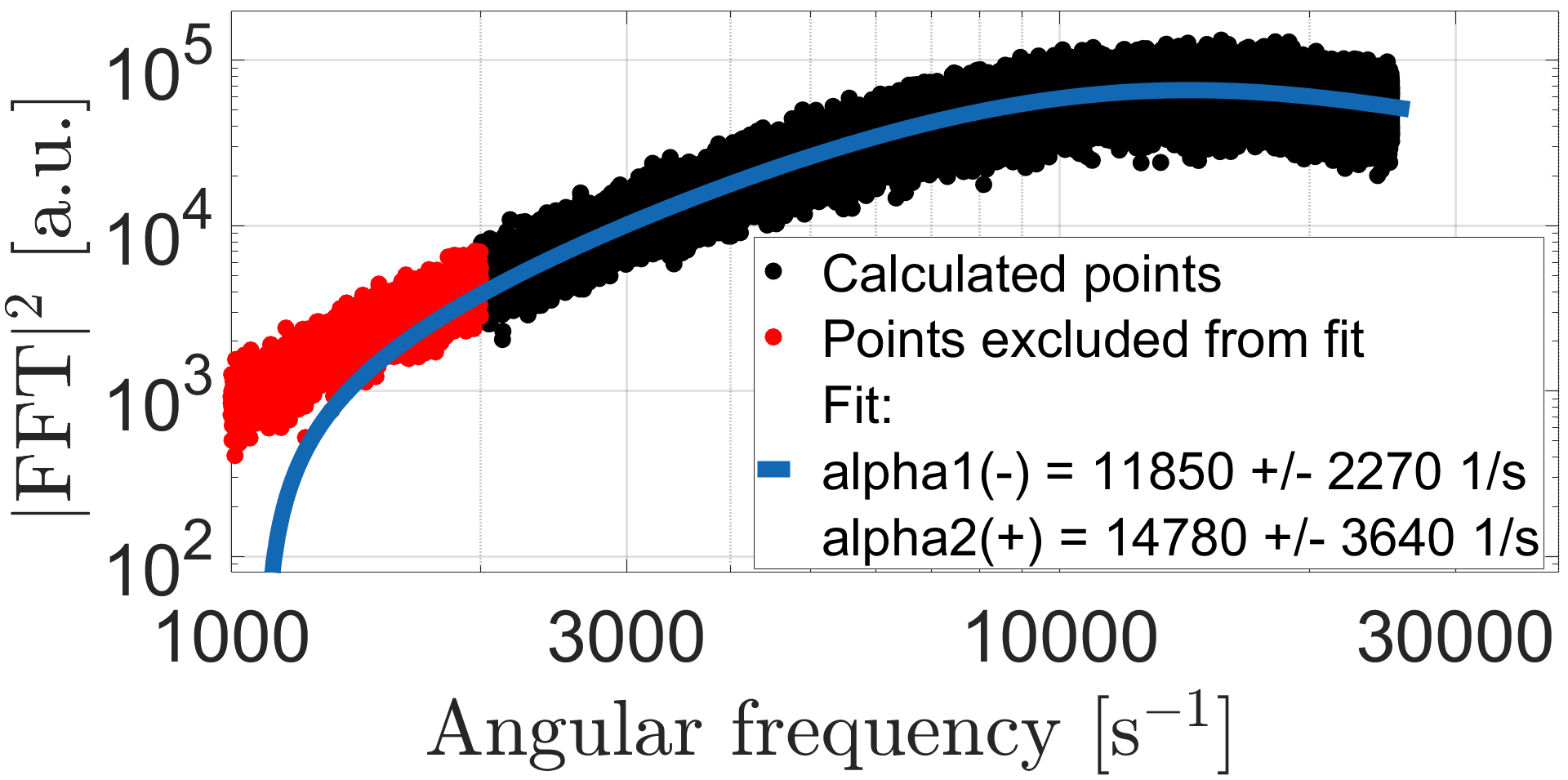}}\\
    \subfloat[ACF from the CR-3 measurement.]{\includegraphics[width=0.45\textwidth]{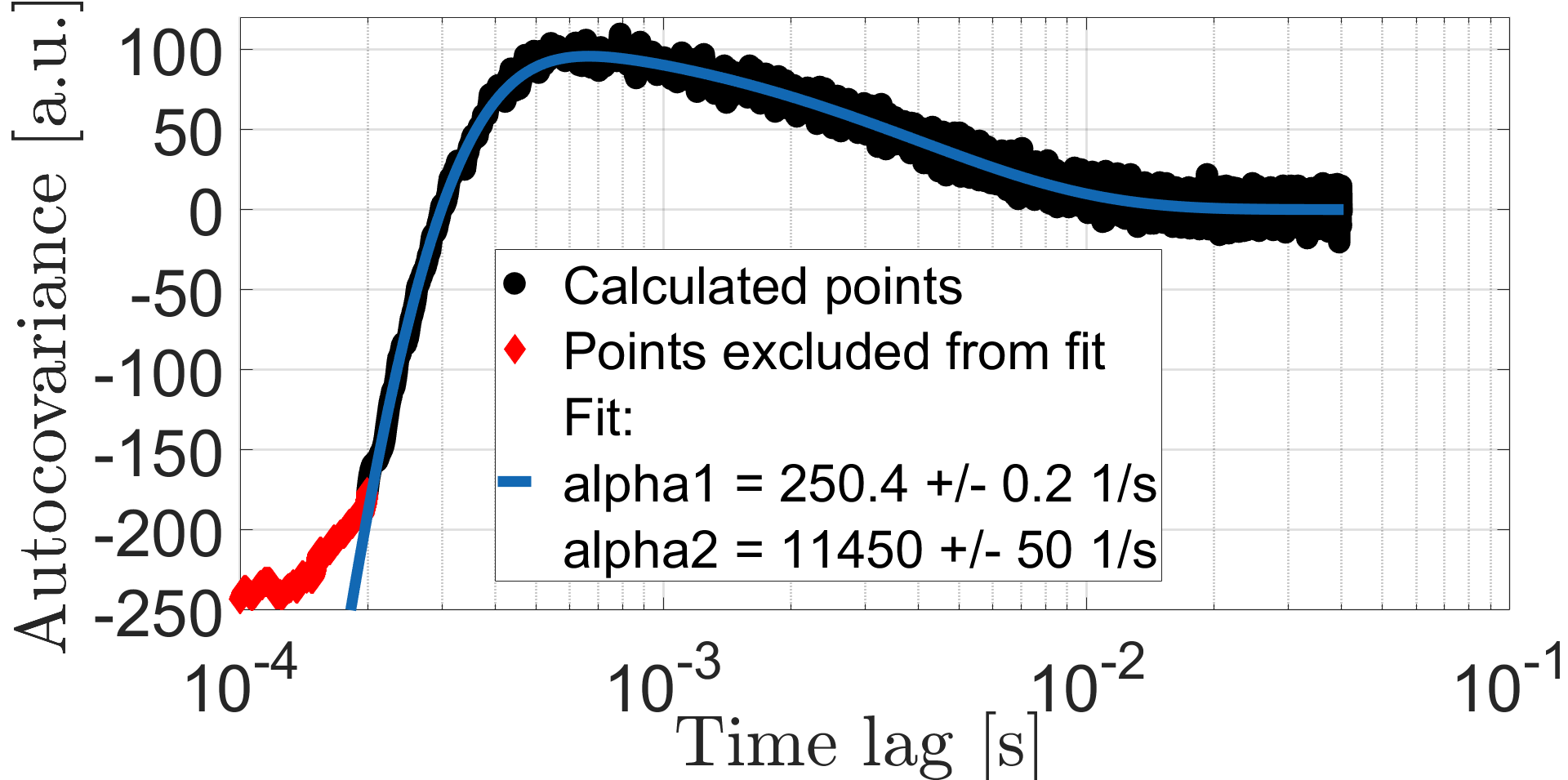}}
    \subfloat[ACF from the CR-2 measurement.]{\includegraphics[width=0.45\textwidth]{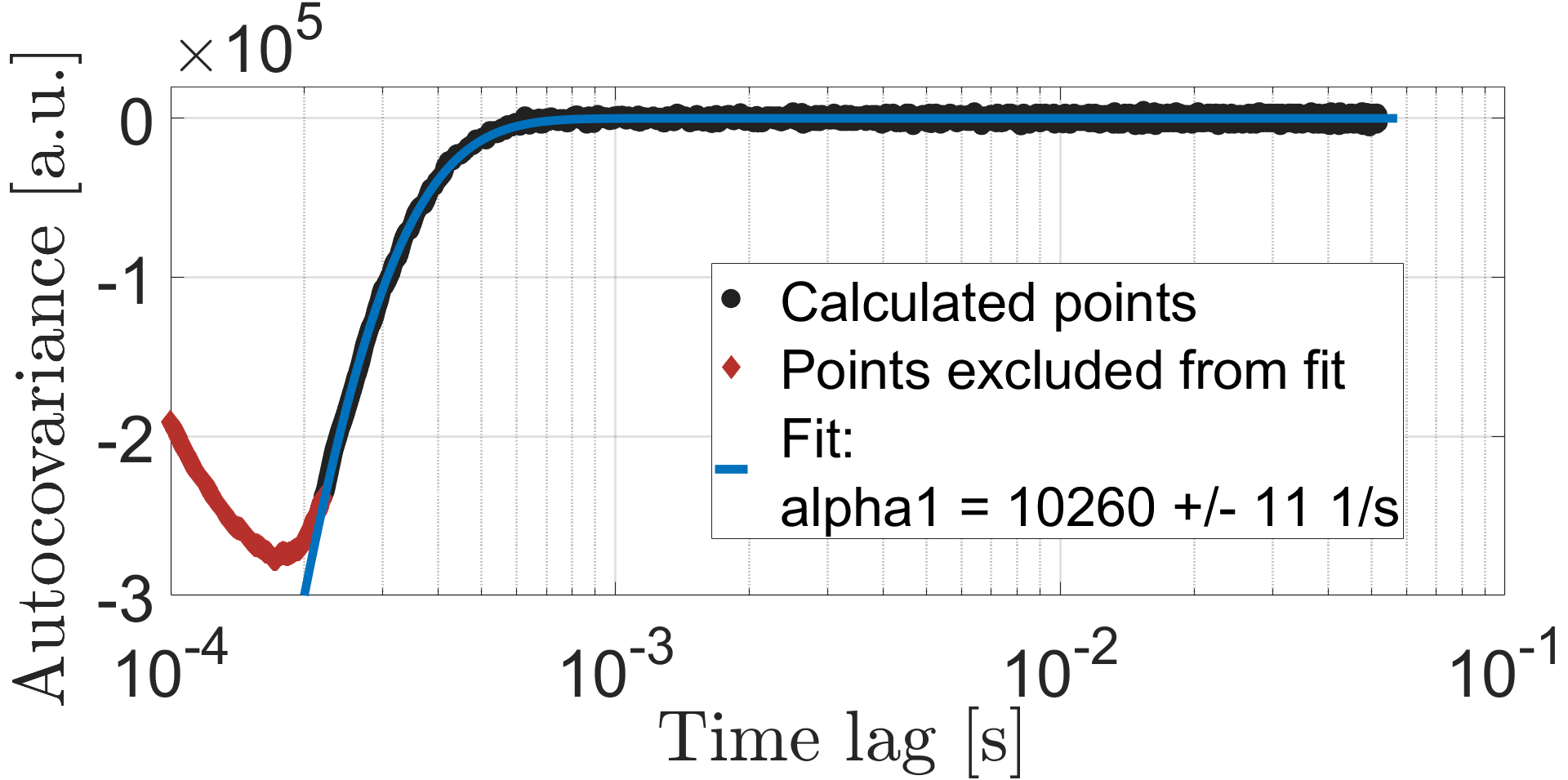}}
    \caption{The squared magnitude of the Fourier transform (proportional to the power spectral density), its relevant section fitted with the appropriate functions defined in eq. \ref{eq:PSD_theoretical_function} and the autocovariance function of the continuous signal of detector 'A' in the CR-3 configuration (compressed recording) and the CR-2 configuration (continuous recording).}
    \label{fig:Negative_correlation_Kuca_ACF_PSD}
\end{figure*}

The results showed that noise suppression had a negligible influence on the covariance functions of the continuous signals. This behaviour can be attributed to a well-known property of the covariance function: as a second-order moment, it inherently reduces the contribution of small-amplitude, uncorrelated signal components, such as the uncorrelated portion of electronic noise. Consequently, it was concluded that the analysis of the time-resolved detector signals is effectively insensitive to uncorrelated electronic noise.

Using the continuous and traditional pulse-based signals, only measurements of subcritical (SCR-1,2) and the lowest power critical (CR-3) configurations could be evaluated successfully. In the remaining two, higher detection rate configurations (CR-1,2), the pulse-based signal could not be used due to too many lost pulses, and the continuous signal could not be recorded in the \emph{compressed} format, as there were no sufficiently long time intervals without detection events for the compression to be feasible. Both the \emph{compressed} and \emph{continuously} recorded continuous signals produced unexpected artifacts given the theoretical and simulation results, most importantly a high amplitude, but quickly decaying (decay constant of $\sim10^4\text{~s}^{-1}$) negative correlation component. This distorted both the autocovariance and variance-to-mean functions. The correlation corresponding to the fundamental generally could be recovered when evaluating the \emph{compressed} signals, but not the \emph{continuously} recorded ones. The problem is suspected to be caused by the nonlinear frequency transfer of the data acquisition chain, and can be understood by looking at the Power Spectral Density (PSD) which is the Fourier transform of the autocovariance function by the Wiener-Khinchin theorem \citep{MILLER2012429} and the ACF itself of different continuous signal types, shown on Figure \ref{fig:Negative_correlation_Kuca_ACF_PSD}. 
The spectra of the two signals differ significantly below 10~kHz. The continuously recorded signal shows a V-shaped band stop filter-like behaviour on a log-log scale, centered between 20 and 30~Hz, which is mostly missing from the compressed signal. The compressed recording of the signal, as expected, also suppresses discrete low-frequency noises such as the 60~Hz ripple and its harmonics. The effect of the fundamental prompt decay constant appears as a low-pass filter in the power spectral density, cutting off at $\alpha/2\pi\approx40$~Hz, which is only visible in the spectrum on the left side. This is suppressed in the other spectrum, as the effect of the V-shaped frequency transfer is greater by several orders of magnitude. However, since this effect is mostly missing from the spectrum of the compressed signal, it is an effect of the frequency transfer of the measurement chain rather than that of neutron detection. The relevant section of the spectra can also be fitted with the appropriate function corresponding to exponentially decaying correlations in the time domain \citep{RPNR}:
\begin{equation}
    F(\omega)=C\left(1+\sum_i\frac{Y_i\alpha_i^2}{\alpha_i^2+\omega^2}\right),
    \label{eq:PSD_theoretical_function}
\end{equation}
where $\omega$ is the angular frequency. Three terms were used for the CR-3 configuration and two terms for the CR-2 configuration, where the prompt decay constant of the reactor could not be found by fitting. The results of the fits are shown on Figure \ref{fig:Negative_correlation_Kuca_ACF_PSD}.
The $\alpha$ value returned by fitting the PSD of the compressed signal recorded in the CR-3 configuration is in good agreement with the decay constants obtained from the Rossi evaluations. In both cases, an additional, slightly faster ($\alpha\approx14000\text{~s}^{-1}$) positive correlation can be found, which was omitted from the ACF evaluations.

In order to make the continuously recorded continuous signal usable, similar to the simulations, the issue of the nonlinear frequency transfer of the measurement chain in the section relevant given the $\alpha$ value of measured system needs to be addressed in future work.
\begin{figure}
  \centering
    \subfloat[Continuous signal.]{\includegraphics[width=0.9\columnwidth]{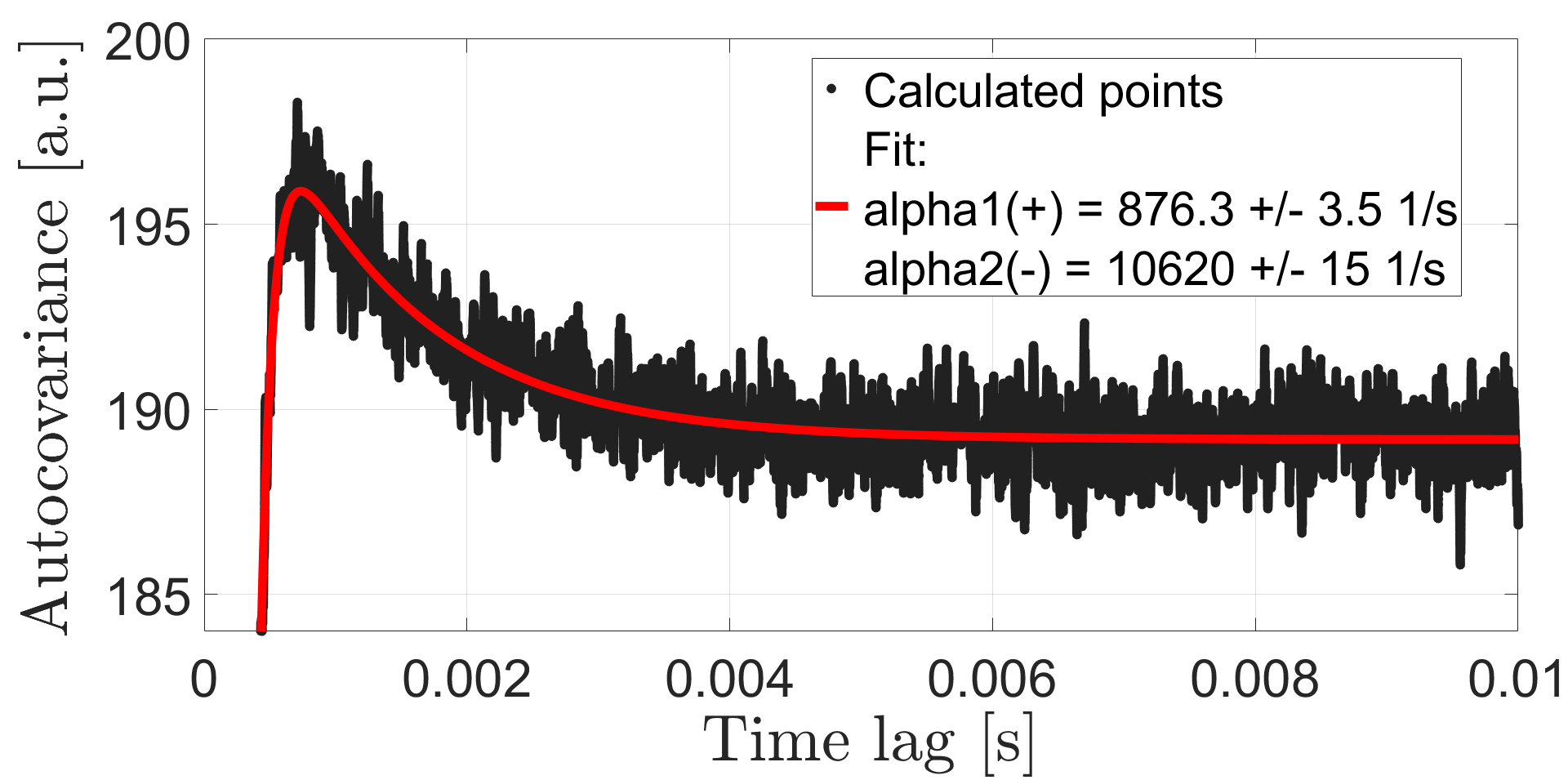}}\\
    \subfloat[Pulse-based signal.]{\includegraphics[width=0.9\columnwidth]{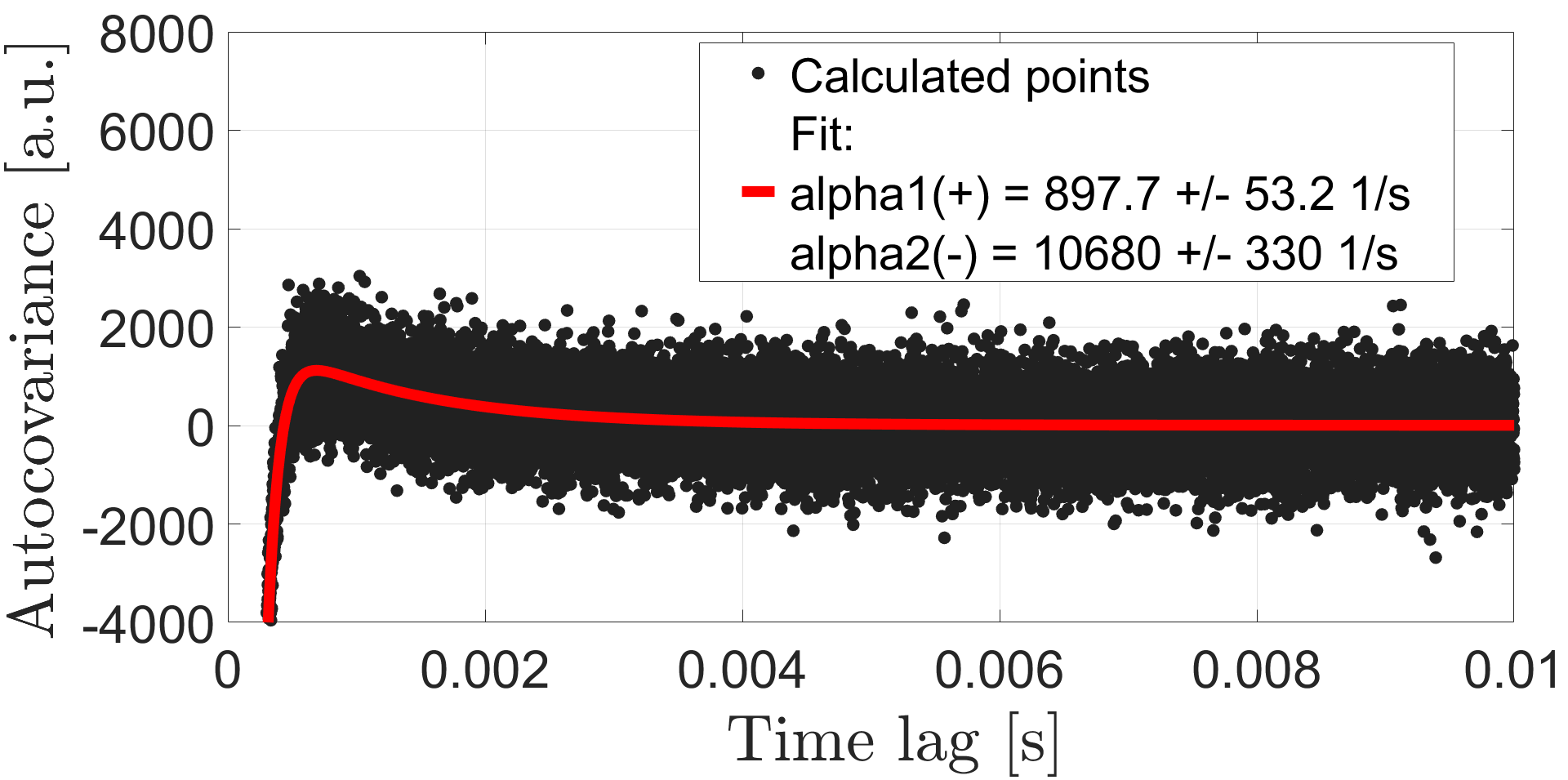}}
  \caption{Rossi-$\alpha$ evaluation of the continuous- and pulse-based signals of detector \textit{'A'} in the SCR-2 configuration.}\label{fig:Rossi_27AA}
\end{figure}

In the two subcritical and lowest power critical configurations, Rossi- and Feynman-$\alpha$ evaluations were performed successfully on both the traditional pulse-based and compressed continuous signals.
Figure \ref{fig:Rossi_27AA} illustrates the results of the Rossi-$\alpha$ analysis for detector \textit{A} in case of the most subcritical configuration (SCR-2). 

Two-detector evaluations were also performed by computing the cross-covariance (Rossi) and covariance-to-mean (Feynman) functions of the \textit{'A-B'} and \textit{'C-D'} detector pairs in the SCR-2 configuration. Detectors \textit{'A'} and \textit{'B'} were attached to one Red Pitaya (RP) device and detectors \textit{'C'} and \textit{'D'} to the other one. Due to the clock speeds and consequently the time measurements slightly drifting in the two RP devices relative to each other, two-detector evaluations were possible only using these very pairs, but not using detectors hooked up to different RP devices. In all configurations except the lowest detection rate SCR-2, only one detector was used per RP device due to bandwidth limitations meaning no possible detector pair-based evaluation.
\begin{figure}
    \centering
    \subfloat[Continuous signal.]{\includegraphics[width=0.9\columnwidth]{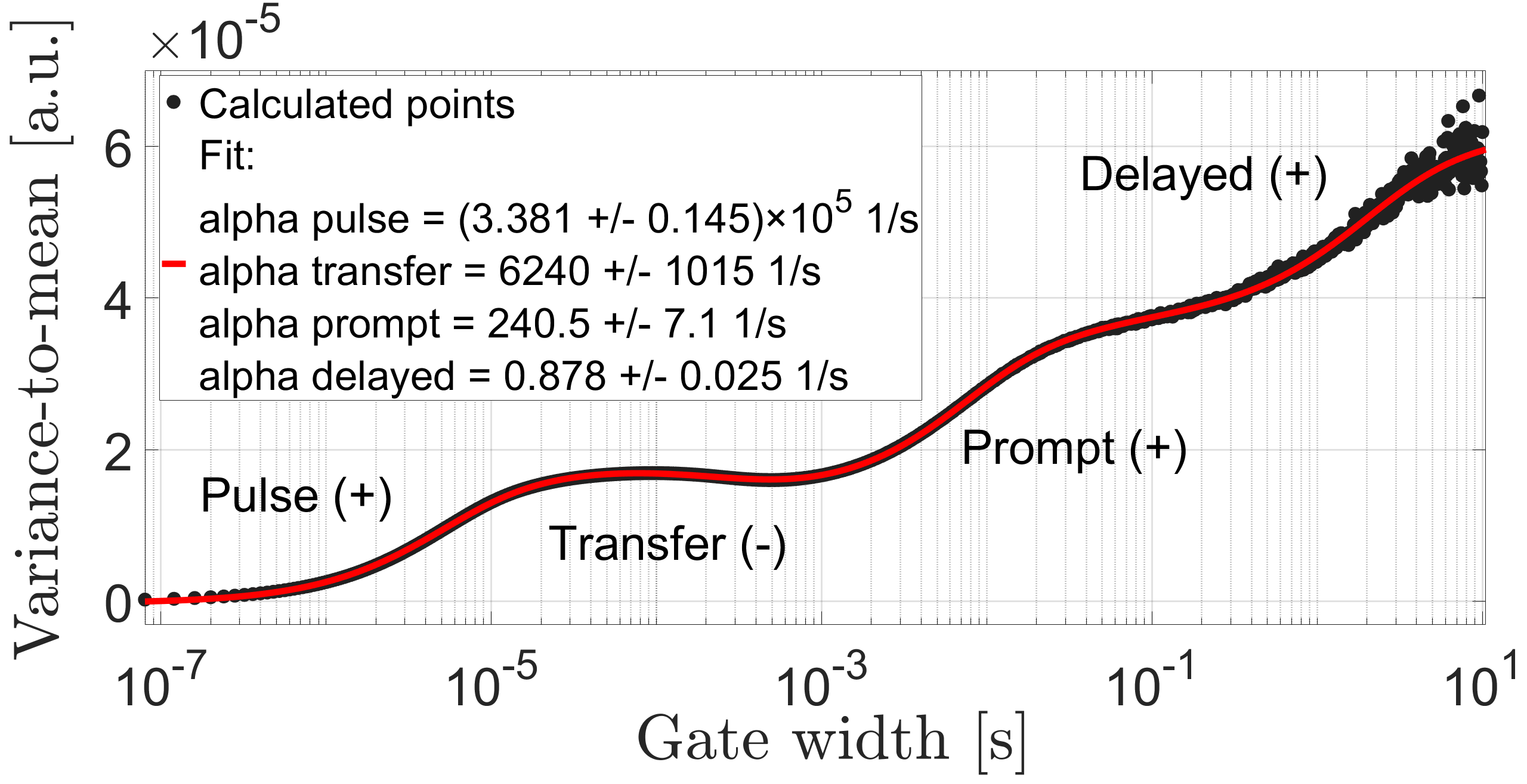}}\\
    \subfloat[Pulse-based signal.]{\includegraphics[width=0.9\columnwidth]{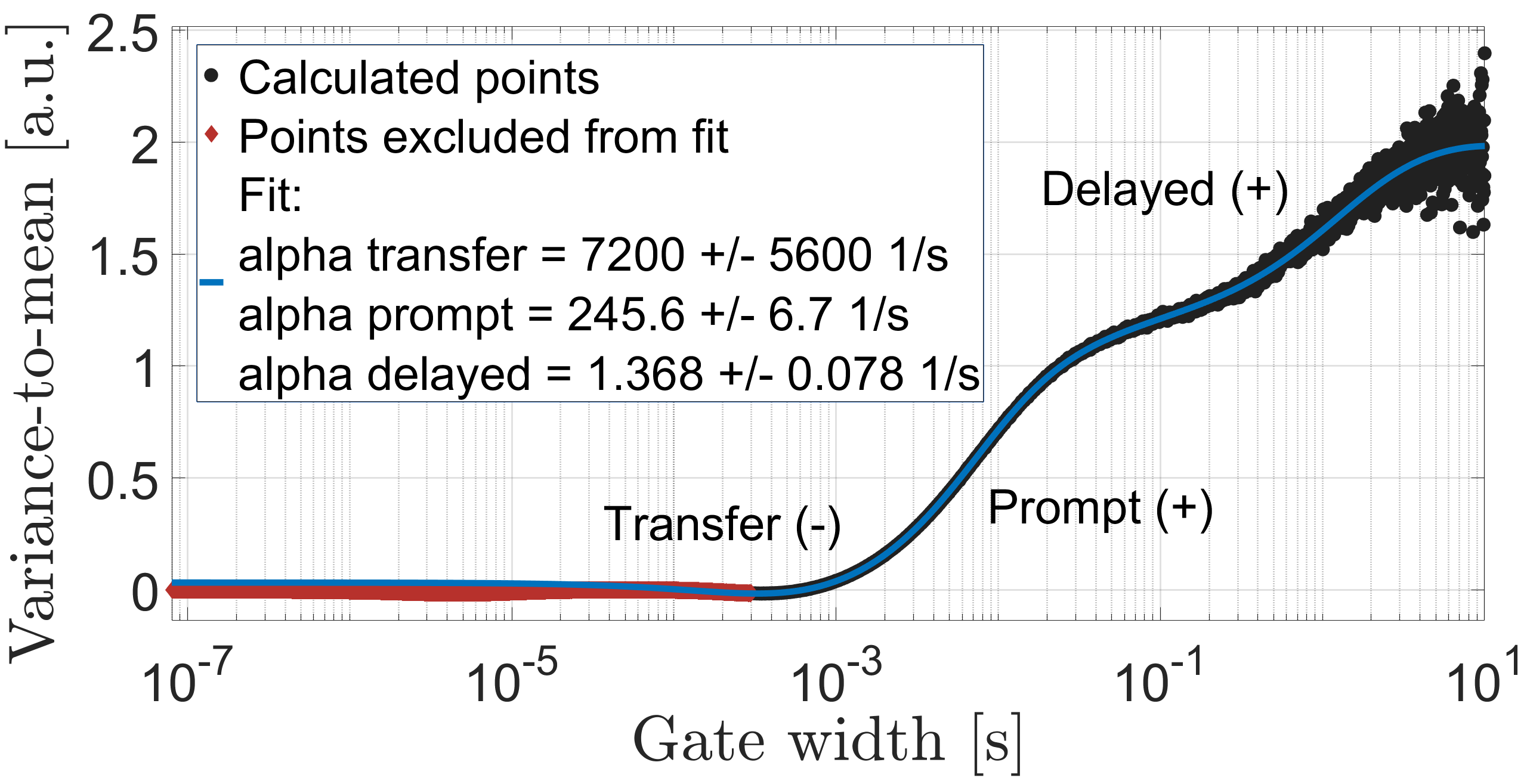}}
    \caption{Variance-to-mean evaluations of the continuous and pulse-based signals using detector \textit{'A'} from the CR-3 measurement.}
    \label{fig:KUCA_CR3_Feynman}
\end{figure}

\begin{figure}  
  \centering
  \includegraphics[width=0.9\columnwidth]{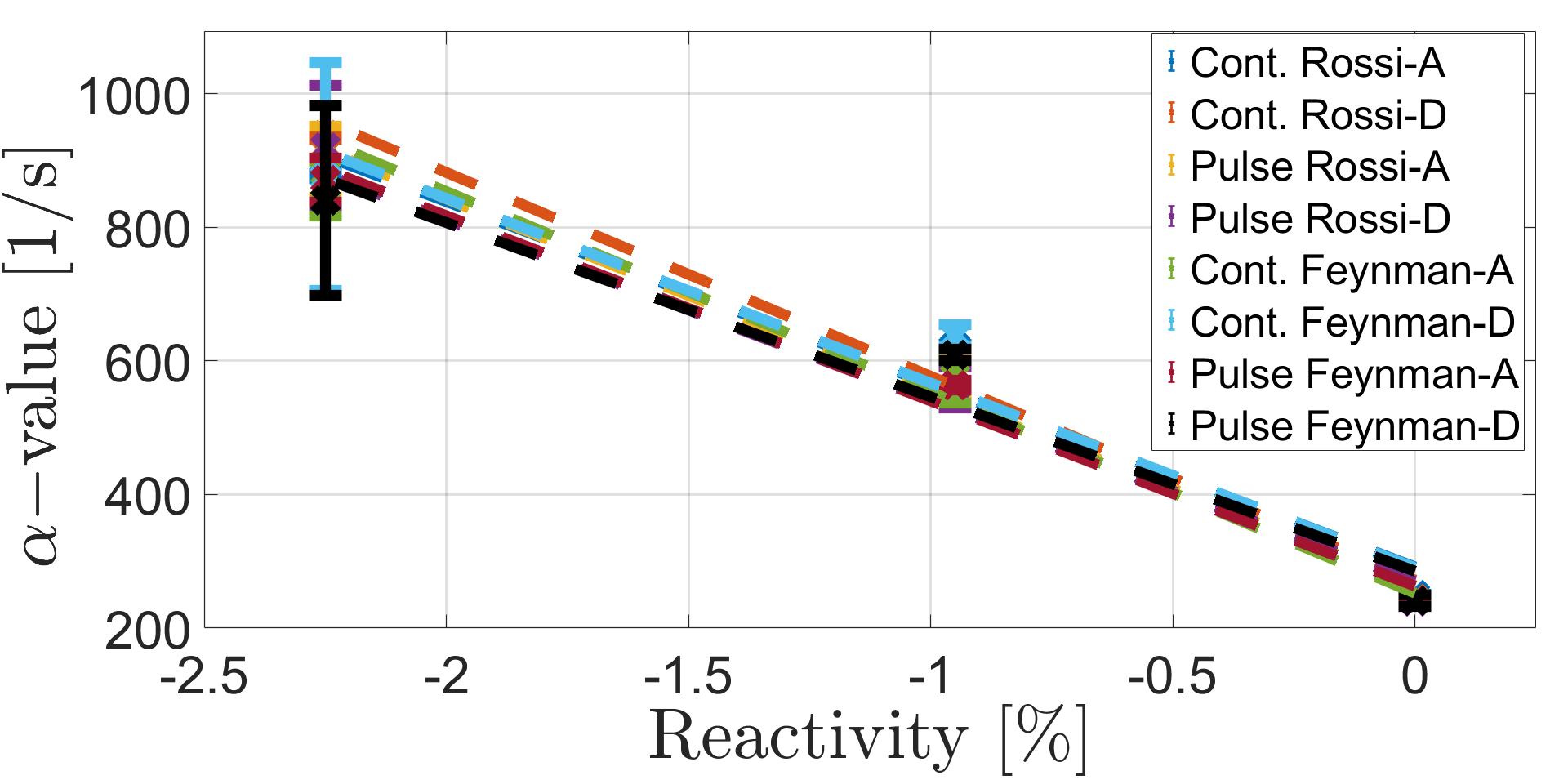}
  \caption{Relation between the estimated $\alpha$-parameters and the reactivity of the system for different detectors and evaluation techniques.}
  \label{fig:alpha_vs_rho}
\end{figure}

To be able to estimate the $\alpha$ values from the variance-to-mean ratio of signals of single detectors, a function containing at least two Feynman-$\alpha$ terms were fitted, the second one corresponding to the phantom negative correlations introduced by the nonlinear frequency transfer of the measurement chain with a corresponding decay constant $\alpha_\text{transfer}\approx10^4\text{~s}^{-1}$. The "real" $\alpha$ values of the system estimated by this technique are in good agreement with those yielded by the traditional pulse counting methods. In fact, the same technique was used in \emph{all} evaluations, including the traditional pulse-based and Rossi-evaluations since this way, a broader section of the calculated points could be included in the fit, and generally, the $\alpha$ values of the system could be estimated with lower uncertainties. In the CR-3 measurement, a third Feynman-$\alpha$ term corresponding to delayed neutrons was included for both signal types. In the case of the continuous signal, a function consisting of four Feynman-terms could be fitted remarkably well on the variance-to-mean curve for gate widths ranging from 80~ns to 10~s, the fourth term corresponding to the decay speed of the detector pulse. This result is in line with the theoretical results of \cite{KITAMURA2018691}, where it was shown that the finite decay constant of the detector pulse will result in an additional Feynman-$\alpha$ term if the pulse-shape is close to an exponential function. Naturally, the fitted function is not exactly matching eq. \ref{eq:AC_VTM_with_pulse_effect} containing three terms, as the pulse-shape-specific third term was omitted and the extra terms corresponding to the frequency transfer and the effect of delayed neutrons are missing in eq. \ref{eq:AC_VTM_with_pulse_effect}. Examples of the Feynman-evaluation of measurement CR-3 are presented on Figure \ref{fig:KUCA_CR3_Feynman}. Table \ref{tab:results_smoothed} lists all estimated $\alpha$ values obtained from the Rossi- and Feynman-$\alpha$ analyses.
Figure \ref{fig:alpha_vs_rho} shows the relation between the estimated Rossi- and Feynman-$\alpha$ parameters using detectors \textit{'A'} and \textit{'D'} and the reactivity of the system.
The results generally show good agreement between values obtained from the continuous signal analysis and the pulse counting methods and using different detectors or detector pairs.

\begin{table*}
\caption{Estimated $\alpha$ values from the Rossi- and Feynman-evaluations of the continuous and pulse-based signals for the low-power critical as well as the subcritical configurations (count rate in descending order).}
\centering
\begin{tabular}[t]{c c *{2}{r@{$\,\pm\,$}l} *{2}{r@{$\,\pm\,$}l} *{2}{r@{$\,\pm\,$}l} *{2}{r@{$\,\pm\,$}l}}
  \toprule
  \multirow{ 2}{*}{Configuration} & \multirow{ 2}{*}{Detector pair} & \multicolumn{2}{c}{$\alpha_{signal~analysis}$} & \multicolumn{2}{c}{$\alpha_{pulse~analysis}$} &\multicolumn{2}{c}{$\alpha_{signal~analysis}$} & \multicolumn{2}{c}{$\alpha_{pulse~analysis}$}\\
        &     & \multicolumn{4}{c}{Rossi [1/s]} &
       \multicolumn{4}{c}{Feynman [1/s]} \\
  \midrule
  CR-3  & A-A &  250.4 & 0.2 &  240.4 & 2.6 &  240.5 & 7.1 &  240.9 & 6.9 \\
        & D-D &  245.9 & 0.2 &  239.4 & 3.5 &  241 & 7.3 &  241.2 & 8.4\\
  \midrule
  SCR-1 & A-A &  622.6 & 2.5 & 579 & 33 &  577 & 41 &  563.7 & 7.5\\
        & D-D &  610 & 2.3 & 560 & 30 &  638 & 16 &  608.6 & 8.8\\
  \midrule
  SCR-2 & A-A &  876.3 & 3.5 &  898 & 53 &  858 & 41 &  870 & 33\\
        & B-B &  847.9 & 4.8 &  871 & 82 &  909 & 78 &  872 & 39\\
        & C-C &  924.7 & 3.1 &  945 & 47 &  964 & 39 &  940 & 23\\
  \vspace{.1cm} 
        & D-D &  937.6 & 6.2 & 919 & 94 &  876 & 171 &  840 & 142\\
        & A-B & 946.1 & 4.1 &  952 & 41 &  1018 & 22 &  964.4 & 3.2\\
        & C-D & 957.1 & 4 &  922 & 37 &  937 & 12 &  976 & 6.5\\
  \bottomrule
\end{tabular}
\label{tab:results_smoothed}
\end{table*}
\begin{figure}
    \centering
    \subfloat[Raw continuous signal.]{\includegraphics[width=0.8\columnwidth]{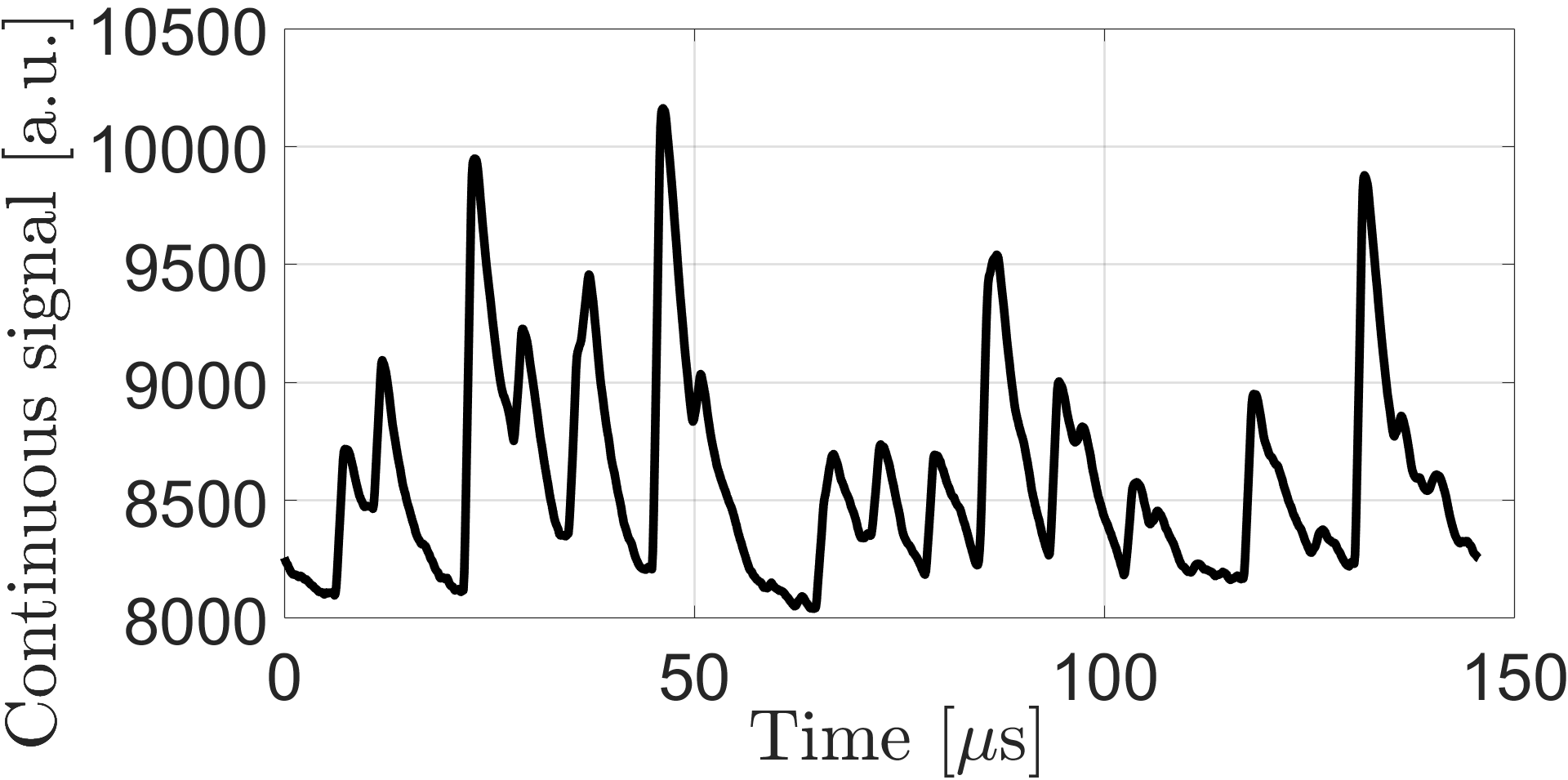}}\\
    \subfloat[Result of the inverse Fourier deconvolution.]{\includegraphics[width=0.8\columnwidth]{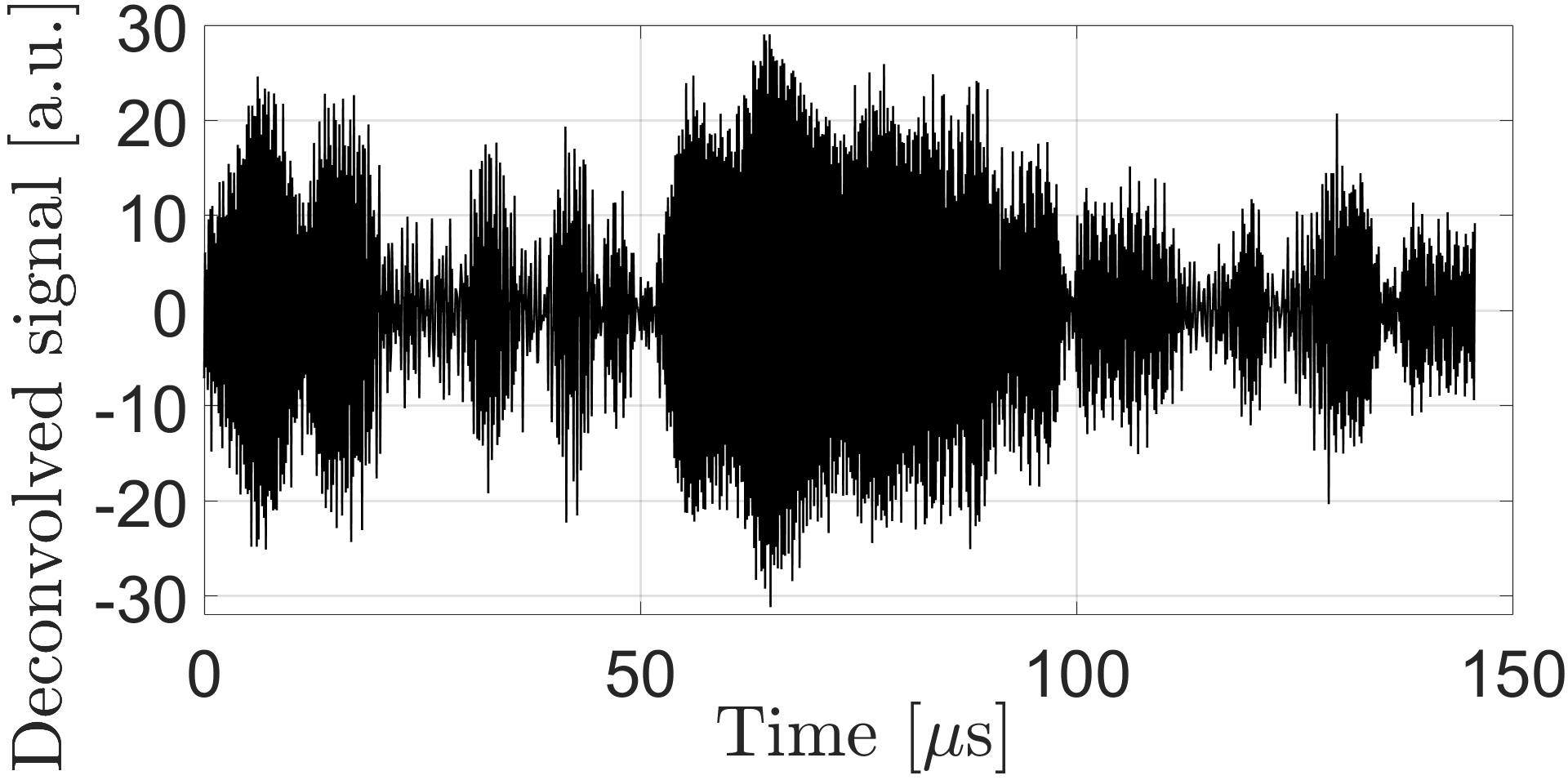}}\\
    \subfloat[Result of the Wiener deconvolution.]{\includegraphics[width=0.8\columnwidth]{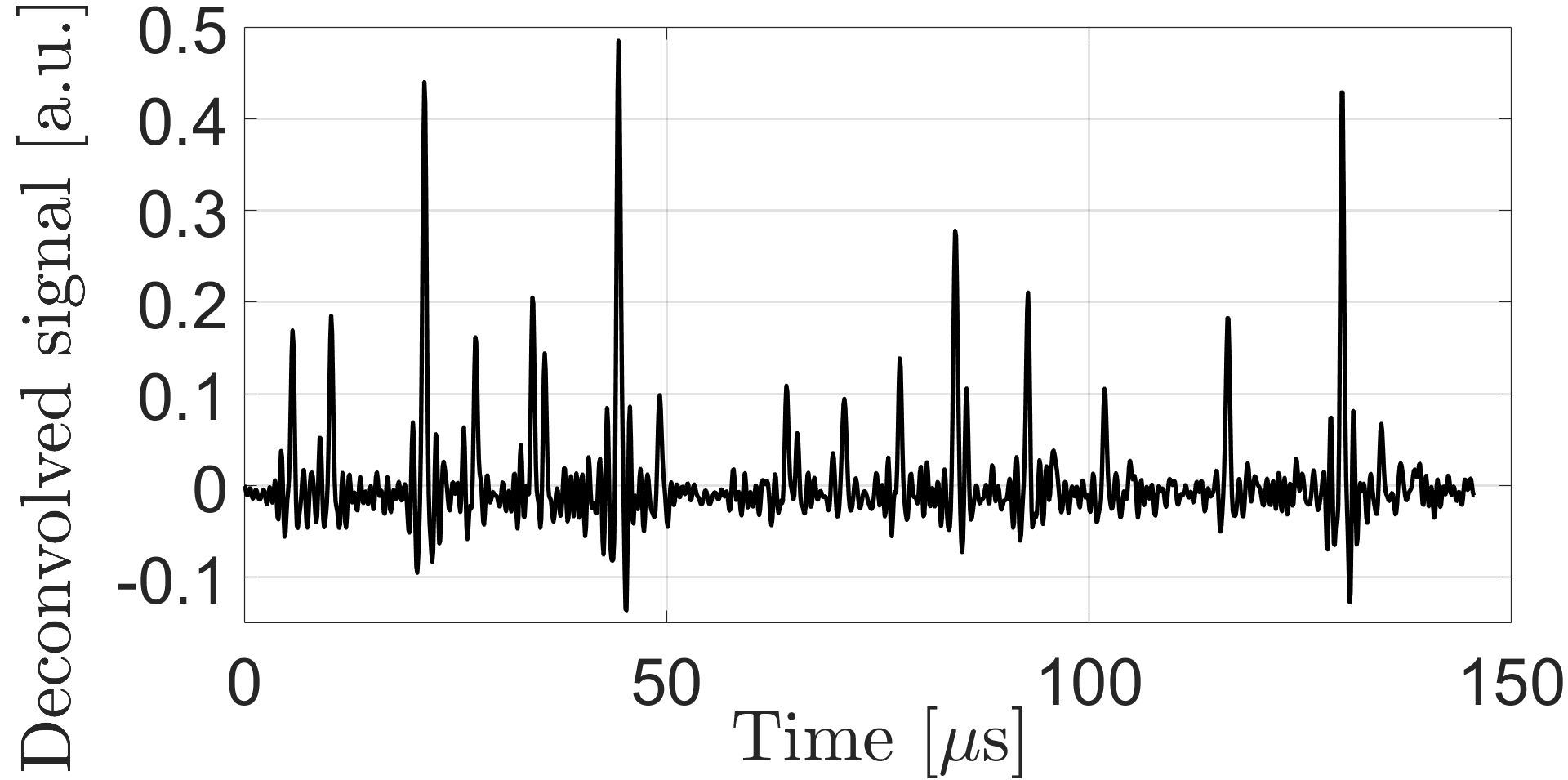}}
    \caption{Short section of the continuous signal of detector 'A' recorded in the CR-2 measurement, the result of the inverse Fourier deconvolution, and the result of the deconvolution using a Wiener filter assuming additive white noise.}
    \label{fig:KUCA_CR1_deconv_sig}
\end{figure}
One should note that the uncertainties of the fitted parameters reported here were obtained without accounting for the variances and covariances of the data points, which may lead to a significant underestimation of the fitting error. 
 As shown by \cite{ENDO2019606}, Rossi‑$\alpha$ and Feynman‑$\alpha$ estimators exhibit significant bin‑to‑bin correlations, which must be included to avoid biasing fit uncertainties, but has not yet been applied to the present evaluations. Applying their recommended bootstrap‑based or analytical error‑propagation methods is expected to increase reported uncertainties, but not to affect the central $\alpha$‑values, which already compare well with reference measurements.
Therefore, the scattering of the results is considered to be largely due to statistical uncertainty. Furthermore, no reference $\alpha$-value is available for the unique core configuration used, which also limits the conclusions from the measurements. According to Table \ref{tab:results_smoothed}, results from the evaluation of the continuous voltage signal show good agreement with the counts-based evaluation, which proves the unbiasedness of the method.


\begin{figure}
    \centering
    \subfloat[Rossi-$\alpha$ evaluation.]{\includegraphics[width=0.9\columnwidth]{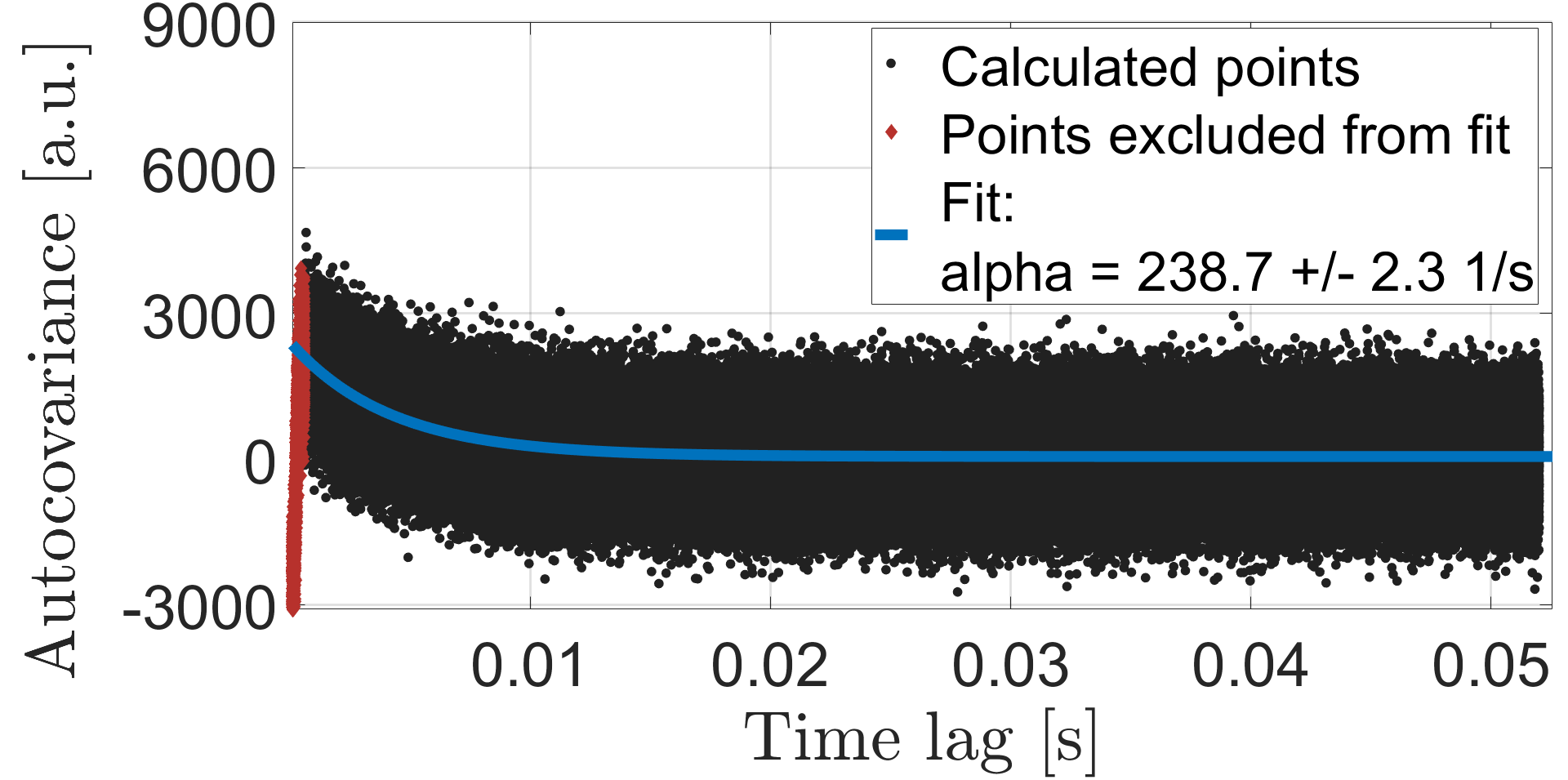}}\\
    \subfloat[Feynman-$\alpha$ evaluation.]{\includegraphics[width=0.9\columnwidth]{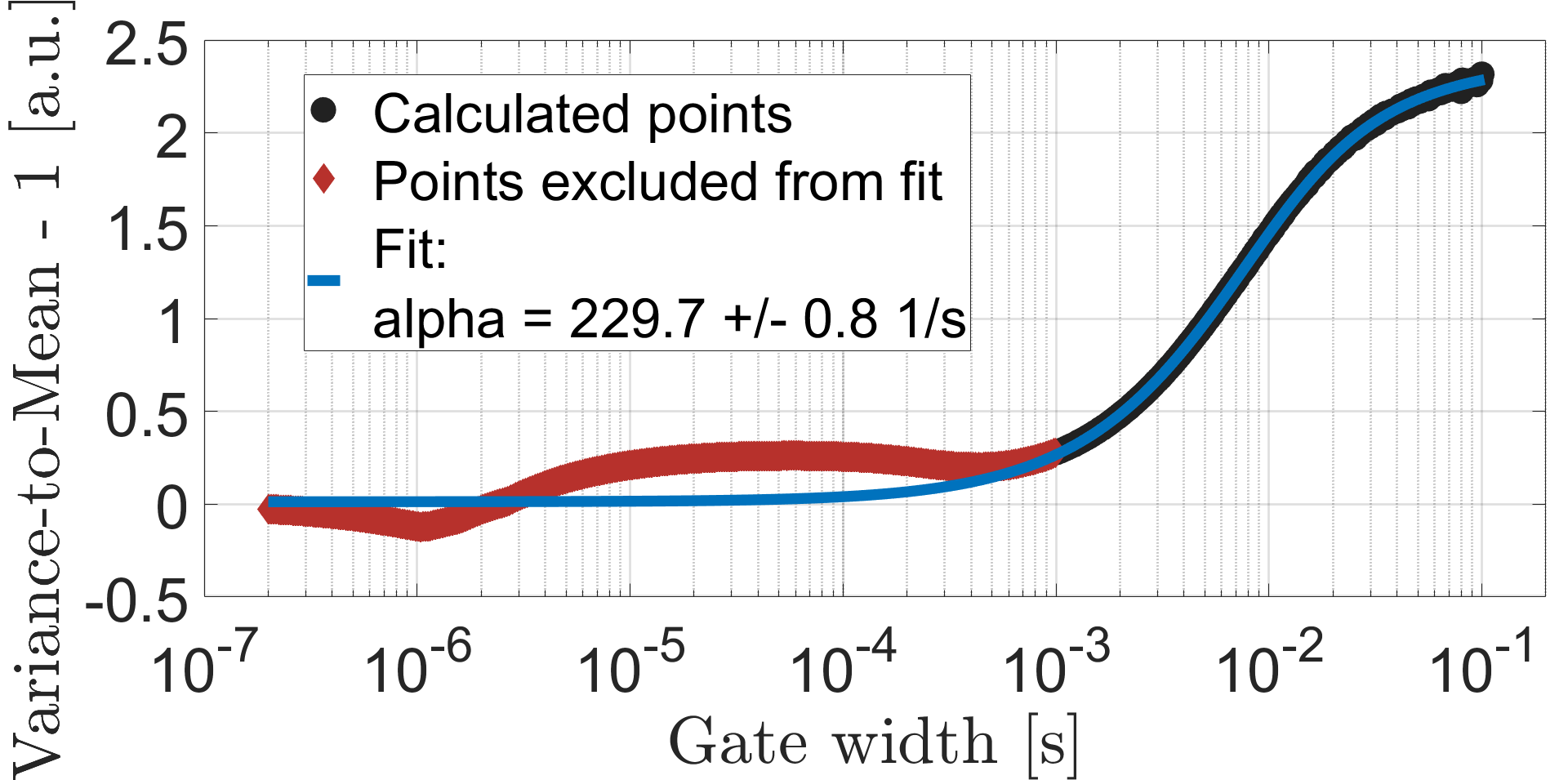}}
    \caption{Rossi- and Feynman-$\alpha$ evaluations of the deconvolved pulse-based signal 'A' of the CR-2 configuration.}
    \label{fig:KUCA_CR1_eval}
\end{figure}
Although the continuous and pulse-based signals could not be evaluated in configurations CR1-2 due to the frequency transfer related artifacts and the pileup of pulses respectively, data of the CR-2 configuration was successfully evaluated using the deconvolution technique. What caused some difficulties is the fact that there was no continuously recorded signal with sufficiently low detection rate where the pile-up of individual pulses was negligible. Therefore, the average pulse-shape used for the deconvolution was obtained from a compressed signal, namely that of the SCR-2 configuration (lowest detection rate). However, since the length of the recorded sections usually was not enough for the pulses to fully decay away, the tail region of the average pulse-shape was extrapolated by an exponential fit. Secondly, as the sampling rate of the compressed and continuously recorded signals differed, the average shape was resampled to match the sampling rate of the continuously recorded signal. Thirdly, the signal recorded in a compressed format used a 21-point moving average smoothing filter, while a 9-point moving average was used for the continuously recorded signals (CR-1, CR-2) having different spectral artifacts in the high-frequency region (see Figure \ref{fig:Negative_correlation_Kuca_ACF_PSD} (a-b)). This resulted in an imperfect estimation of the average pulse-shape, and made the deconvolution more difficult. 
\begin{figure}
    \centering
    \includegraphics[width=0.9\columnwidth]{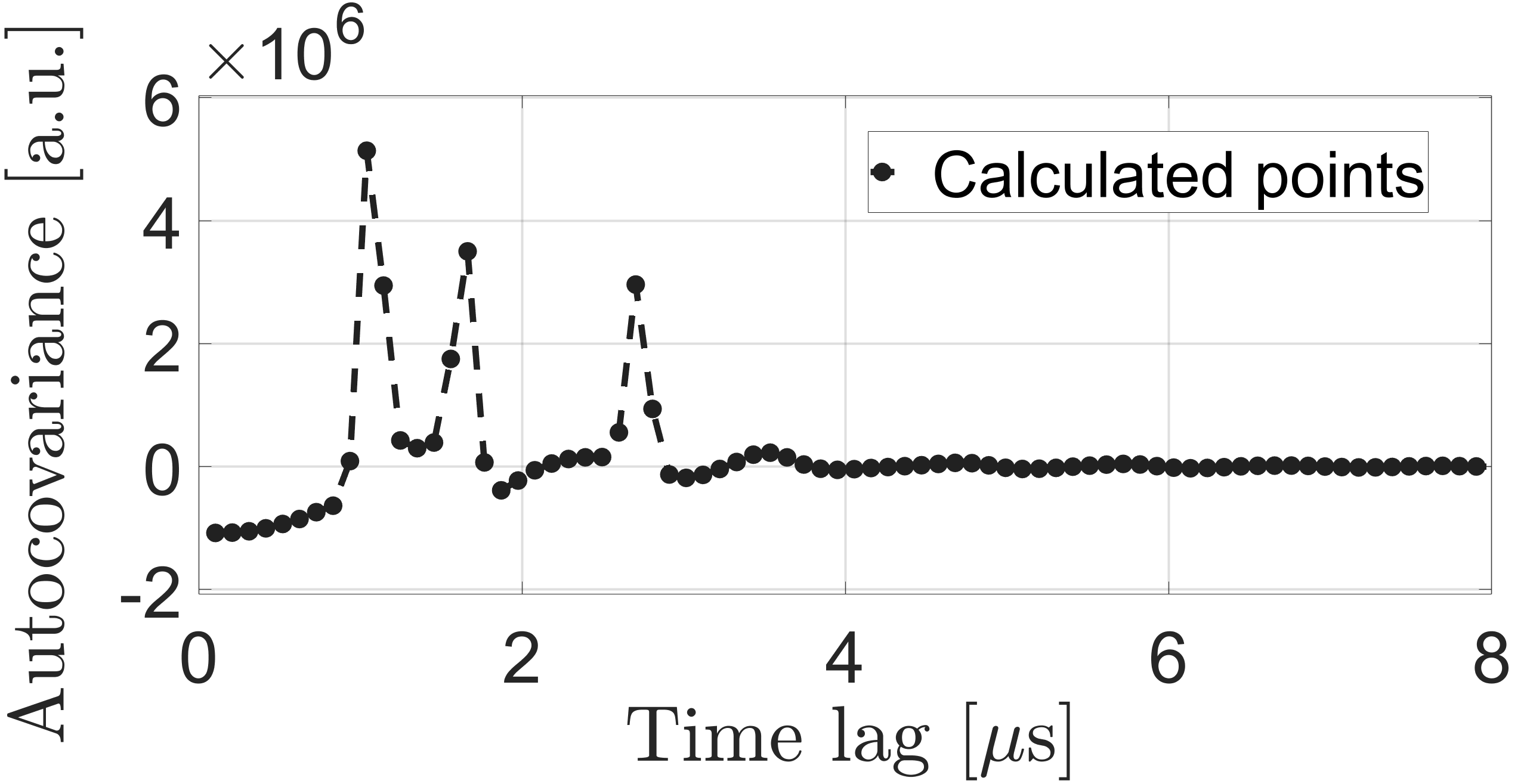}
    \includegraphics[width=0.9\columnwidth]{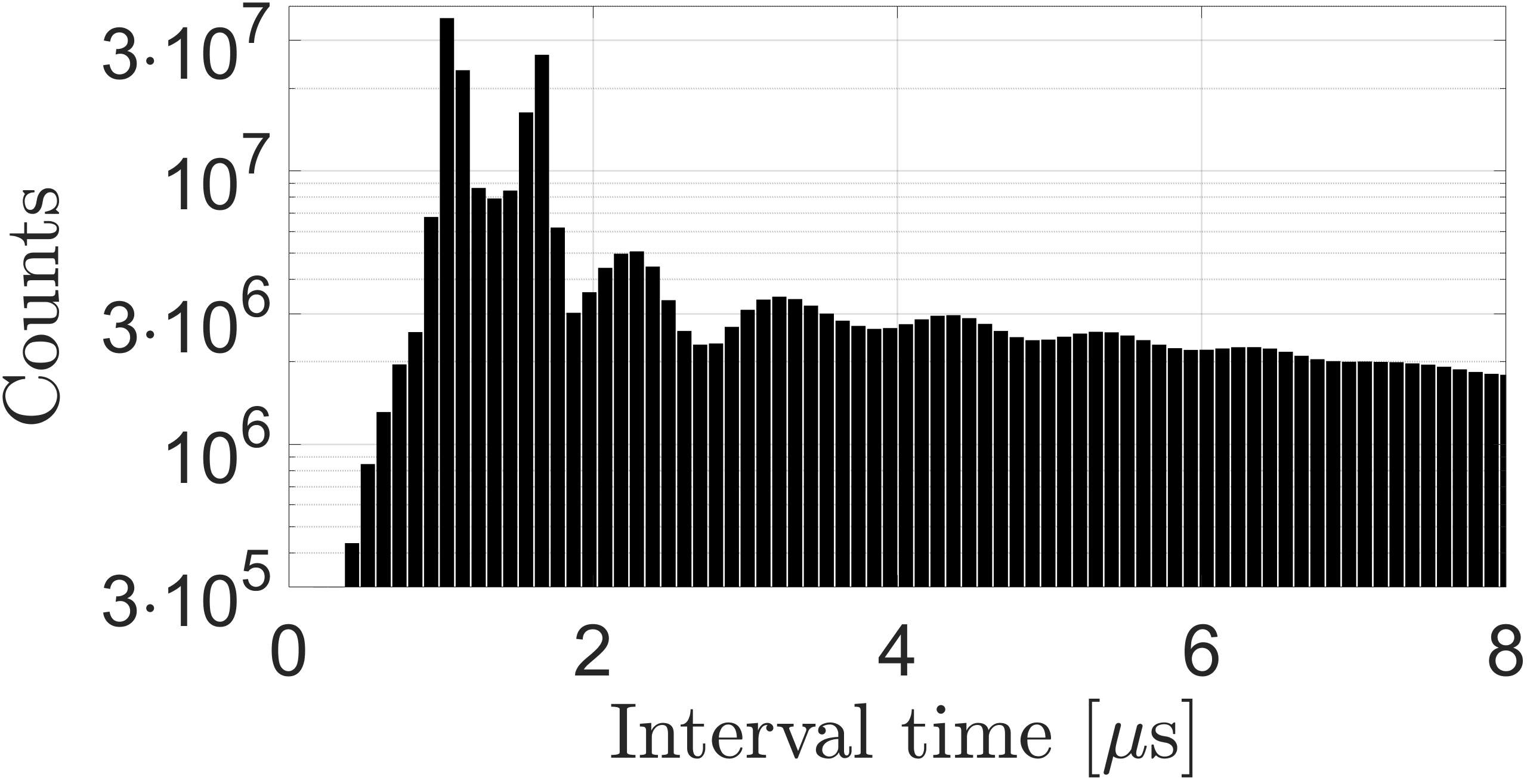}
    \caption{The autocovariance function and the ITD of the deconvolved pulse-based signal of detector 'A' in the CR-2 configuration.}
    \label{fig:KUCA_CR1_ACF_TID}
\end{figure}
The imperfect estimation of the average pulse-shape meant that the simple inverse Fourier technique could not be applied. Therefore, a custom Wiener filter was applied during the deconvolution assuming additive white noise for simplicity. The strength of this filter was tuned to suppress the unwanted high-frequency oscillations at each pulse, but do not widen the pulses in the deconvolved signal too much. Figure \ref{fig:KUCA_CR1_deconv_sig} shows a short section of the original recorded signal, the result of the simple inverse Fourier deconvolution and the result of the deconvolution using the custom Wiener filter.

The result of the deconvolution was converted into a pulse-based signal using an appropriate threshold level, which was successfully evaluated using both the Rossi-$\alpha$ and Feynman-$\alpha$ techniques. The results are in good agreement with those of the continuous and pulse-based signals recorded in the lower power CR-3 configuration.

One can notice an unexpected oscillatory behaviour in the Variance-to-Mean function for short Gate widths (<~1~ms). This is the result of the imperfect high frequency oscillation (ringing) at each detection event in the deconvolved signal, which is the effect of the imperfectly recorded average pulse-shape. These ringing oscillations are generally not powerful enough to be recorded as counts, but they increase the probability for counts to be recorded at certain interval times. This effect can be observed well when looking at the interval time distribution (ITD) of the recorded counts as well as the initial section of the Autocovariance function (Time lag~<~10~$\mu$s), shown in Figure \ref{fig:KUCA_CR1_ACF_TID}.
Given the difficulties in recording the average detector pulse-shape correctly, it is remarkable that the measurement of the CR-2 configuration could be evaluated accurately using the Wiener deconvolution technique. Together with the limited time resolution of 104~ns, the highest power CR-1 measurement could not be evaluated, even using deconvolution as the power (and therefore the detection rate) was increased $\sim30$ times from CR-2. This meant that the average time between detections was comparable to the time resolution of the recorded signal, making reliable pulse detection impossible even after the pulse-shape deconvolution.

\subsection{Measurements at BME TR}
More recently, in April 2024, another set of measurements were conducted at the Training Reactor of the Budapest University of Technology and Economics (BME TR).
\subsubsection{Instrumentation}
The BME TR is a light water moderated and cooled reactor of 100 kW nominal thermal power. The core consists of EK-10 type fuel assemblies, containing 10\% enriched UO$_2$ in a magnesium metallic matrix. The fuel region is surrounded by graphite reflector assemblies.

For the measurements, two KNT-31-type fission chambers were placed in the so-called vertical dry channel of the Training Reactor. The center of the dry channel is approximately 25~cm away from the reactor core and 40~cm away from the nearest fuel element. This meant that the detection efficiency achieved was much lower than that of the simulations. Consequently, the measurement duration was chosen in each configuration to be 30 minutes. One of the Red Pitaya STEMLab 125-14 devices and the in-house built preamplifiers were reused from the KUCA measurements.

Due to the lack of available space, only one of the preamplifiers could be placed near its detector in the dry channel. The other preamplifier was on the reactor deck, connected to its detector by a 6~m long cable. This resulted in a much noisier signal. This signal was eventually omitted from the evaluations, and only the less noisy one was used.

One subcritical (SCR) and three critical (CR1-3) configurations were used, at power levels ranging from 0.01~W (SCR) to 10~W (CR3).
\begin{table}
\caption{The approximate reactivity, positions of the automatic (A) and manual (M) control rods, the reactor power and the approximate dead time-free detection rate per detector at each measured state.}
\centering
\begin{tabular}{cccccc}
\hline
Name & $\rho$ [\textcent] & A [mm] & M [mm] & Power [W] & Det. rate [1/s] \\ \hline
SCR  & -10         & 502                 & 360                    & 0.01          & 500               \\ \hline
CR1  & 0           & 505                 & 385                    & 0.1           & 5000              \\ \hline
CR2  & 0           & 507                 & 385                    & 1             & 50000             \\ \hline
CR3  & 0           & 509                 & 385                    & 10            & 500000            \\ \hline
\end{tabular}
\label{tab:meas_parameters}
\end{table}

\subsubsection{Results}
\begin{figure}
    \centering
    \subfloat[CR1 configuration (compressed recording.)]{\includegraphics[width=0.9\columnwidth]{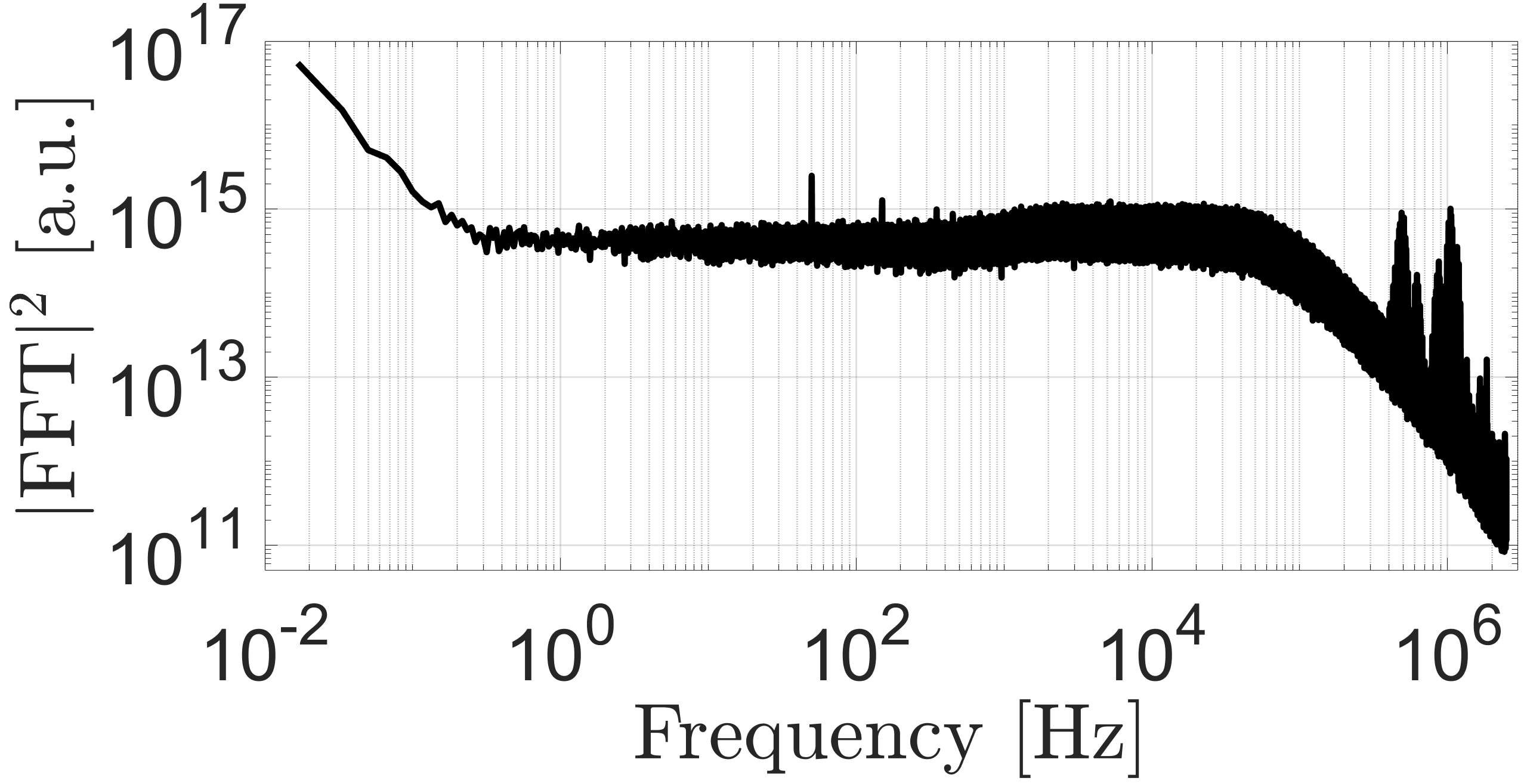}}\\
    \subfloat[CR3 configuration (continuous recording.)]{\includegraphics[width=0.92\columnwidth]{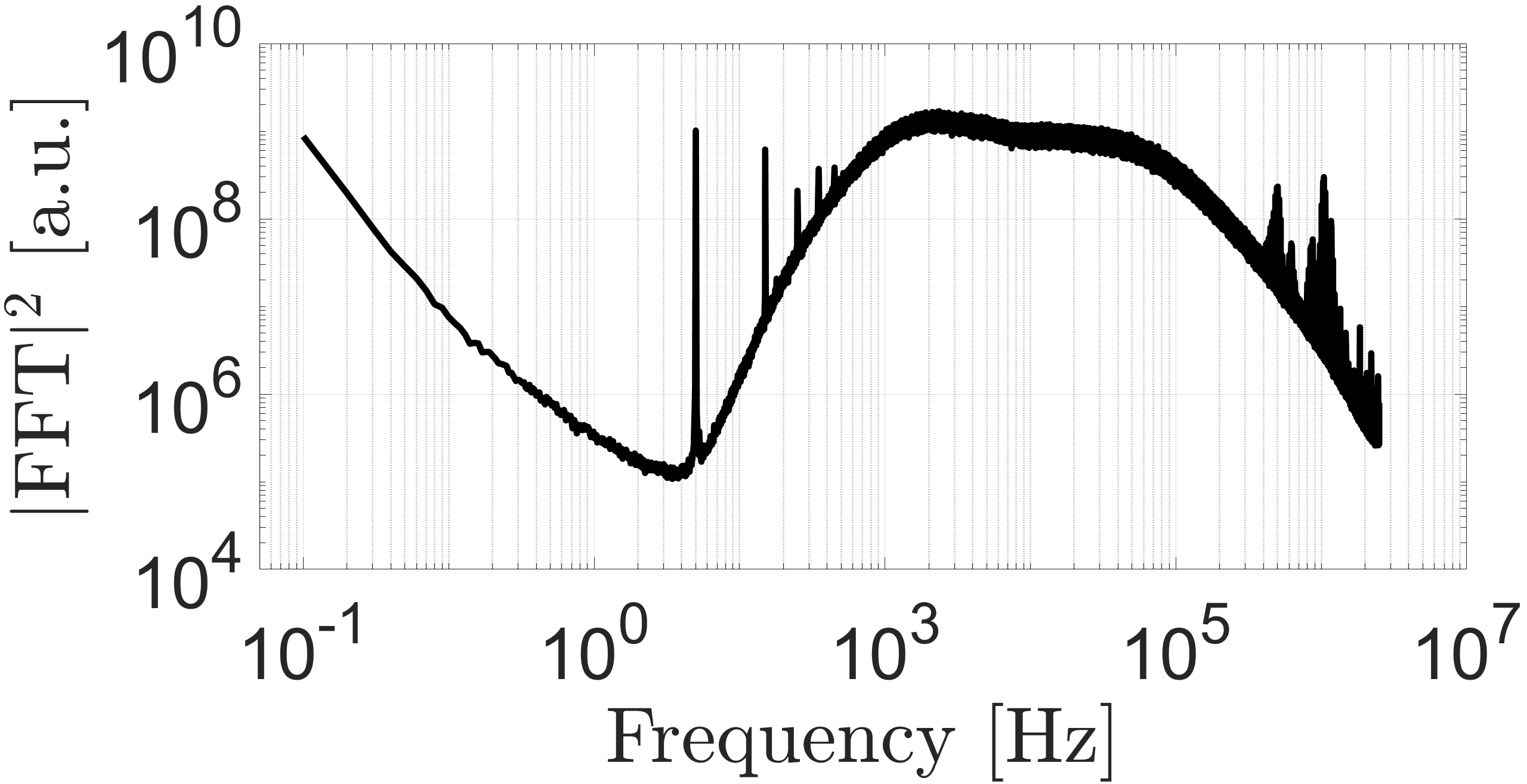}}
    \caption{Squared magnitude of the Fourier transform of signals recorded in the CR1 (compressed recording) and the CR3 (continuous recording) configurations.}
    \label{fig:BME_PSD}
\end{figure}

Similarly to the KUCA measurements, the frequency transfer of the measurement chain produced significant spectral artifacts, making the evaluation difficult. These artifacts were, to some degree, mitigated by the compressed recording of the continuous signal and by creating a better pulse-based signal using the deconvolution of the average pulse-shape. The power spectra of the compressed and continuously recorded signals follow a similar behaviour to those of the KUCA measurements, as illustrated on Figure \ref{fig:BME_PSD}.
The V-shaped bandstop filter-like behaviour of the continuously recorded signal can be observed again, which is mostly missing from the compressed recording.

\begin{figure}
    \centering
    \subfloat[Raw continuous signal.]{\includegraphics[width=0.9\columnwidth]{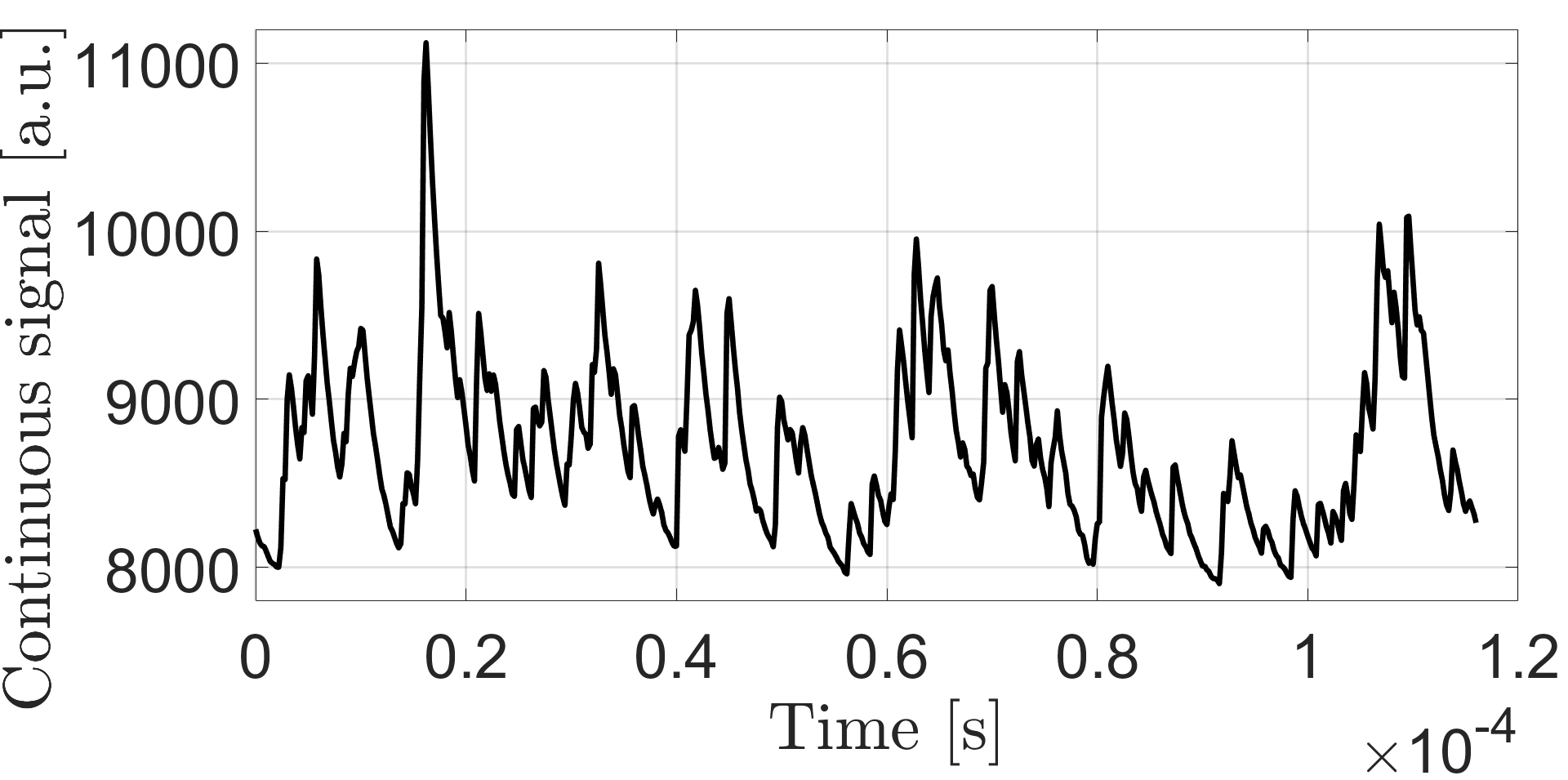}}\\
    \subfloat[Result of the inverse Fourier deconvolution.]{\includegraphics[width=0.9\columnwidth]{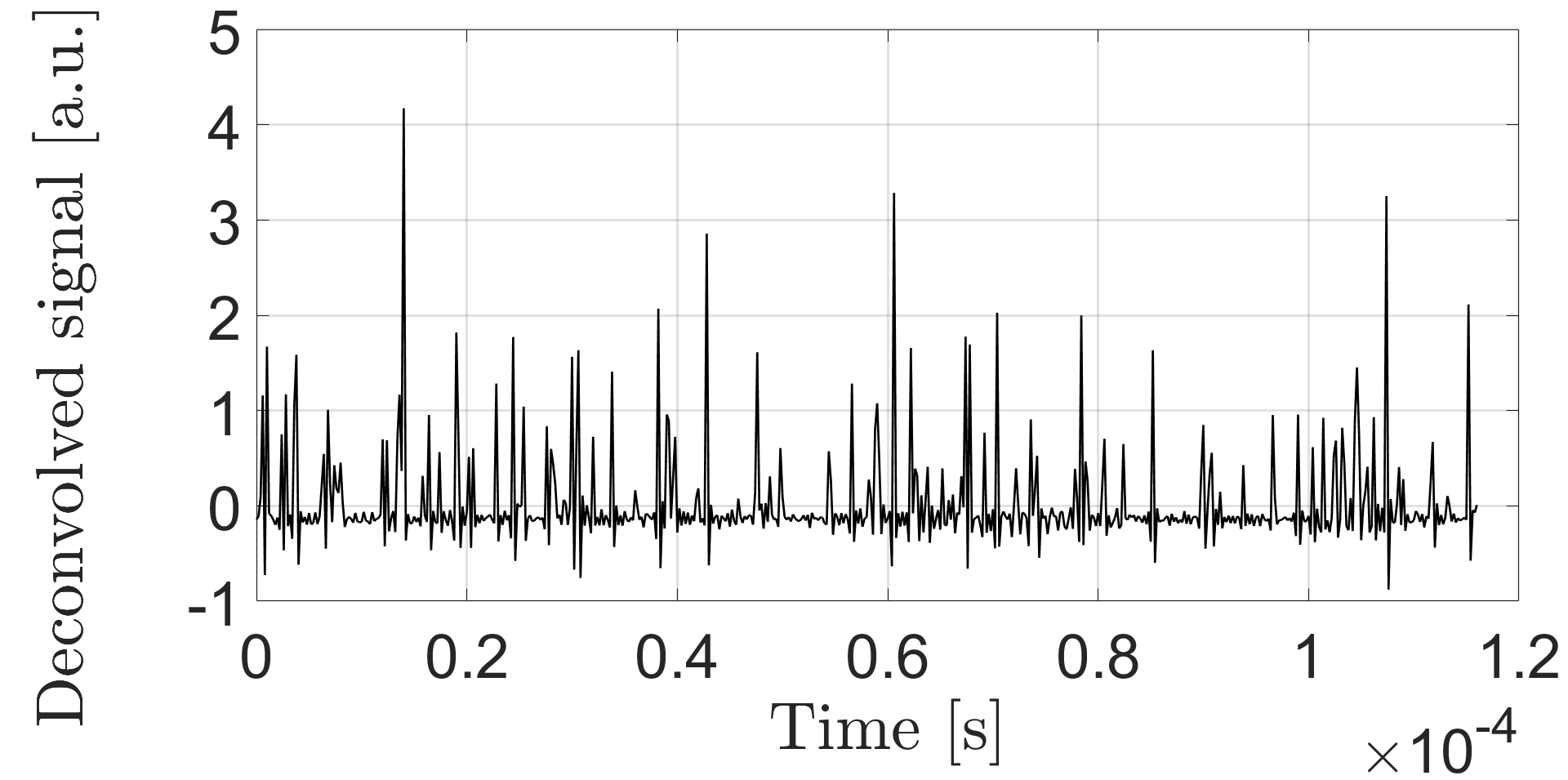}}
    \caption{A short section of the continuous signal recorded in configuration CR3, before and after the deconvolution of the average pulse-shape.}
    \label{fig:meas_cont_vs_deconv}
\end{figure}

The deconvolution of the average pulse-shape from the measured signal was performed successfully, even by the simple inverse Fourier method. A short section of the original continuous signal and the deconvolved signal, recorded in configuration CR3, are shown in Figure \ref{fig:meas_cont_vs_deconv}.
Similarly to the simulations and KUCA measurements, the deconvolved signal was used to produce a pulse-based signal with much shorter dead time than that of the traditional pulse-based signal. This technique is also resistant to the distortion effects introduced by the frequency transfer of the measurement chain. The results obtained from the different signal types are collected in Table \ref{meas_results}. 

\begin{table}
\caption{$\alpha$ values yielded by different signal types and noise techniques for the different critical measurements in s$^{-1}$ units.}
\centering
\begin{tabular}{cccccc}
\hline
Eval. type & CR1    & CR2      & CR3            \\ \hline
\multicolumn{1}{c}{Cont. ACF}    & \multicolumn{1}{c}{\begin{tabular}[c]{@{}l@{}}1490$\pm$141\\ 138.7$\pm$20\end{tabular}}    & \multicolumn{1}{c}{1260$\pm$52}     & \multicolumn{1}{c}{1314$\pm$57}       \\ \hline
\multicolumn{1}{c}{Pulsed ACF} & \multicolumn{1}{c}{254.7$\pm$196} & \multicolumn{2}{c}{\multirow{2}{*}{\begin{tabular}[c]{@{}c@{}}Unsuccessful evaluation\\ due to too many lost counts\end{tabular}}}      \\ \cline{1-2}
\multicolumn{1}{c}{Pulsed VTM}   & \multicolumn{1}{c}{219$\pm$90}     & \multicolumn{2}{c}{}          \\ \hline
\multicolumn{1}{c}{Deconv. ACF}  & \multicolumn{1}{c}{1177$\pm$249}     & \multicolumn{1}{c}{174.8$\pm$188} & \multicolumn{1}{c}{1930$\pm$710} \\ \hline
Deconv. VTM  & 121$\pm$31  & 163.1$\pm$41    & \multicolumn{1}{c}{296$\pm$320}          \\ \hline
\end{tabular}

\label{meas_results}
\end{table}

It can be said that most of the evaluations did not give a high-quality estimation of the expected $\alpha$ value, which can be attributed to the low detection efficiency given the position of the detectors. As expected, in the CR2 and CR3 configurations, the traditional pulse-based signal did not produce any evaluable data due to the dead time effect, whereas both the continuous and the deconvolved pulse-based signal did. In addition, the estimated $\alpha$ had considerably lower uncertainty when using the continuous and deconvolved signals in the CR1 configuration. In some cases, especially at higher power, the correlations in the signals were dominated by a decay constant of $\sim1500\text{~s}^{-1}$, and not the expected fundamental prompt decay factor of $\sim100\text{~s}^{-1}$. In one instance, namely the evaluation of the continuous signal of the CR1 configuration, both the fundamental and this $\sim1500\text{~s}^{-1}$ decay constant could be estimated by fitting the sum of two exponentials. In all the other cases, only one exponential term could be fitted successfully, resulting in the more dominant decay constant of the two or some intermediary value.
\begin{figure}
    \centering
    \includegraphics[width=0.32\columnwidth]{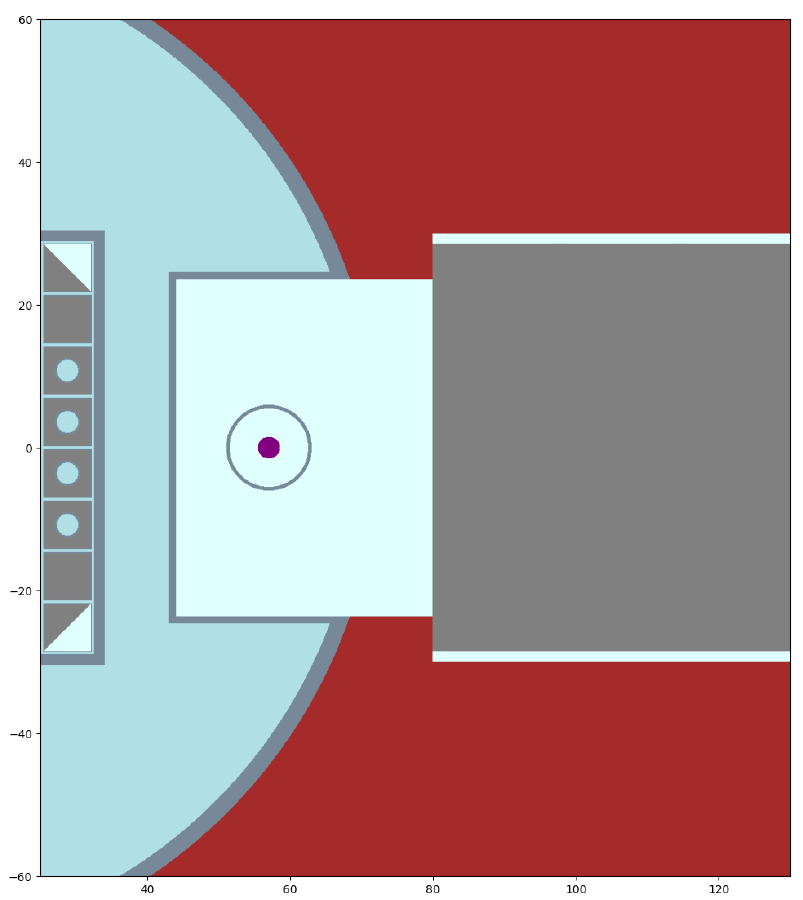}
    \includegraphics[width=0.63\columnwidth]{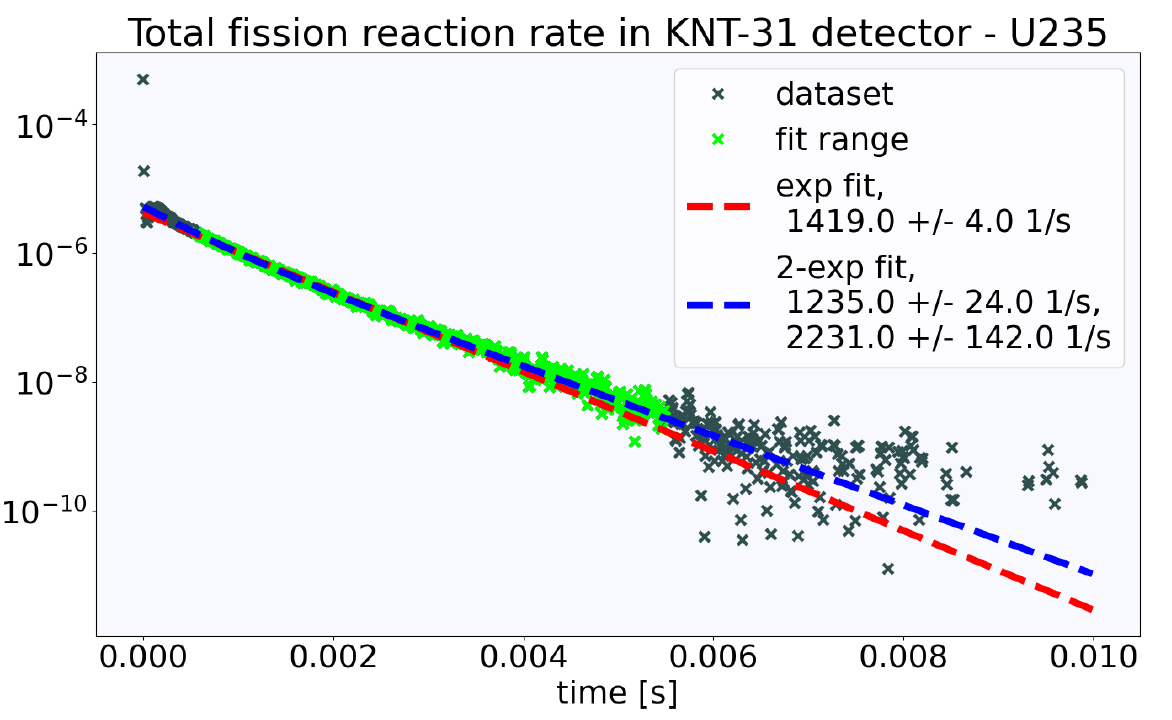}
    \caption{OpenMC model of the vicinity of the detector position, and the time distribution of detection events after starting a pulse of neutrons from the detector position.}
    \label{fig:1500_alpha_origin}
\end{figure}
\begin{figure}
    \centering
    \subfloat[Rossi-$\alpha$ evaluation of the continuous signal.]{\includegraphics[width=0.9\columnwidth]{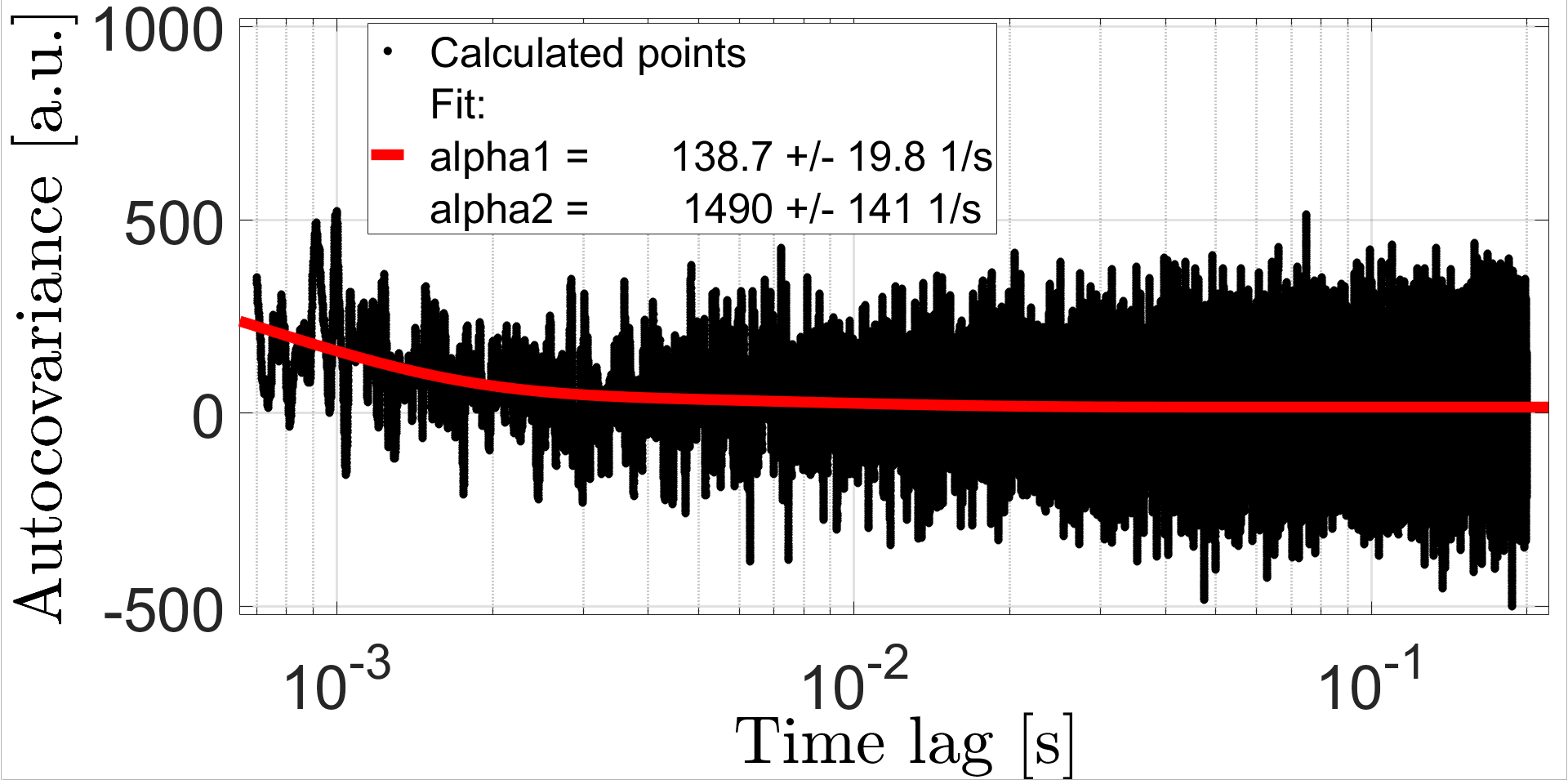}}\\
    \subfloat[Feynman-$\alpha$ evaluation of the deconvolved, pulse-based signal.]{\includegraphics[width=0.9\columnwidth]{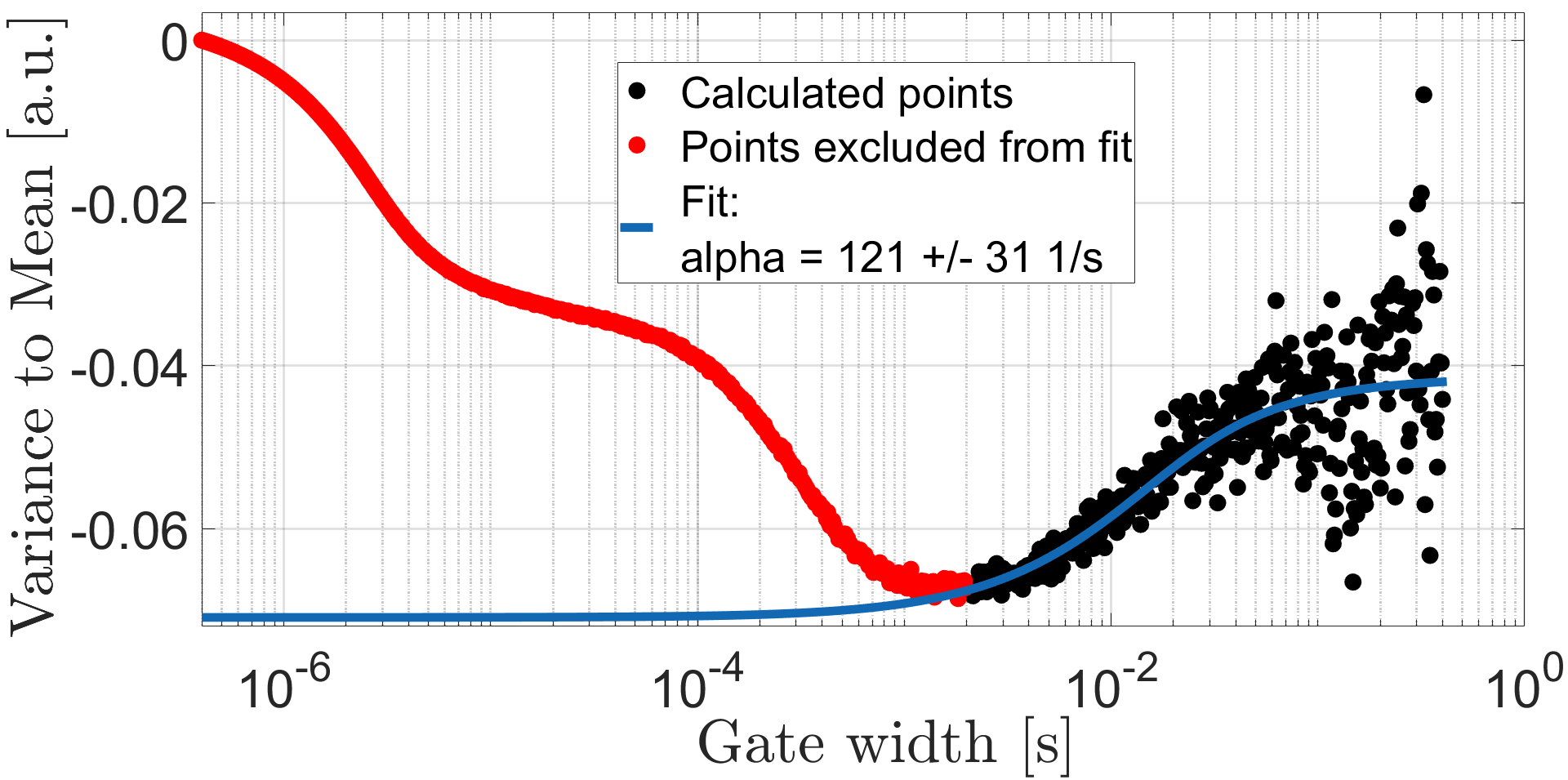}}
    \caption{Rossi-$\alpha$ evaluation (ACF, two fitted $\alpha$ values) of the continuous signal, and the Feynman-$\alpha$ evaluation (VTM, one fitted $\alpha$ value) of the deconvolved pulse-based signal of the CR1 measurement.}
    \label{fig:BME_succ_eval}
\end{figure}
The origin of this higher decay constant was investigated by performing OpenMC \citep{ROMANO201590} simulations of the geometry around the position of the fission chambers. Again, it is important to note that the detectors were a considerable distance away from the reactor core, and were mostly surrounded by moderator and reflector materials such as graphite and water. In the simulations, a pulse of neutrons was emitted from the detector position, and the time evolution of the detection rate was recorded. The top view of the layout near the detector position and the detection rate over time are shown on Figure \ref{fig:1500_alpha_origin}.

The results of the OpenMC simulation indicate that the higher decay constant of $\alpha\approx1500\text{~s}^{-1}$  indeed corresponds to the neutrons emitted from the fission chamber detector scattering back from the surrounding moderator/reflector material. It is quite remarkable that this effect is made visible by using both the continuous (compressed and continuously recorded) signal and the deconvolved pulse-based signal.

Examples of the Rossi-$\alpha$ evaluation of the compressed continuous signal and the Feynman-$\alpha$ evaluation of the deconvolved pulse-based signal are shown on Figure \ref{fig:BME_succ_eval}.

Although the determination of the reactor's fundamental $\alpha$-mode was difficult, it was possible in several configurations using the compressed continuous and deconvolved pulse-based signals. The overall weak statistics of these measurements can be attributed to the low detection efficiency achieved by the detector position being well outside the reactor core. In future work, this issue is planned to be mitigated by using miniature fission chambers that can be placed inside the reactor core. In addition, the usability of the continuous signal of gamma-sensitive neutron detectors such as $^3$He chambers or scintillators having much higher internal neutron detection efficiency compared to fission chambers is being investigated by the authors.
\section{General discussion}
The simulations and measurements consistently show that continuous‑signal neutron noise analysis offers clear advantages over traditional pulse counting at elevated detection rates. In the simulated cases, continuous‑signal Rossi‑ and Feynman‑$\alpha$ evaluations remained accurate well beyond the point where pulse‑based methods were affected by dead time and pulse pile‑up, and they enabled the estimation of significantly larger $\alpha$‑values. This makes the method suitable for fast or deeply subcritical systems where higher prompt decay constants must be resolved.

A key factor influencing performance is the detector pulse-shape. When the detector decay constant approaches the intrinsic $\alpha$ of the system, pulse-shape–induced short‑lag correlations appear as predicted by theory. These effects were effectively mitigated either by using detector pairs, which suppress pulse-shape distortions via cross‑covariance, or by deconvolving the average pulse-shape from the signal. The latter produced signals close to the ideal Dirac-delta sequence and improved single‑detector evaluations, provided that the average pulse-shape and high‑frequency noise were well controlled.
The experiments at KUCA and BME TR confirmed these trends but also highlighted practical limitations stemming from the data‑acquisition system. Nonlinear frequency‑transfer characteristics of the DAQ introduced artificial short‑lag correlations in continuously recorded signals, while low detection efficiency at BME TR restricted statistical precision. Compressed recording and Wiener‑filtered deconvolution mitigated some of these effects, and the appearance of a $\sim1500\text{~s}^{-1}$ component at BME TR was attributed to local neutron backscattering, as supported by OpenMC simulations.

Overall, the results demonstrate that continuous‑signal neutron noise analysis is a robust and unbiased method that extends the usable operating range of traditional noise techniques. Most remaining limitations arise from DAQ hardware rather than from the methodology itself, and planned improvements in electronics and pulse‑shape characterization are expected to further enhance performance.
\section{Conclusions}
This work demonstrated that neutron noise analysis based on the continuous signal of fission chambers is a viable and robust alternative to traditional pulse‑counting methods. Simulations and experiments at KUCA and BME TR consistently showed that continuous signal-based Rossi‑ and Feynman‑$\alpha$ evaluations remain accurate in high count rate-regimes where pulse counting is limited by dead time, and that they enable access to substantially higher prompt decay constants relevant to fast and deeply subcritical systems.
pulse-shape deconvolution and detector‑pair evaluation were found to be effective in mitigating distortions arising from the detector pulse-shape and electronics, further extending the usable operating range. The remaining limitations are dominated by properties of the data‑acquisition chain rather than by the methodology itself.
Ongoing improvements in the acquisition hardware and pulse-shape characterization are expected to further enhance the precision and applicability of the approach. Future work will focus on validating the method with upgraded electronics, exploring higher‑efficiency detectors, and extending the technique to the measurement of higher‑order kinetic modes.
\section*{Acknowledgements}
The project supported by the Doctoral Excellence Fellowship Programme (DCEP) is funded by the National Research Development and Innovation Fund of the Ministry of Culture and Innovation and the Budapest University of Technology and Economics.






\bibliographystyle{elsarticle-harv}
\bibliography{Bibliography.bib}
\end{document}